\documentclass{emulateapj}
\usepackage{amsmath}
\usepackage{subfigure}
\usepackage{color}

\usepackage{mathrsfs}
\usepackage{graphicx}
\usepackage{cases}
\usepackage{txfonts}
\usepackage{epsfig}
\usepackage{multirow}
\usepackage{dcolumn}
\usepackage{bm}
\usepackage{amssymb}
\usepackage{multirow}
\usepackage{enumitem}
\usepackage{enumerate}
\usepackage[colorlinks,linkcolor=blue,citecolor=blue,urlcolor=blue]{hyperref}
\graphicspath{{figure/}}
\usepackage{lineno}

\def\prd{Phys. Rev. D}

\def\prl{Phys. Rev. Lett.}

\def\jcap{JCAP}
\def\apj{Astrophys. J.}
\def\apjl{Astrophys. J. Lett.}
\def\araa{Annu. Rev. Astron. Astrophys.}
\def\mnras{Mon. Not. Roy. Astron. Soc.}
\def\apjs{Astrophys. J. Suppl. Ser.}

\def \nat{Nature}

\def\pasa{PASA}

\def\dif{\mathrm{d}}

\def\be{\begin{equation}}
\def\ee{\end{equation}}

\def\bea{\begin{eqnarray}}
\def\eea{\end{eqnarray}}
\newcommand{\bes}{\begin{equation*}}
\newcommand{\ees}{\end{equation*}}
\newcommand{\beqa}{\begin{eqnarray}}
\newcommand{\eeqa}{\end{eqnarray}}

\newcommand{\lsim}{\mathrel{\hbox{\rlap{\lower.55ex\hbox{$\sim$}} \kern-.3em \raise.4ex \hbox{$<$}}}}
\newcommand{\gsim}{\mathrel{\hbox{\rlap{\lower.55ex\hbox{$\sim$}} \kern-.3em \raise.4ex \hbox{$>$}}}}

\renewcommand{\arraystretch}{2.0}

\begin{document}

\title{Measuring the Hubble constant with coalescences of binary neutron star and neutron star-black hole: bright sirens \& dark sirens}

\author{
Jiming Yu$^{1,2,3,4\, \ast}$, 
Zhengyan Liu$^{3,4\, \dagger}$,
Xiaohu Yang$^{1,2,5}$, 
Yu Wang$^{3,4}$,
Pengjie Zhang$^{1,2,5}$,
Xin Zhang$^{6,7,8}$
and Wen Zhao$^{3,4\, \ddagger}$,
\\
$^{1}$\,Department of Astronomy, School of Physics and Astronomy, Shanghai Jiao Tong University, Shanghai 200240, China; $^\ast$\, jimingyu@sjtu.edu.cn\\
$^{2}$\,Key Laboratory for Particle Astrophysics and Cosmology (MOE) / Shanghai Key Laboratory for Particle Physics and Cosmology, China\\
$^{3}$\,CAS Key Laboratory for Researches in Galaxies and Cosmology, Department of Astronomy, University of Science and Technology of China, Chinese Academy of Sciences, Hefei 230026, China; $^\dagger$\,ustclzy@mail.ustc.edu.cn, $^\ddagger $\,wzhao7@ustc.edu.cn \\
$^{4}$\,School of Astronomy and Space Science, University of Science and Technology of China, Hefei 230026, China \\
$^{5}$\,Tsung-Dao Lee Institute, Shanghai Jiao Tong University, Shanghai 200240, China\\
$^{6}$\,Key Laboratory of Cosmology and Astrophysics (Liaoning Province) \& Department of Physics, College of Sciences, Northeastern University, Shenyang 110819, China\\
$^{7}$\,Key Laboratory of Data Analytics and Optimization for Smart Industry (Ministry of Education), Northeastern University, Shenyang 110819, China\\
$^{8}$\,National Frontiers Science Center for Industrial Intelligence and Systems Optimization, Northeastern University, Shenyang 110819, China
}

\begin{abstract}

{The observations of gravitational wave (GW) provide us a new probe to study the universe. GW events can be used as standard sirens if their redshifts are measured. Normally, stardard sirens can be divided into bright/dark sirens according to whether the redshifts are measured by electromagnetic (EM) counterpart observations. Firstly, we investigate the capability of the 2.5-meter Wide-Field Survey Telescope (WFST) to take follow-up observations of kilonova counterparts. For binary neutron star (BNS) bright sirens, WFST is expected to observe 10-20 kilonovae per year in the second-generation (2G) GW detection era. As for neutron star-black hole (NSBH) mergers, when a BH spin is extremely high and the NS is stiff, the observation rate is $\sim10$ per year.  Combining optical and GW observations, the bright sirens are expected to constrain the Hubble constant $H_0$ to $\sim2.8\%$ in five years of observations. As for dark sirens, tidal effects of neutron stars (NSs) during merging time provide us a cosmological model-independent approach to measure the redshifts of GW sources. Then we investigate the applications of tidal effects in redshift measurements. We find in 3G era, the host galaxy groups of around 45\% BNS mergers at $z<0.1$ can be identified through this method, if the EOS is ms1, which is roughly equivalent to the results from luminosity distant constraints. Therefore, tidal effect observations provide a reliable and cosmological model-independent method of identifying BNS mergers' host galaxy groups. Using this method, the BNS/NSBH dark sirens can constrain $H_0$ to 0.2\%/0.3\% over a five-year observation period.}

\end{abstract}

\section{Introduction}
\label{intro}
The advanced LIGO and Virgo collaborations' (LVC) detection of a number of compact binary mergers \citep{2016PhRvL.116f1102A, 2016PhRvL.116x1103A, 2016PhRvX...6d1015A, 2017PhRvL.118v1101A, 2017PhRvL.119n1101A, 2017PhRvL.119p1101A, 2017ApJ...848L..13A, 2019PhRvX...9c1040A, 2020ApJ...892L...3A, 2020ApJ...896L..44A, 2020PhRvL.125j1102A, 2020PhRvD.102d3015A, 2021PhRvX..11b1053A, 2021arXiv211103606T} directly confirms the existence of gravitational waves (GWs) and provides us with a new probe to study the universe. Since the observation of GW waveforms can measure the luminosity distance ($d_L$) of the GW sources directly \citep{1986Natur.323..310S}, when the redshifts of GW sources can be measured by other methods, such GW events can then be applied as standard sirens for studying the evolution of the universe \citep{2005ApJ...629...15H, 2010CQGra..27u5006S, 2011PhRvD..83b3005Z, 2020ApJ...889...79Y, 2020ApJS..250....6W}. On 2017 August 17, LVC observed a GW event produced by a binary neutron star (BNS) merger (GW170817; \citealt{2017PhRvL.119p1101A}). Soon afterwards, its electromagnetic (EM) counterparts, gamma-ray burst GRB 170817A \citep{2017ApJ...848L..14G, 2017ApJ...848L..15S} and kilonova emission AT2017gfo \citep{2017PASA...34...69A, 2017ApJ...848L..33A, 2017ApJ...848L..19C, 2017Sci...358.1556C, 2017ApJ...848L..17C, 2017ApJ...848L..29D, 2017Sci...358.1570D, 2017Sci...358.1565E, 2017SciBu..62.1433H, 2017Sci...358.1559K, 2017ApJ...850L...1L, 2017ApJ...848L..32M, 2017ApJ...848L..18N, 2017Natur.551...67P, 2017Sci...358.1574S, 2017Natur.551...75S, 2017ApJ...848L..16S, 2017PASJ...69..101U, 2017ApJ...848L..24V, 2017ApJ...848L..27T} were also observed. Combing the redshift information from its host galaxy NGC 4993, for the first time LVC took GW event as a bright standard siren and constrain the Hubble constant to $H_0=70_{-8.0}^{+12.0}\ \mathrm{km}\ \mathrm{s}^{-1}\ \mathrm{Mpc}^{-1}$ \citep{2017Natur.551...85A}. Besides AT2017gfo, during the second and third observing (O2/O3) runs of the advanced LIGO and advanced Virgo, many telescopes, including the Zwicky Transient Facility (ZTF; \citealt{2019PASP..131a8002B, 2019PASP..131g8001G}), the Dark Energy Camera (DECam; \citealt{2015AJ....150..150F}), etc., have made follow-up observations based on the early warning localization results of GW events \citep{2019ApJ...881L..16A, 2020ApJ...890..131A, 2020A&A...643A.113A, 2020ApJ...905..145K, 2021NatAs...5...46A, 2022ApJ...929..115T}. However, none of them have observed significant evidence of kilonova emission other than AT2017gfo.\par

In May 2023, LIGO/VIRGO and the Kamioka Gravitational Wave Detector (KAGRA) will start the fourth observing (O4) run with a higher detection sensitivity \citep{2018LRR....21....3A}. And in 2025, the fifth observing (O5) will begin, with the A+ upgrade and the addition of an intrument in India. At the same time, there will be a variety of advanced telescopes, such as the Large Synoptic Survey Telescope (LSST; \citealt{2019ApJ...873..111I}), in operation. These will significantly improve the detection capability of kilonova and broaden the application of bright sirens in cosmology. Among them, the 2.5-meter Wide-Field Survey Telescope (WFST), which is located at the Saishiteng Mountain in the Lenghu area ($93^\circ53'$ E, $38^\circ36'$ N), is scheduled to be installed in the early of 2023. It has a field-of-view (FOV) of 6.55 deg$^2$, $5\sigma$ limiting magnitudes of (22.40, 23.35, 22.95, 22.59, 21.64, 22.96) in the ($u,\ g,\ r,\ i,\ z,\ w)$ for a 30s exposure time \citep{2023arXiv230200246L}, and a median seeing of 0.75 arcseconds in the Lenghu site \citep{2021Natur.596..353D}. With these parameters, WFST is expected to be one of the major follow-up telescopes in the northern hemisphere during the O4 and O5 runs of the advanced LIGO, advanced Virgo and KAGRA. In the first half of this paper, we will discuss the capability of WFST to perform follow-up observations during the O4/O5 runs and foresee the application of the observed BNS/NSBH bright sirens in the Hubble constant constraints.\par
 
Multi-messenger observations are a very direct and effective method of measuring redshift. However, among the other GW events observed by LIGO/VIRGO so far, only GW190521/ZTF19abanrhr \citep{2020PhRvL.125j1102A, 2020PhRvL.124y1102G} is a probable multi-messenger event. For those `dark standard sirens', \cite{2008PhRvD..77d3512M} and \cite{2012PhRvD..86d3011D} proposed that their redshifts could be statistically obtained from the comparison of GW events' localization areas and the survey data. The ability of localizing GW events has been improved significantly with the operation of VIRGO, and this method has been widely discussed in recent years \citep{2017PhRvD..96j1303N, 2018PhRvD..98b3502N, 2018Natur.562..545C, 2019ApJ...871L..13F, 2019ApJ...876L...7S, 2019BAAS...51c.310P, 2020MNRAS.498.1786Y, 2020PhRvD.101l2001G, 2020ApJ...900L..33P, 2020ApJ...902..149V, 2020ApJ...905L..28B, 2020SciBu..65.1340Z, 2020JCAP...03..051J, 2022CoTPh..74j5404J, 2022arXiv220211882J, 2023arXiv230106722J, 2021ApJ...909..218A, 2021JCAP...08..026F, 2021arXiv211103604T, 2021arXiv211106445P, 2022SCPMA..6510411W, 2022arXiv221200531S}. In particular, this method will be most effective when the number of probable host galaxies, groups or clusters in the localization region is the only one. However, since the localization regions from GW detectors are related to $d_L$, in order to compare them with the survey catalog, these localization regions need to be converted to redshift coordinates under priors of specific cosmological models. Therefore, due to the indeterminacy in the $d_L-z$ relation, the existence of other galaxies/groups/clusters with different redshifts in the same sky region will have a significant impact on redshift statistics. \par

In order to overcome the reliance on cosmological models, a natural way is to measure redshifts directly from GW data. In GW waveforms, the redshift of a binary system is usually tightly coupled to its physical mass $m_\mathrm{phy}$. If $m_\mathrm{phy}$ could be measured, the coupling would break down and the redshift could also be measured. Neutron stars (NSs) under the effect of strong tidal fields will have a tidal deformation, which will have an effect on the waveform \citep{2011PhRvD..83h4051V, 2012PhRvD..85l4034B, 2013PhRvD..88b4046V, 2018PhRvD..98j4046A, 2018arXiv180501882L, 2020PhRvD.101f4003B}. Since the tidal deformability $\lambda$ is related to the equation of state (EOS) as well as the physical mass of the NS,  if the EOS is known, then the physical mass of the NS can be measured by tidal effect and thus the redshift can be determined \citep{2012PhRvL.108i1101M, 2017PhRvD..95d3502D, 2020ApJS..250....6W, 2021PhRvD.104h3528C}. \par

Currently, GW170817 has given a loose constraint on $\lambda$ \citep{2017PhRvL.119p1101A} and excluded some of the EOSs \citep{2018PhRvL.121p1101A}. In the future, with the third generation (3G) of GW detectors CE \citep{2015PhRvD..91h2001D} and ET \citep{2010CQGra..27s4002P} in operation, the accuracy of $\lambda$ measurements will improve significantly. Meanwhile, observations of EM counterparts such as kilonova \citep{2021MNRAS.505.3016N} and X-ray plateau \citep{2014PhRvD..89d7302L, 2016PhRvD..93d4065G, 2016PhRvD..94h3010L, 2020A&A...641A..56R, 2020JHEAp..27...33L, 2020PhRvL.125n1103B, 2021RAA....21...47L}, and the search for massive NSs in the Milky Way \citep{2013Sci...340..448A, 2018MNRAS.478.1377A, 2020NatAs...4...72C}, will impose strong constraints on the NS EOS. Thus, in the 3G era, tidal effect observations will be a powerful and cosmologically model-independent method of measuring the dark siren redshifts \citep{2020ApJS..250....6W}.\par

In our previous work \citep{2020MNRAS.498.1786Y}, we have discussed the ability of the GW detector network to identify stellar mass binary black hole (SBBH) mergers' host galaxy groups by comparing $d_L$ localization results with the SDSS DR7 halo-based group catalog \citep{2005MNRAS.358..217Y, 2007ApJ...671..153Y, 2009ApJS..182..543A}, and estimated the potential of these SBBH mergers to constrain the Hubble constant. Compared to galaxies, groups represent a larger physical structure and are therefore more easily identified. In addition, the use of group catalog can partially overcome the influence of peculiar velocity of the galaxies. For these two redshift measurement methods, we follow the calculations in \cite{2020MNRAS.498.1786Y} and  compare their ability of identifying the BNS/neutron star-black hole (NSBH) mergers' host groups and constraining the Hubble constant. To exclude the influence of external factors such as edge effect and incompleteness effect, we imply a mock galaxy group catalog based on the Uchuu simulation \citep{2021MNRAS.506.4210I} instead in this work, which covers the whole sky and has a much higher upper redshift limit.\par

Our paper is organized as follows. In Section \ref{section_method}, we introduce the GW waveform, the detector response and the Fisher information matrix method. We introduce the kilonova model used in the analysis of bright sirens in Section \ref{section_kilonova} and then estimate the application of bright sirens in the Hubble constant constraints in Section \ref{section_ls}. In Sections \ref{section_ds} and \ref{section_dsH0}, we analyze the dark sirens and their cosmological applications. And in Section \ref{section_conclusion}, we summarize our results.\par

Throughout the paper, we adopt the standard $\Lambda$CDM model with parameters from the Planck satellite's latest results, $\Omega_\mathrm{m}=0.3089$, $\Omega_\mathrm{b}=0.0486$, $\Omega_{\Lambda}=0.6911$, $h=0.6774$, $n_\mathrm{s}=0.9667$, and $\sigma_8=0.8159$ \citep{2020A&A...641A...6P}, as a fiducial model in the mock data simulation.

\section{GW detection}
\label{section_method}
\subsection{The detector's response}
In the paper, we mainly consider the array with $N_d$ GW detectors, whose spatial locations are denoted as $\bm{r}_I$ with $I=1, 2,\cdots,N_d$. For an incoming GW signal propagating along $\bm{n}$, the response of the $i$-th detector in the frequency domain can be written as
\begin{equation}
	d_I(f)=F^{+}_{I}h_{+}(f)+F^{\times}_{I}h_{\times}(f),
\end{equation}
where $h_{+}(t)$ and $h_{\times}(t)$ are two wave polarizations in the transverse traceless gauge, $F_I^+$ and $F_I^\times$ are two beam-pattern functions, which depend on the source's right ascension $\alpha$, declination, $\delta$, the polarization angle, $\psi$, the position and the orientation of the interferometer's arms. In this paper, for bright sirens, we discuss the network of LIGO (Livingston), LIGO (Handford) and Virgo, and the network of five 2G GW detectors with the addition of KAGRA and LIGO-India \citep{2013IJMPD..2241010U}. We denote these two networks as LHV and LHVKI, respectively. For dark sirens, we focus on a network of three 3G GW detectors, which consists of ET, CE and an assumed CE-type detector located in Australia. Their parameters and noise curves can be found in our pervious paper, \cite{2020MNRAS.498.1786Y} \footnote{The noise curve files can be found in \url{https://github.com/jimingyu-1996/GWnoisecurve.git}.}.  {For ET, we use the proposed ET-D project noise curves \citep{2010CQGra..27s4002P}}. For CE and assumed CE-type detector, we consider the proposed noise curve in \cite{2017CQGra..34d4001A} and \cite{2015PhRvD..91h2001D}. We denote this network as CE2ET.\par
In this paper, we adopt the restricted post-Newtonian (PN) waveform for the nonspinning systems in an inspiralling stage \citep{1993PhRvL..70.2984C, 2001PhLB..513..147D, 2001PhRvD..63f4038I, 2002LRR.....5....3B, 2002PhRvD..65f1501B, 2004PhRvD..69l4007B, 2004PhRvL..93i1101B, 2003PhRvD..68l1501I, 2004PhRvD..69f4018I, 2004CQGra..21S.529I, 2005PhRvD..71b4004B, 2009LRR....12....2S}. For a binary system with component masses $m_1$ and $m_2$, total mass $M=m_1+m_2$, mass ratio $\eta\equiv m_1 m_2/M^2$, chirp mass $M_c\equiv M\eta^{3/5}$, inclination angle $\iota$ and luminosity distance $d_L$, the waveform is given by
\bea
	F^{+}_{I}h_+(f)+F_I^\times h_\times(f)=&\mathcal{A}_{I}f^{-7/6}\exp(i[2\pi f t_c-\pi/4+\nonumber\\
	                                                    &2\varphi(f/2)-\phi_{I,(2,0)}+\Phi_\mathrm{tidal}(f)]),
\eea       
where $t_c$ is the merging time. The amplitude $\mathcal{A}_I$ is given by
\be
	\mathcal{A}_I=\frac{1}{d_L}\sqrt{(F_{I}^{+}(1+\cos^2\iota))^2+(2F_{I}^{\times}\cos\iota)^2}\sqrt{\frac{5\pi}{96}}\pi^{-7/6}\mathcal{M}_c^{5/6},
\ee
where $\mathcal{M}_c\equiv(1+z)M_c$ is the ``redshift chirp mass". Meanwhile, after defining $\mathcal{M}=(1+z)M$, the functions $\varphi$ and $\phi_{I,(2,0)}$ are given by
\be
	\varphi(f)=-\varphi_c+\frac{3}{256\eta}\sum_{i=0}^{7}\varphi_i(2\pi\mathcal{M} f)^{i/3},
\ee
\\
\be
	\phi_{I,(2,0)}=\tan^{-1}\left(-\frac{2\cos\iota F_{I}\times}{(1+2\cos^2\iota)F_{I}^{+}}\right),
\ee
where we use the 3.5 PN approximation for the phase, and $\varphi_c$ is the merging phase. The parameters $\varphi_i$ can be found in \cite{2009LRR....12....2S}. For a BNS merger, the tidal deformation term $\Phi_\mathrm{tidal}(f)$ is given by \citep{2011PhRvD..83h4051V, 2012PhRvL.108i1101M}
\be
\begin{aligned}
\Phi_{\mathrm{tidal}}(f) &=\sum_{a=1}^{2} \frac{3 \lambda_{a}}{128 \eta M^{5}}\left[-\frac{24}{\chi_{a}}\left(1+\frac{11 \eta}{\chi_{a}}\right)(\pi \mathcal{M} f)^{5 / 3}\right.\\
&\left.-\frac{5}{28 \chi_{a}}\left(3179-919 \chi_{a}-2286 \chi_{a}^{2}+260 \chi_{a}^{3}\right)(\pi \mathcal{M} f)^{7 / 3}\right],
\end{aligned}
\label{tidal_phase}
\ee
where $\chi_a\equiv m_a/(m_1+m_2)$, and $\lambda_a(m_a)$ characterizes the changes of the induced quadrupole with an external tidal field. In this paper, we use the $\lambda-m$ relation under the linear approximation mentioned in \cite{2020ApJS..250....6W},
\be
	\lambda=B m+C,
\ee
where $B$ and $C$ are two parameters determined by the EOS of NSs. Following \cite{2020ApJS..250....6W}, we discuss a sample of 13 NS EOSs, alf2, ap3, ap4, bsk21, eng, H4, mpa1, ms1, ms1b, qmf40, qmf60, sly, wff2, and 4 quark star EOSs, MIT2, MIT2cfl, pQCD800, sqm3 in this paper \citep{2016ARA&A..54..401O, 2018ApJ...862...98Z, 2018PhRvD..97h3015Z, 2021ChPhC..45e5104X}. The corresponding fitted values of B and C can be found in Table 2 of \cite{2020ApJS..250....6W}. The maximum stable NS mass, $M_\mathrm{TOV}$ (Tolman–Oppenheimer–Volkoff mass) and the radius of a $1.4\ M_\odot$ NS, $R_{1.4}$ can also be found in this paper. For a NSBH merger, the tidal deformation term $\Phi_\mathrm{tidal}(f)$ can be obtained by simply setting $\lambda_1=0$ in Eq. (\ref{tidal_phase}), where $a=1$ represents the BH component.\par
From the PN waveforms above, we can see that in all terms except $\Phi_\mathrm{tidal}(f)$ , redshift $z$ and mass $M$ are tightly coupled through the redshift mass $\mathcal{M}=(1+z)M$. With the addition of the tidal deformation term, the coupling between redshift and mass will break down, and we can obtain the redshift of GW source from the phase observations. \par
Therefore, for a BNS/NSBH merger, the response of the interferometer depends on twelve parameters, $\bm{\theta}=\{\alpha,\delta, \psi,\iota, M, \eta, t_c, \varphi_c, \log(d_L), z, B, C\}$. Using the method in \cite{2010PhRvD..81h2001W}, we can obtain the Fisher matrix $\Gamma_{ij}$ and covariance matrix $(\Gamma^{-1})_{ij}$ for these 12 parameters. The signal-to-noise ratio (SNR) of the GW signal can also be obtained from the GW detector's response (see \citealt{2018PhRvD..97f4031Z} for details of the Fisher matrix and SNR). In this work, for dark sirens we choose SNR $>12$ as the threshold for GW detection. For the bright sirens, we relax this threshold to SNR $>8$ due to the existence of EM counterparts.

\subsection{Samples of GW events}
In this paper, we adopt the BNS mergers' redshift distribution model in our previous work \cite{2021ApJ...916...54Y}, with the star formation rate (SFR) model from \cite{2008ApJ...683L...5Y}, the log-normal time delay model from \cite{2015MNRAS.448.3026W}, and the local merger rates updated to GWTC-3, $R_\mathrm{BNSmergers,0}=13-1900\ \mathrm{Gpc}^{-3}\ \mathrm{yr}^{-1}$, and $R_\mathrm{NSBHmergers,0}=7.4-320\ \mathrm{Gpc}^{-3}\ \mathrm{yr}^{-1}$\citep{2021arXiv211103634T}. Since there is a large difference between the upper and lower bounds of the merger rates of GWTC-3, we use the geometric mean of the upper and lower bounds, $R=\sqrt{R_\mathrm{low}R_\mathrm{high}}$ in our calculations to simplify the analysis, which are $R_\mathrm{BNSmergers,0}=157.2\ \mathrm{Gpc}^{-3}\ \mathrm{yr}^{-1}$ and $R_\mathrm{NSBHmergers,0}=48.7\ \mathrm{Gpc}^{-3}\ \mathrm{yr}^{-1}$.  {After adopting these two local merger rates, the total numbers of $z<0.2$ BNS and NSBH mergers per year in the whole sky $N_\mathrm{yr}$ are about 600 and 180, respectively.} These two numbers are roughly in line with the predictions in \cite{2022ApJ...941..208I}.\par
We assume a uniform distribution of NS masses between $1.2\ M_\odot$ and $2.0\ M_\odot$. This assumption is consistent with the observational constraints of GWTC-3 \citep{2021arXiv211103634T}, and keeps the mass of the sample below the maximum stable NS mass $M_\mathrm{TOV}$ of the EOS mentioned in \cite{2020ApJS..250....6W} (except for sqm3, $M_\mathrm{TOV}=1.99\ M_\odot$). For BH, we follow \cite{2021arXiv211103634T} and use a power-law mass distribution model in the interval $[2.5\ M_\odot,50\ M_\odot]$ with an index of -3.4, supplemented by a Gaussian peak at $34_{-3.0}^{+4.0}\ M_\odot$. The distribution of inclination angles $\iota$ is $\propto\sin\iota$. All of the BNS and NSBH mergers in our simulation are nonspinning If no special instructions.

\section{Kilonova Model}
\label{section_kilonova}
During the merger of BNS and part of NSBH binaries, neutron rich ejecta will be ejected through tidal interactions \citep{1999A&A...341..499R, 2015PhRvD..91f4059S}, material squeezed at the contact interface \citep{2007A&A...467..395O, 2013ApJ...773...78B, 2013PhRvD..87b4001H} and disk outflows \citep{2008ApJ...679L.117S, 2008ApJ...676.1130M, 2019ApJ...886L..30N}. Rapid neutron capture ($r$-process) nucleosynthesis will happen in these ejecta and produce heavy elements \citep{1982ApL....22..143S} and the decay of these heavy elements will power a rapidly evolving, roughly isotropic thermal transient named as `kilonova' \citep{2010MNRAS.406.2650M}. Since kilonovae have a much larger viewing angle compared to GRBs, they are believed to be the main EM counterpart search targets for low redshift BNS and NSBH mergers.\par
POSSIS \citep{2019MNRAS.489.5037B} is a three-dimensional Monte Carlo code for predicting spectra and light curves of supernovae and kilonovae. With this code, \cite{2020Sci...370.1450D} and \cite{2021NatAs...5...46A} provide kilonova simulation grids based on a three-component BNS model and a two-component NSBH model, repectively. In this section, we use POSSIS code to obtain the kilonova light curves.\par

\subsection{Kilonova emission from BNS mergers}
For the three-component BNS model in \cite{2020Sci...370.1450D}, the first and second components are ejected dynamically through tidal interactions \citep{1999A&A...341..499R, 2015PhRvD..91f4059S} and material squeezed at the contact interface \citep{2007A&A...467..395O, 2013ApJ...773...78B, 2013PhRvD..87b4001H}, respectively. The former component is neutron rich and concentrated near the equatorial plane, with a high opacity and named as `red' component. The second component has a much lower neutron abundance and opacity due to the $e^\pm$ captures and neutrino irradiation \citep{2015PhRvD..91f4059S, 2016MNRAS.460.3255R}, and named as `blue' component. The third component comes from the post-merger disk outflows \citep{2008ApJ...679L.117S, 2008ApJ...676.1130M, 2019ApJ...886L..30N}. Its velocity is much slower than dynamical ejecta, and its opacity lies between the first two components. After giving different values of opacities to the three components and characterizing the velocity with ejecta mass, \cite{2020Sci...370.1450D} simulate Spectral Energy Distributions (SEDs) and light curves with four free parameters, which are the mass of the dynamical ejecta $M_\mathrm{dyn}$, the mass of post-merger disk-wind ejecta $M_\mathrm{wind}$, the half-opening angle of the tidal dynamical ejecta $\Phi$, and the viewing angle $\theta_\mathrm{obs}$. \par
For $M_\mathrm{dyn}$, we use the fitting formula from \cite{2019MNRAS.489L..91C}, 
\begin{equation}
    \log_{10}M_\mathrm{dyn}=\left [a\frac{(1-2C_1)m_1}{C_1}+b\ m_2\left( \frac{m_1}{m_2}\right )^n + \frac{d}{2}\right ]+[1\leftrightarrow 2],
\end{equation}
with fitting parameters $a=-0.0719,\ b=0.2116,\ d=-2.42,\ n=-2.905$, and 
\begin{equation}
    C_i\equiv Gm_i/r_i
\end{equation}
is the 
$C_i$ is the compactness of NS with radius $r_i$. The term $[1\leftrightarrow2]$ represents exchanging the subscripts. \par
When binary masses are unequal, the less massive NS will be disrupted by the tidal forces before the collision, which will suppress the production of shocks \citep{2013PhRvL.111m1101B, 2013PhRvD..87b4001H, 2016CQGra..33r4002L, 2017CQGra..34j5014D}. Here we use the simulations by \cite{2016PhRvD..93l4046S} to estimate the fractions of the red and blue components. \cite{2021MNRAS.505.3016N} presented a polynomial fit to these simulation results. For BNS mergers with $q\gtrsim0.8$, the fit can be written as
\begin{equation}
    f_\mathrm{red}=\min([1,a\ q^2 + b\ q +c]),
\end{equation}
with fitting parameters $a=14.8609,\ b=-28.6148,\ c=13.9597$, and for mergers with $q\lesssim0.8$, $f_\mathrm{red}\approx1$. In the POSSIS model, the half-opening angle of the red component $\pm\Phi$ is left as a free parameter. Here we assume that the volume ratio of the red and blue components is proportional to their mass ratio. Therefore, after using the estimations of AT2017gfo, $\Phi=49.50^\circ$ \citep{2020Sci...370.1450D} as a calibration, $\Phi$ can be expressed in terms of $f_\mathrm{red}$.\par
For disk mass, we adopt the fitting formula from \cite{2020Sci...370.1450D},
\begin{equation}
    \log_{10}\left (\frac{M_\mathrm{disk}}{M_\odot}\right )=\max\left[-3,a\left\{1+b\tanh\left(\frac{c-M/M_\mathrm{thr}}{d}\right)\right\}\right],
\label{m_disk}
\end{equation}
where $M_\mathrm{thr}$ is the threshold mass of a NS prompt collapsing into a BH, which can be parameterised as 
\begin{equation}
    M_\mathrm{thr}=\left(2.38-3.606\frac{M_\mathrm{TOV}}{R_{1.4}}\right)M_\mathrm{TOV},
\end{equation}
$a$ and $b$ in Eq. (\ref{m_disk}) are given by

\begin{equation}
\begin{split}
    a&=a_o+\delta a\cdot\zeta,\\
    b&=b_o+\delta b\cdot\zeta,
\end{split}
\end{equation}

where 
\begin{equation}
    \zeta=\frac{1}{2}\tanh(\beta(q-q_\mathrm{trans})),
\end{equation}
the best fitting values of parameters in Eq. (\ref{m_disk}) can be found in \cite{2020Sci...370.1450D}, which are $a_o=-1.581$, $\delta a=-2.439$, $b_o=-0.538$, $\delta b=-0.406$, $c=0.953$, $d=0.0417$, $\beta=3.910$, $q_\mathrm{trans}=0.900$. Approximately 10-50\% of the disc mass will be ejected by viscously-driven winds \citep{2018ApJ...858...52S, 2018ApJ...869..130R}, we choose a typical fraction, $\epsilon_\mathrm{disk}\approx0.2$ in this work.\par

For a BNS merger and a given NS EOS, we can then use the above equations to calculate its $M_\mathrm{dyn}$, $M_\mathrm{wind}$ and $\Phi$. With these parameters and the values of $\theta_\mathrm{obs}$ and $d_L$, we obtain the light curves by interpolating the grids in \cite{2020Sci...370.1450D}.

\subsection{Kilonova emission from NSBH mergers}
For NSBH systems, their BH components differ greatly from the properties of NSs during merger. So they will produce kilonova emissions that are quite different compared to BNS mergers. Firstly, NSs only produce ejecta when they are tidally disrupted outside the radius of the innermost stable circular orbit (ISCO) $R_\mathrm{ISCO}$, or they will be swallowed whole. The normalized ISCO radius, $\tilde{R}_\mathrm{ISCO}=R_\mathrm{ISCO}/M_\mathrm{BH}$ is given by \citep{1972ApJ...178..347B}
\begin{equation}
    \tilde{R}_\mathrm{ISCO}=3+Z_2-\mathrm{sign(\chi_\mathrm{BH})}\sqrt{(3-Z_1)(3+Z_1+2Z_2)},
\label{eq_isco}
\end{equation}

with
\begin{equation}
\begin{split}
    &Z_1=1+(1-\chi^2_\mathrm{BH})^{1/3}[(1+\chi_\mathrm{BH})^{1/3}+(1-\chi_\mathrm{BH})^{1/3}],\\
    &Z_2=\sqrt{3\chi^2_\mathrm{BH}+{Z_1}^2},
\end{split}
\end{equation}
where $M_\mathrm{BH}$ and $\chi_\mathrm{BH}$ are the mass and the dimensionless spin parameter of a BH, respectively. The tidal disrupted radius of the NS in Newtonian theory is approximately equal to \citep{2020ApJ...897...20Z}
\begin{equation}
    R_\mathrm{tidal}\sim R_\mathrm{NS}\left(\frac{3M_\mathrm{BH}}{M_\mathrm{NS}}\right)^{1/3},
\end{equation}
where $R_\mathrm{NS}$ and $M_\mathrm{NS}$ are the radius and mass of a NS. It can be seen that the significant kilonova emission occurs only when the BH has a small mass and a large spin.\par
Secondly, the dynamical ejecta are primarily produced by tidal disruption and concentrated around the equatorial plane for NSBH mergers\citep{2015PhRvD..92b4014K}, so \cite{2021NatAs...5...46A} consider only two components in their NSBH's kilonova model and the value of $\Phi$ is fixed at $30^\circ$. \par
For the masses of dynamical ejecta from NSBH mergers, we adopt the \cite{2020ApJ...897...20Z}'s formulation based on 66 numerical relativistic simulation results,
\begin{equation}
\begin{split}
    \frac{M_\mathrm{dyn}}{M^\mathrm{b}_\mathrm{NS}}=&\max\left [ \left(0.273\frac{1-2C_\mathrm{NS}}{\eta^{1/3}}\right.\right.\\
    &\left.\left.-0.034\tilde{R}_\mathrm{ISCO}\frac{C_\mathrm{NS}}{\eta}-0.153\right )^{1.491},0\right],
\end{split}
\end{equation}
where $C_\mathrm{NS}$ is the compactness of the NS, and we apply the fitting results of \cite{2020FrPhy..1524603G} for the NS's baryonic mass $M^\mathrm{b}_\mathrm{NS}$ in this work,
\begin{equation}
    M^\mathrm{b}_\mathrm{NS}=M_\mathrm{NS}+0.08 M_\mathrm{NS}^2/M_\odot.
\end{equation}

\cite{2018PhRvD..98h1501F} presented a fit of the remnant baryon mass outside of the BH after a NSBH  merger based on 75 simulations, 
\begin{equation}
\begin{split}
    \frac{M_\mathrm{rem}}{M^\mathrm{b}_\mathrm{NS}}=&\max\left [ \left( 0.406\frac{1-2C_\mathrm{NS}}{\eta^{1/3}}\right.\right.\\
    &\left.\left.-0.139\tilde{R}_\mathrm{ISCO}\frac{C_\mathrm{NS}}{\eta}+0.255\right)^{1.761},0\right],
\end{split}
\end{equation}
The disc mass $M_\mathrm{disk}$ can therefore be obtained by 
\begin{equation}
    M_\mathrm{disk}=M_\mathrm{rem}-M_\mathrm{dyn}.
\end{equation}
Then, similar to the BNS mergers, the kilonova light curves generated by NSBH merging can be obtained by interpolating the grids from \cite{2021NatAs...5...46A}.

\section{bright Sirens}
\label{section_ls}
In our previous work \cite{2021ApJ...916...54Y}, we have discussed the applications of GW-GRB multi-messenger observations in constraining the EOS of dark energy. In this section, we continue the discussion of the bright sirens in cosmology. For kilonovae, which are much less luminous than GRBs but have much larger observable angles, they will be the main EM counterpart search targets for low redshift BNS and NSBH mergers. Due to limitations in the detection capabilities of GW detectors and telescopes, in this section we simulate 10000 low redshift ($z<0.2$) BNS and NSBH mergers, respectively, and calculate the strength of their GW and kilonova emissions, as well as the probability of being detected by multi-messenger. Subsequently, by multiplying the normalization factor $N_\mathrm{yr}/10000$, we calculate the multi-messenger detection rates of the BNS and NSBH mergers and their potentials in the Hubble constant constraints.

\subsection{Kilonova searching}
For each BNS and NSBH merger, we calculate the SNR of its GW signal, derive its covariance matrix $\mathbf{Cov}[\alpha,\ \delta,\ \log(d_L)]$ through the Fisher matrix method, and assume that its actual location is normally distributed. For each GW event with SNR $>8$, we pixelate its localization area with the HEALPix pixelization algorithm \citep{2005ApJ...622..759G, 2019JOSS....4.1298Z}. Through the HEALPix code, the whole sky is divided into $12\times$nside$^2$ pixels, we choose nside $=512$ and obtain the probability of the source lying at each pixel from the covariance matrix. \par
In the follow-up observations of GW events during the O2 and O3 of the advanced LIGO and advanced Virgo, many telescopes, including ZTF, DECam, etc., have designed their observation plans based on the early warning localization results of GW events, and prioritized the observation of high probability sky region to improve the efficiency of follow-up observations. In this work, we consider the effect of the telescope's observation plan, and generate a WFST observation plan for each simulated merger by combining the results of GW source localization and other properties of the event. Then we estimate the multi-messenger observation rate by combining the observation plan of each event.\par
To generate the observation plan, we use the Python-based package $gwemopt$, which was first developed by \cite{2018MNRAS.478..692C} to optimize the search for EM counterparts of the GW events, and has been refined in the subsequent observations \citep{2021MNRAS.504.2822A}. In O3, many follow-up observation plans are developed with the help of $gwemopt$ for both single-telescope observations and joint observations with multiple telescopes (e.g., \citealt{2019ApJ...885L..19C, 2020ApJ...905..145K, 2020MNRAS.497.5518A, 2021NatAs...5...46A, 2022ApJ...926..152F}). Given a position of a telescope, $gwemopt$ takes into account the influence from the Sun, the Moon and the Milky Way disk. In our analysis, we adopt the following settings: i) it is night when the sun's altitude is below $-15^\circ$; ii) the maximum airmass for telescopic observations is 2.0; iii) the telescope only observes sky areas with galactic latitude $|b|>15^\circ$. In addition, to reduce the influence of the Moon, we subtract the observation area near the Moon, the size of which is determined by the lunar phase.\par
$gwemopt$ provides various algorithms in the following computational processes, i) skymap tiling, ii) time allocations and iii) scheduling. \cite{2018MNRAS.478..692C} discussed in detail the efficiency of various combinations of these algorithms, among which the combination of MOC (multiorder coverage) algorithm, power law algorithm and greedy algorithm is the most efficient, so we choose this combination of algorithms in the simulations. The details of these algorithms can be found in \cite{2018MNRAS.478..692C}. \par
For each GW event, we set the observation window to three days after merging, during which the telescope will repeatedly cover the localization area. At the end of the $gwemopt$ running time, the code will output a list with the coordinates of each exposure, the observation time, and the probability of observing a kilonova in the $i$-th tile $P^{i}_\mathrm{kn}$. For the $j$-th observation in the $i$-th tile, its detection depth $d^{i,\ j}_{L,\ \mathrm{obs}}$ can be obtained from the absolute magnitude of the kilonova and the limiting magnitude of the telescope. Then $P^{i}_\mathrm{kn}$ can be obtained from 
\begin{equation}
    P^i_\mathrm{kn}=P^{i}_\mathrm{sky}\times \mathrm{CDF}^{i}(d^{i}_{L,\ \mathrm{obs}}),
\end{equation}
where $P^{i}_\mathrm{sky}$ is the probability that the GW source is located at the $i$-th tile, $\mathrm{CDF}^{i}(d^{i}_{d_L})$ is the CDF of kilonova's luminosity distance at the $i$-th tile which can be obtained from the Fisher matrix, $d^{i}_{L,\ \mathrm{obs}}$ is the maximum value of $d^{i,\ j}_{L,\ \mathrm{obs}}$. Finally, the probability of this kilonova's observation is
\begin{equation}
    P_\mathrm{kn}=\sum_i^{n} P^{i}_\mathrm{kn}.
\label{eq_p_kn}
\end{equation}

\subsection{BNS mergers}
For the LHV and LHVKI networks, there are about 7.4\% and 15.1\% $z<0.2$ BNS can be detected, respectively. In the left panels of Figure \ref{figure_bns_kilonova}, we show the distributions of these observable BNS mergers' redshifts, $\Delta \Omega$ and $\Delta\ln(d_L)$. For LHV, the observed BNS mergers are mainly distributed around $z\sim0.08$, and for LHVKI, $z\sim0.11$. The typical localization areas given by LHV are about 40 deg$^2$ (90\% CL). With a longer baseline and higher detection SNR, the typical localization areas of the LHVKI array has improved to $\sim$10 deg$^2$. \cite{2018LRR....21....3A} predicted that during the O4, the median 90\% credible region for localization area of BNS is about $33\ \mathrm{deg}^2$. The LHV results of ours are very close to them. After normalization, the detection numbers of LHV and LHVKI are 44 and 91 in one year of full-operation period, respectively.\par

However, a recent work, \cite{2022ApJ...924...54P}, estimated this value with a data-driven method. They found that with the improvement of data analysis methods in O3, many signals that were only triggered by a single detector or had a low SNR were also identified, which on the one hand increased the number of detected events, but on the other hand also decreased the average localization accuracy. In their estimation, the median 90\% credible region for localization area of BNS is 1820 deg$^2$ for O4, which is far greater than our results as well as those of \cite{2018LRR....21....3A}. Thus our simulation may lose these `bad-localized' events. However, on the orther hand, they also found these method improvements would basically not change the rate of the well-localized events. Considering that the EM counterparts of those bad located events are difficult to be searched, we neglect their effects in this work and leave them for future work.\par

For each BNS merger with SNR $>8$, we draw its kilonova light curve and calculate the value of $P_\mathrm{kn}$ through the observation introduced above. The sums represent the expectation of the number of kilonova that WFST can detect through follow-up observations of BNS mergers. Table \ref{kilonova_psum1}, \ref{kilonova_psum2} and \ref{kilonova_psum3} show the sum of $P_\mathrm{kn}$ which normalized to one year's observation, with different EOSs, telescope's bands and exposure times. Here we list the results with the ms1 and wff2 models, which are the stiffest and softest EOSs mentioned in this paper. In addition, we pick the wff2 model as a comparison. As for observation bands, since the light curves for kilonovae in the $u$ band change very rapidly and are less likely to be searched, WFST has a poor observation capability in the $z$ band, we only consider single-band observations in the $g$, $r$, and $i$ bands, and multi-band observations in the $g$ and $r$/$i$ bands in this work. Since the mass of the ejecta is related to the EOS of the NS, the stiffer EOS model ms1 corresponds to a lower observation probability, which are about 9.9 and 14.3 per year for the LHV and LHVKI networks, respectively, if the exposure time is 30s in the multi-band observations. The results are approximately the same for the bsk21 and wff2 models.  \par

In the left panel of Figure \ref{figure_bns_band}, we compare the difference between the $g$ band and $i$ band $P_\mathrm{kn}$ in the case of bsk21 model, 30s exposure time and the LHV network. For some of the kilonovae, the detection probability in the $g$ band is significantly lower than in the $i$ band, due to the light curves change more rapidly in the $g$ band. Finally, the overall observation rates in the $g$ band appear to be $\sim20\%$ and $\sim 30\%$ lower than that in the $i$ band for the LHV and LHVKI arrays, respectively. \par

Generally, a longer exposure times allow the telescope to have a larger observation depth, but on the other hand this will also decrease the area of the sky covered by the telescope when searching for kilonovae. Therefore, we also consider the results with two different exposure times, 60s and 90s. For these three cases, their readout time between each exposure is all set to 10 s. In the left panel of Figure \ref{figure_bns_exp}, we compare the probability of each kilonova being observed with 30s and 90s exposure times. It can be seen that for most BNS mergers, increasing the telescope's single exposure time to 90s can help the search for kilonovae. In Table \ref{kilonova_psum1}, \ref{kilonova_psum2} and \ref{kilonova_psum3}, we compare the expectation of detection rates when using different exposure times. For the LHV network, using exposure times of 60s and 90s in multi-band observations will improve detection rates by $\sim8\%$ and $\sim10\%$, respectively. As for the LHVKI array, it can be seen from Figure \ref{figure_bns_kilonova} that the improvement in the localization capability of the GW network will be significantly greater than the $d_L$ constraints. Therefore, the use of long exposure times facilitates the telescope to cover more of the observation volume. At this time, 60s and 90s exposure times improve the observation rate by $\sim15\%$ and $\sim20\%$, respectively, compared to 30s exposure time.\par

\subsection{NSBH mergers}

\begin{figure*}[htbp]
\centering
\subfigure{
	\includegraphics[width=8.5cm]{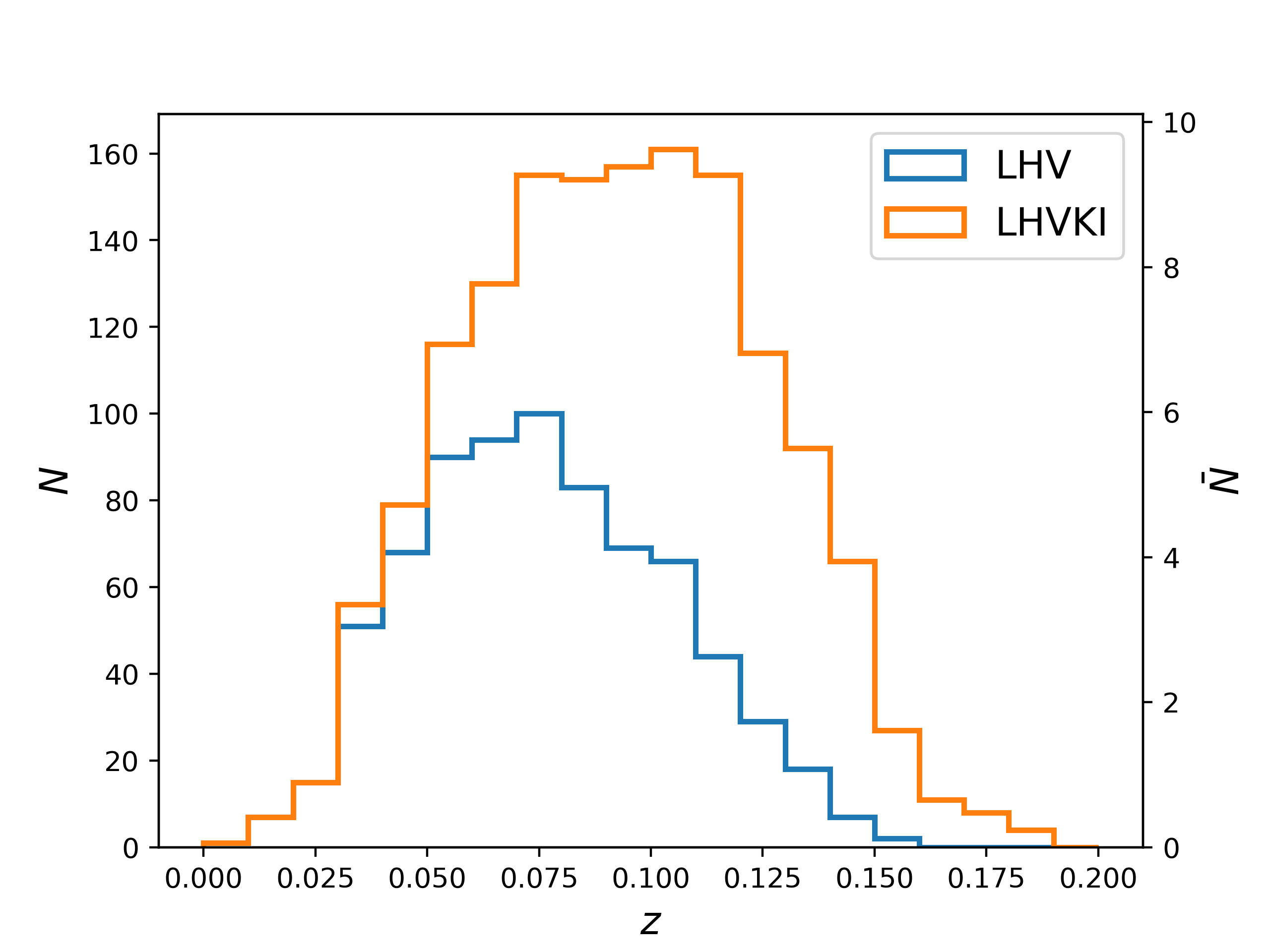}
}
\hspace{1.0pt}
\subfigure{
	\includegraphics[width=8.5cm]{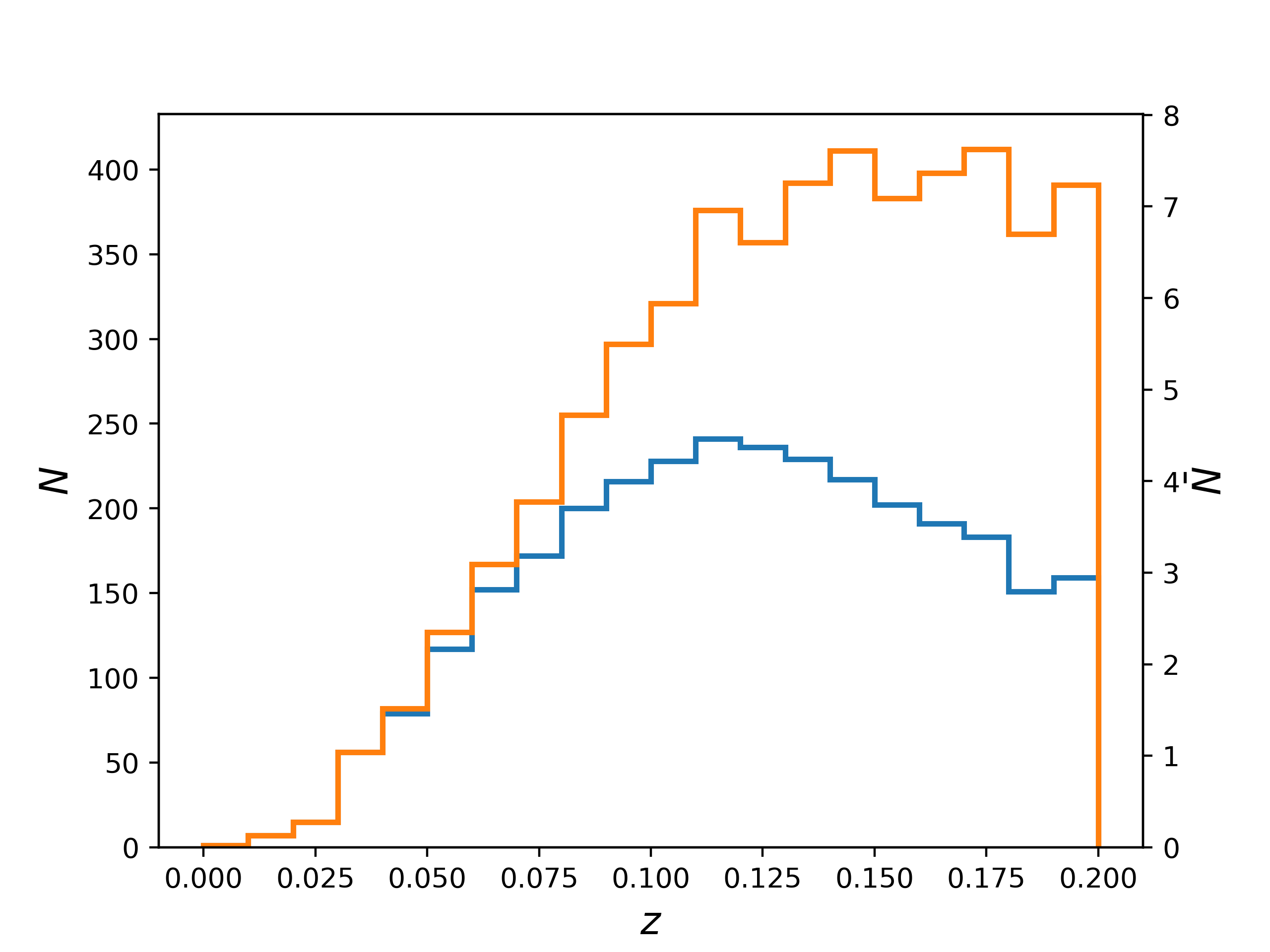}
}
\hspace{1.0pt}
\subfigure{
	\includegraphics[width=8.5cm]{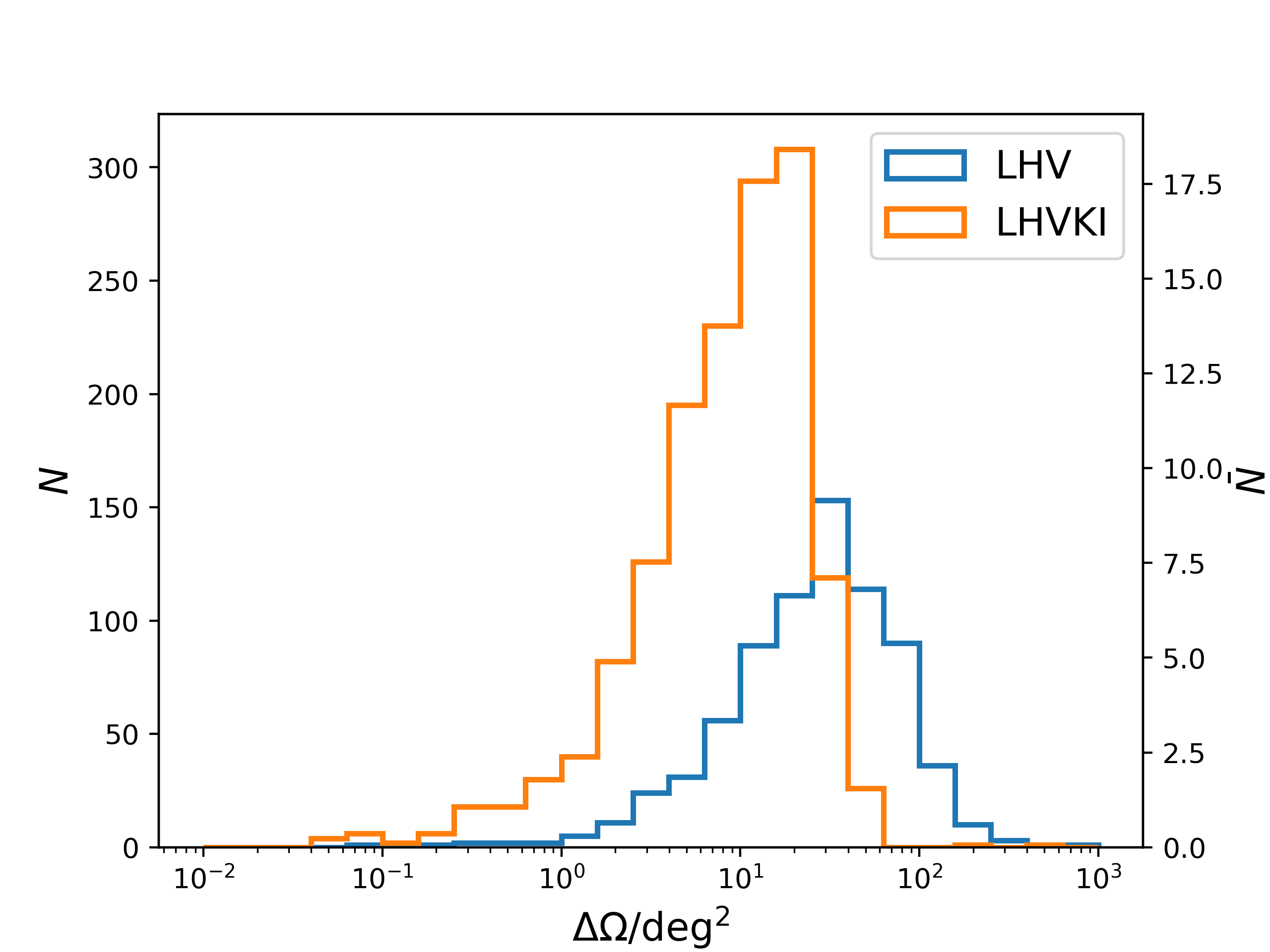}
}
\hspace{1.0pt}
\subfigure{
	\includegraphics[width=8.5cm]{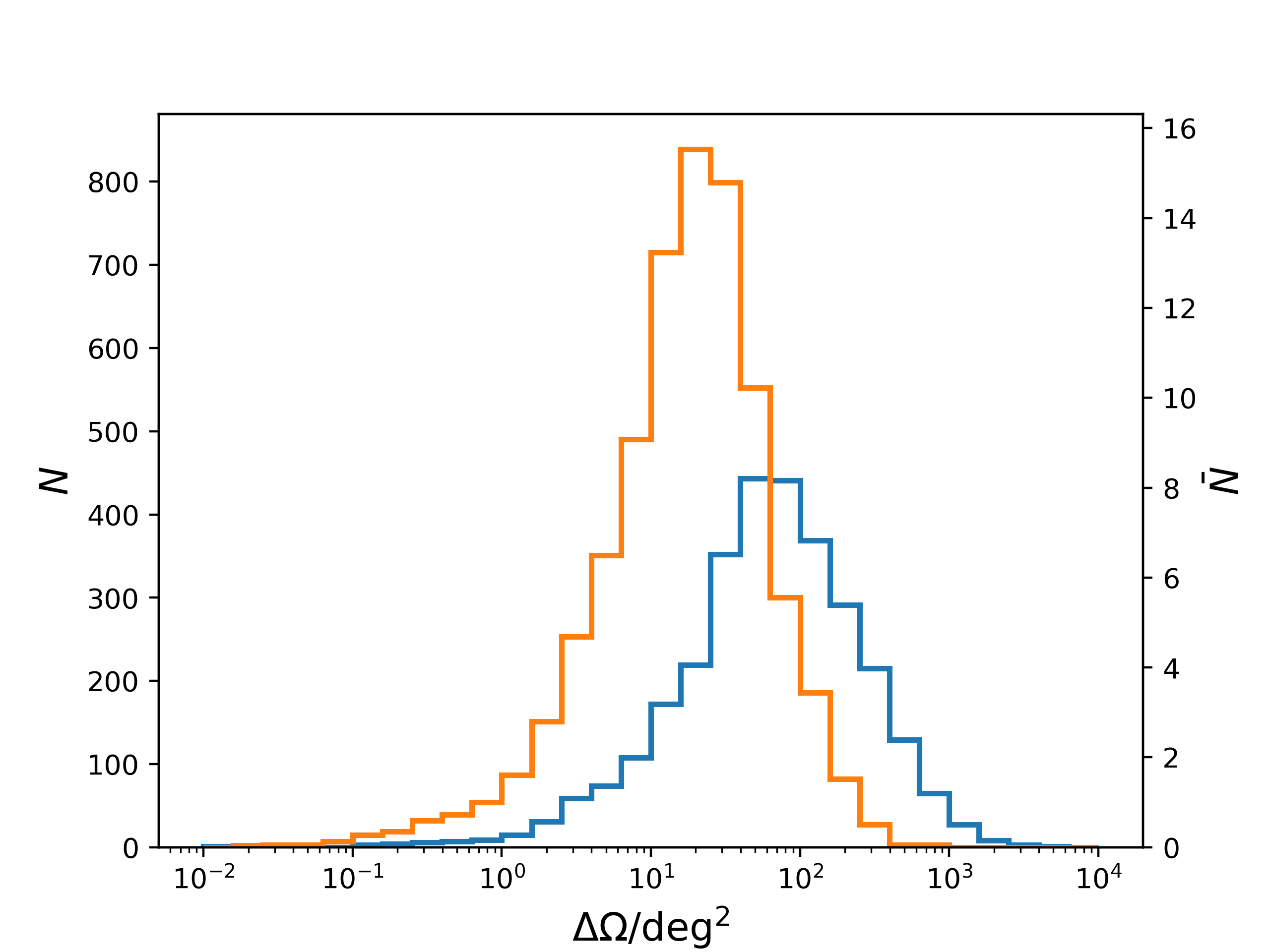}
}
\subfigure{
	\includegraphics[width=8.5cm]{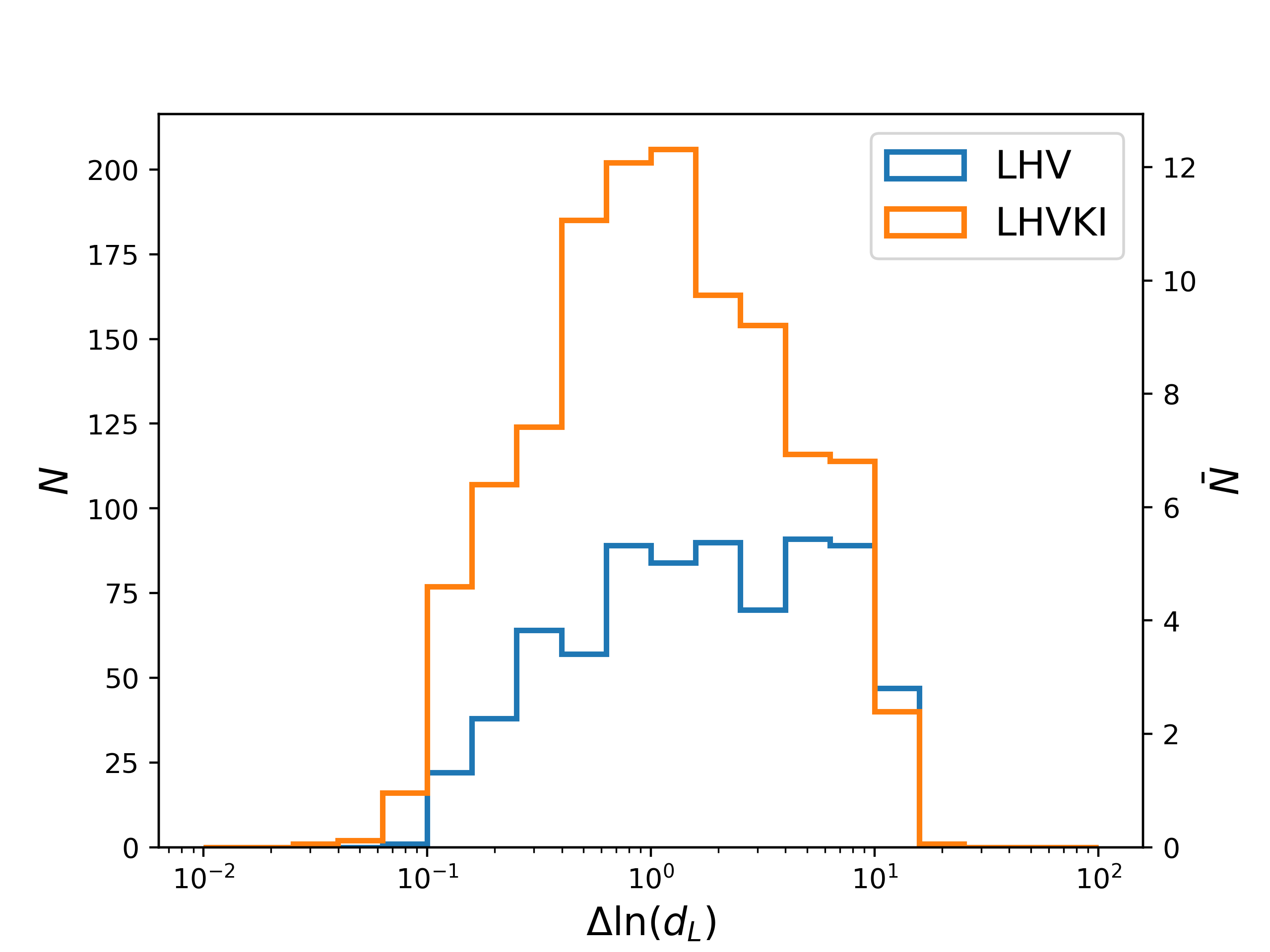}
}
\hspace{1.0pt}
\subfigure{
	\includegraphics[width=8.5cm]{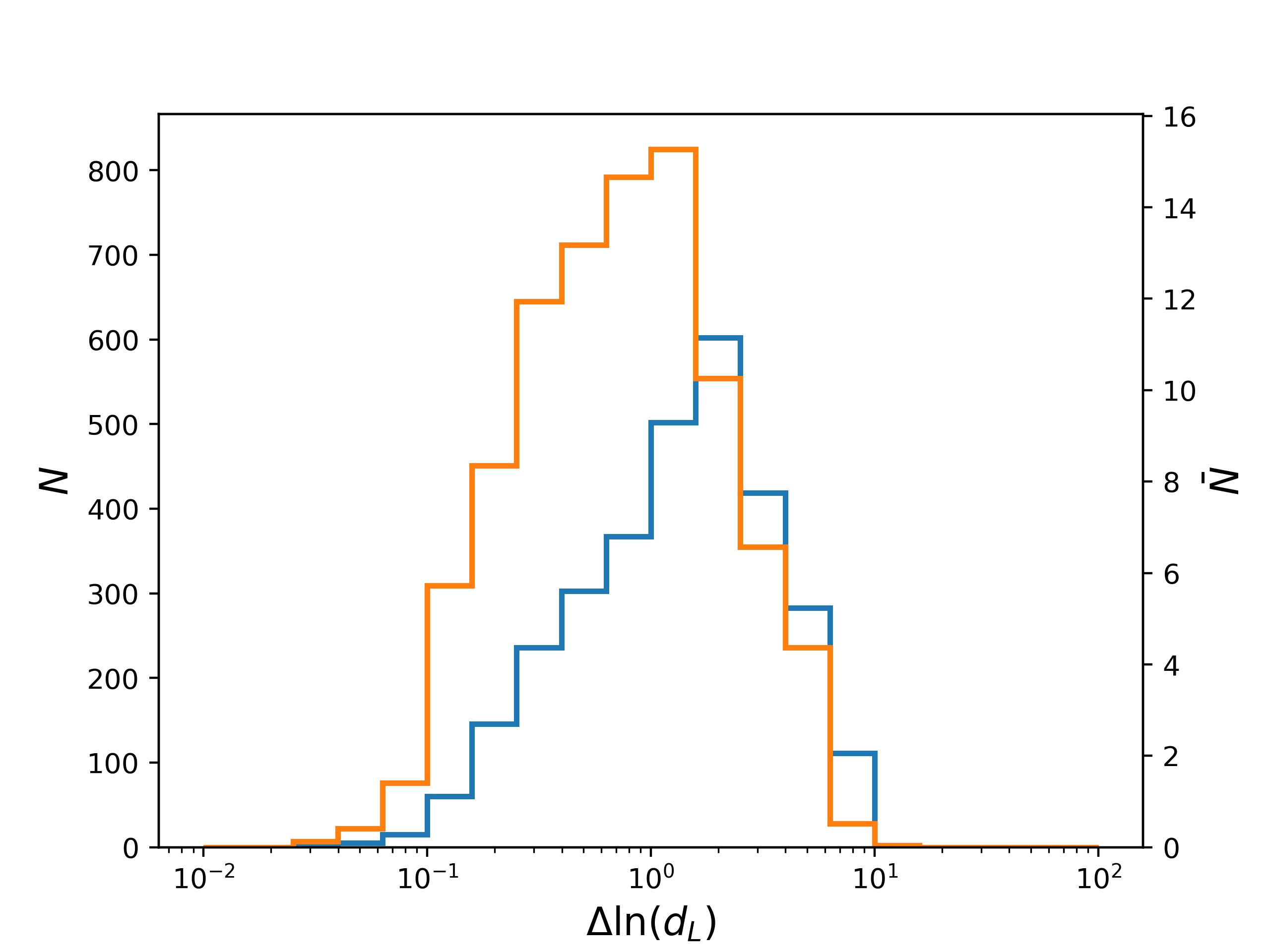}
}
\caption{The distributions of the redshifts, $\Delta\Omega$ with 90\% CL and $\Delta\ln(d_L)$ for BNS and NSBH mergers with SNR$>8$. The left and right panels show the distributions of BNS and NSBH mergers, respectively.  {In each panel, the left y-axis represents the number distributions of 10000 samples obtained from the full populations of BNS/NSBH mergers, and the right y-axis $\bar{N}$ represents the distributions normalized to one-year's observation time.}}
\label{figure_bns_kilonova}
\end{figure*}

Compared with BNS, NSBH will produce more powerful GW signals when it merges because of its larger mass. Therefore, for $z<0.2$ NSBH mergers, about 30.5\% and 50\% of them can be detected by LHV and LHVKI network, respectively. We show the distributions of these observable NSBH mergers' redshifts, $\Delta \Omega$ and $\Delta\ln(d_L)$ in the right panels of Figure \ref{figure_bns_kilonova}. Since most of the observable NSBH mergers are at $z>0.1$, the typical localization area given by LHV and LHVKI at 90\% CL are $\sim 60$ deg$^2$ and $\sim 20$ deg$^2$, which are a bit larger than the results of BNS mergers. The normalized detection numbers of LHV and LHVKI per year are 55 and 90, respectively.\par

According to Eq. (\ref{eq_isco}), the spin of the BH will have a significant effect on the ISCO radius of the BH and further affect the matter ejection during NSBH merging. When $\chi_\mathrm{BH}$ deviates from 0, NSBH mergers will likely eject more matter and generate brighter kilonova emission. In general, the spins of binary mergers are related to the angular momentum transport in their stellar progenitors \citep{2020A&A...636A.104B, 2019MNRAS.485.3661F, 2019ApJ...881L...1F}, the environment where binary formed, and the tides and mass transfer during the orbit \citep{2018PhRvD..98h4036G, 2019ApJ...870L..18Q, 2020A&A...635A..97B}. The results of GWTC-3 suggested that the effective spins of BH $\chi_\mathrm{eff}$ distribute around a very small value \citep{2021arXiv211103634T}. However, if the binaries are located in the AGN accretion disk, they can gain spins by the accretion of surrounding gas\citep{2019PhRvL.123r1101Y, 2020ApJ...894..129S, 2022ApJ...928...75H} and the observations of BH X-ray binaries suggest the existence of high spin BHs \citep{2011CQGra..28k4009M, 2011ApJ...731L...5M, 2019ApJ...870L..18Q}. Thus, we follow the \cite{2021ApJ...917...24Z} and discuss the spin distribution for two special cases, $\chi_\mathrm{BH}\sim\mathcal{N}(0, 0.15^2)$ and $\chi_\mathrm{BH}\sim\mathcal{N}(0.85, 0.15^2)$ for the low-spin case and high-spin case, respectively. It is worth to mention that in the previous calculation we used the waveform for the nospinning system. Although the spin of the BH affects the waveform, it has no significant effect on the localization of the GW source (see \citealt{2017PhRvL.119p1101A} for example), so for the high-spin NSBH mergers, we used the same error bar as in the low-spin case for convenience.\par

In the low-spin case, we find that NSBH mergers can hardly produce strong kilonova emission. This situation is particularly evident in the soft EOS, for wff2 model, no NSBH merger can produce observable kilonovae in our simulations. For the stiffer EOS ms1, the multi-messenger observations rate for NSBH mergers is only $\sim1/15$ compared with BNS mergers.\par

And in the high-spin case, the NSBH mergers have greater probabilities to produce stronger kilonova emissions due to a smaller ISCO radius. For the stiffer model ms1, about 7 and 11 NSBH mergers per year can be multi-messenger observed for the LHV and LHVKI networks, respectively. Since the merging rate of NSBH is only about 1/3 that of BNS, these results are approximately 30\% worse than BNS. In addition to $\chi_\mathrm{BH}$, the EOS of the NS also has a large impact on the detection rate. In the case of bsk21 model, the observation rates are $\sim4$ and $\sim7.2$ per year for two networks. For the wff2 model, the detection rate of NBSH mergers is only $\sim1/5$ as much as the detection rate of BNS.\par

\begin{figure*}[htbp]
\centering
\subfigure{
	\includegraphics[width=8.5cm]{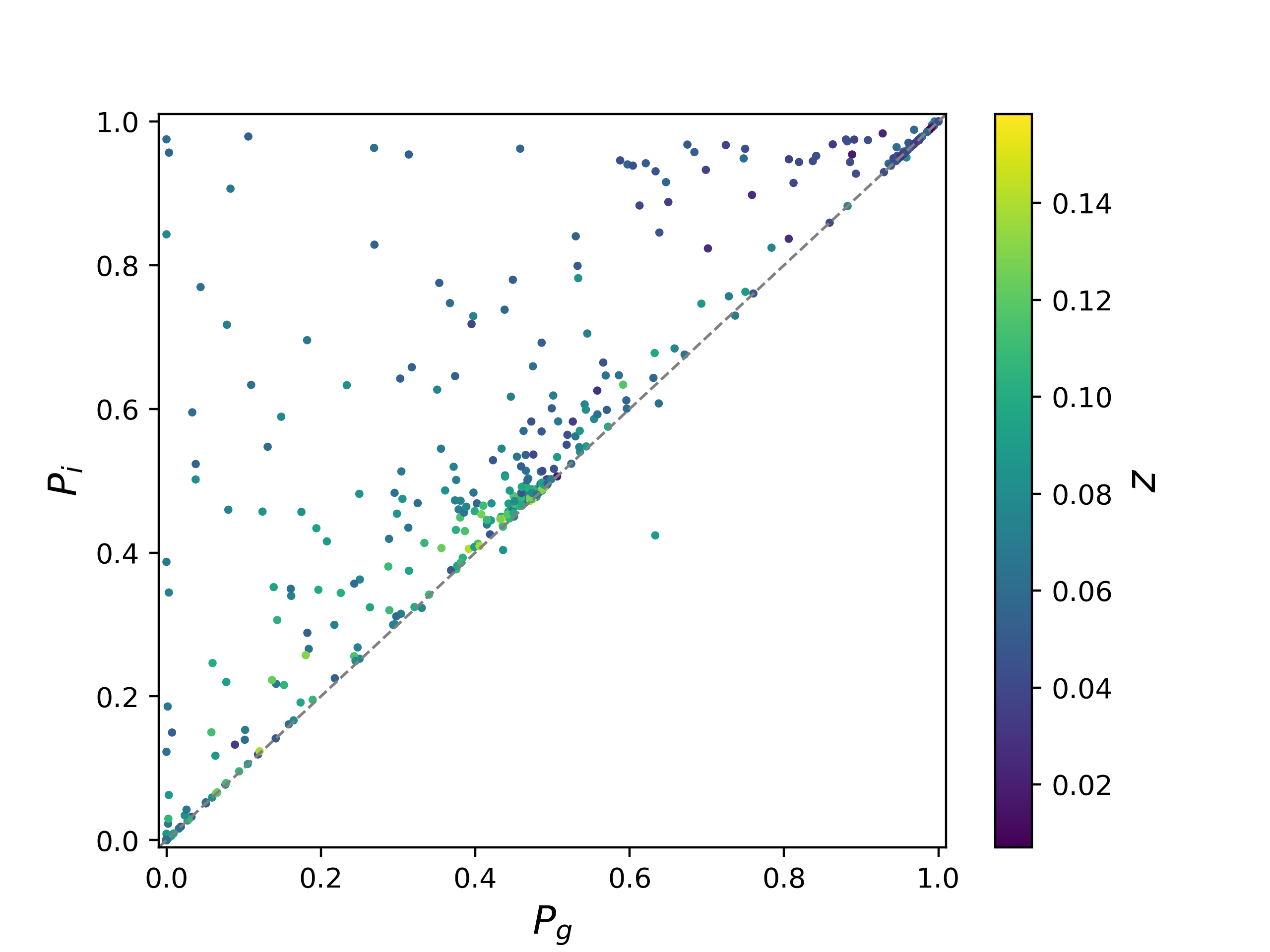}
}
\hspace{1.0pt}
\subfigure{
	\includegraphics[width=8.5cm]{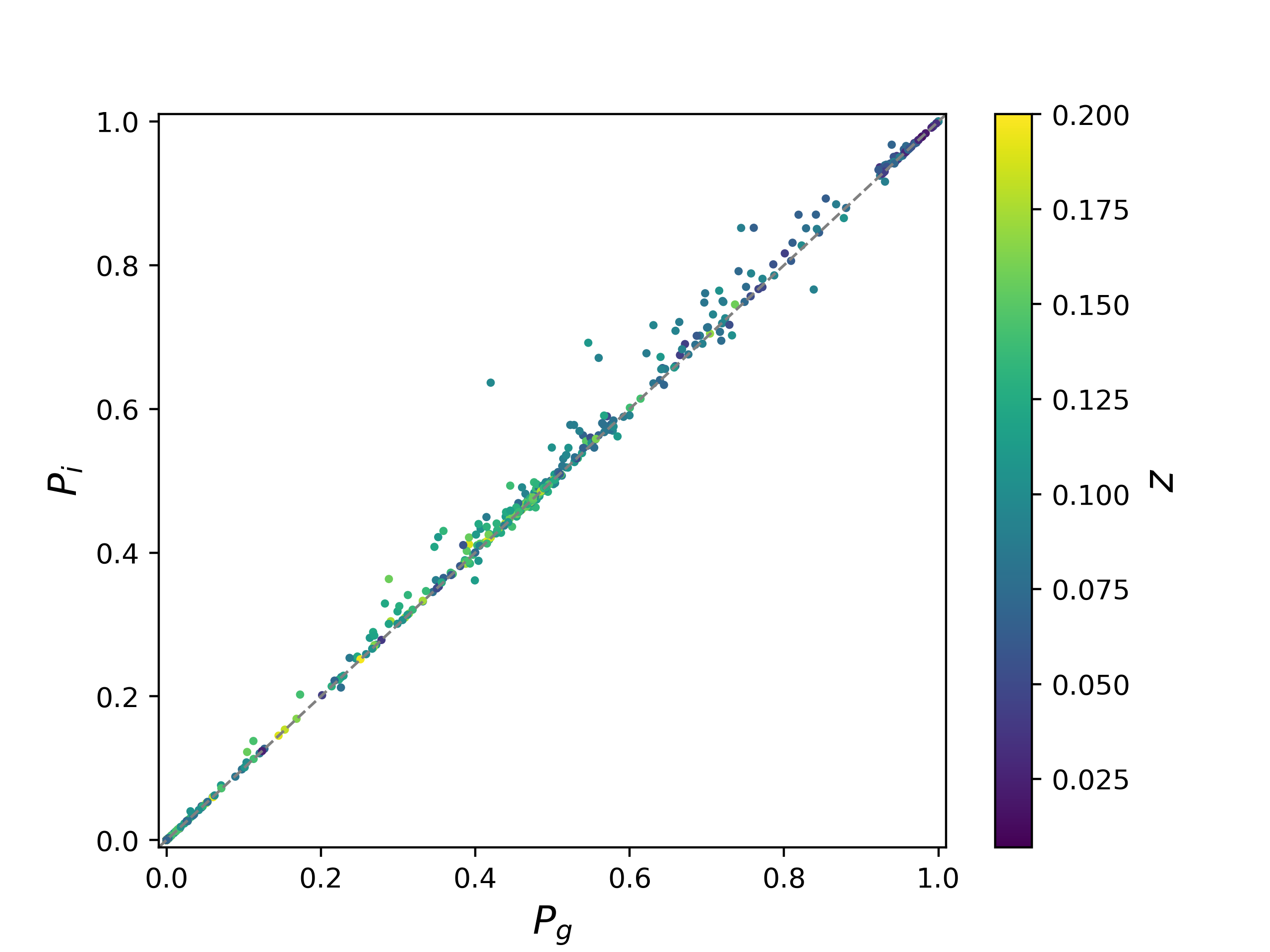}
}
\caption{Comparison of the probabilities of observing kilonova between the $g$ band and $i$ band with the bsk21 model and 30s exposure time. The left and right panels show the distributions of BNS and NSBH mergers, respectively.}
\label{figure_bns_band}
\end{figure*}

Since the kilonova emission from NSBH mergers is mainly contributed by the `red' component,
the light curves change slowly in all bands. Although the luminosity in $g$ band is lower, WFST has a deeper detection depth in this band, so the detection probability of the kilonova emission from NSBH mergers in the $g$ band and $i$ band is approximately the same as seen in the right panel of the Figure \ref{figure_bns_band}. \par

\begin{figure*}[htbp]
\centering
\subfigure{
	\includegraphics[width=8.5cm]{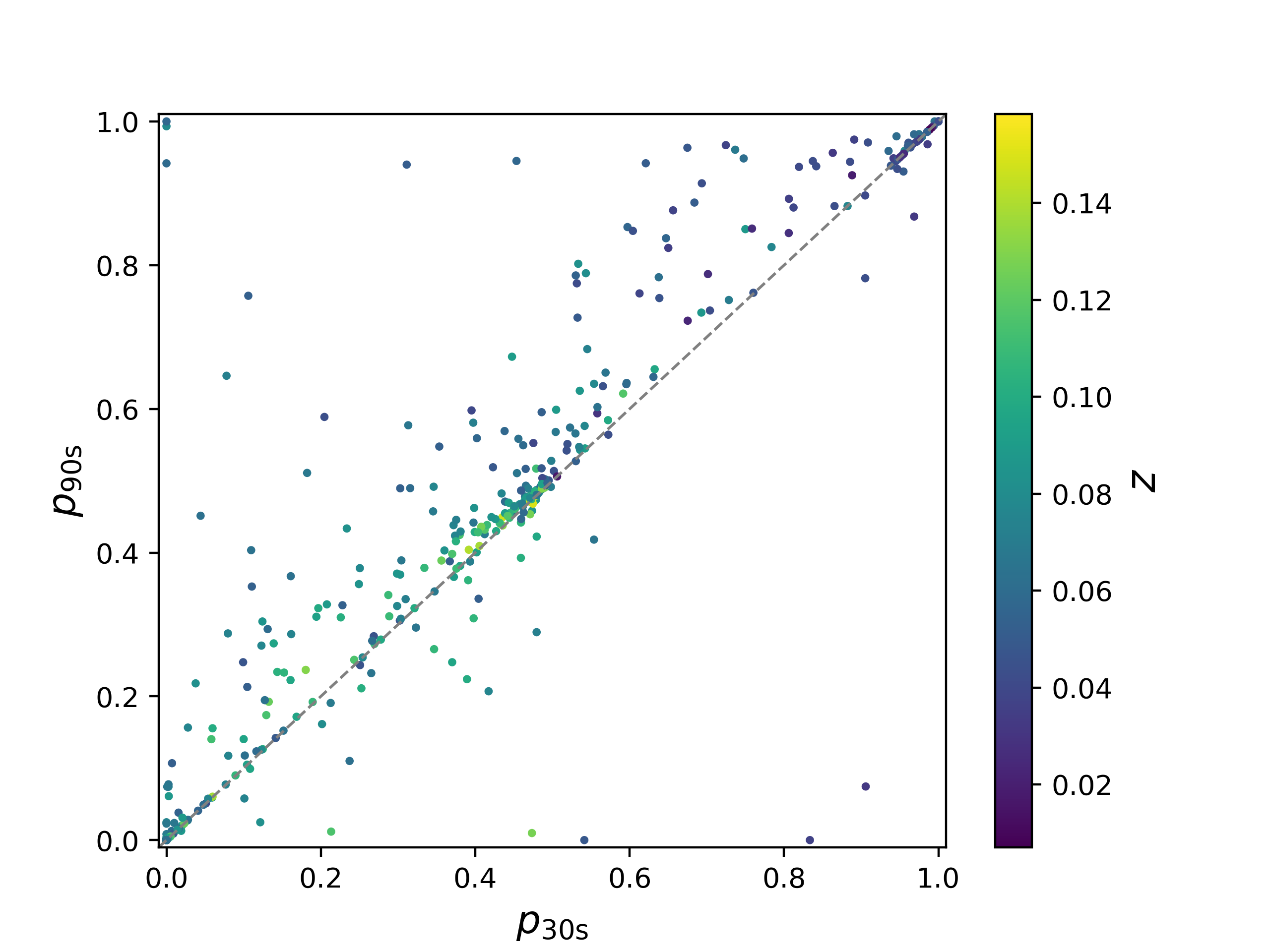}
}
\hspace{1.0pt}
\subfigure{
	\includegraphics[width=8.5cm]{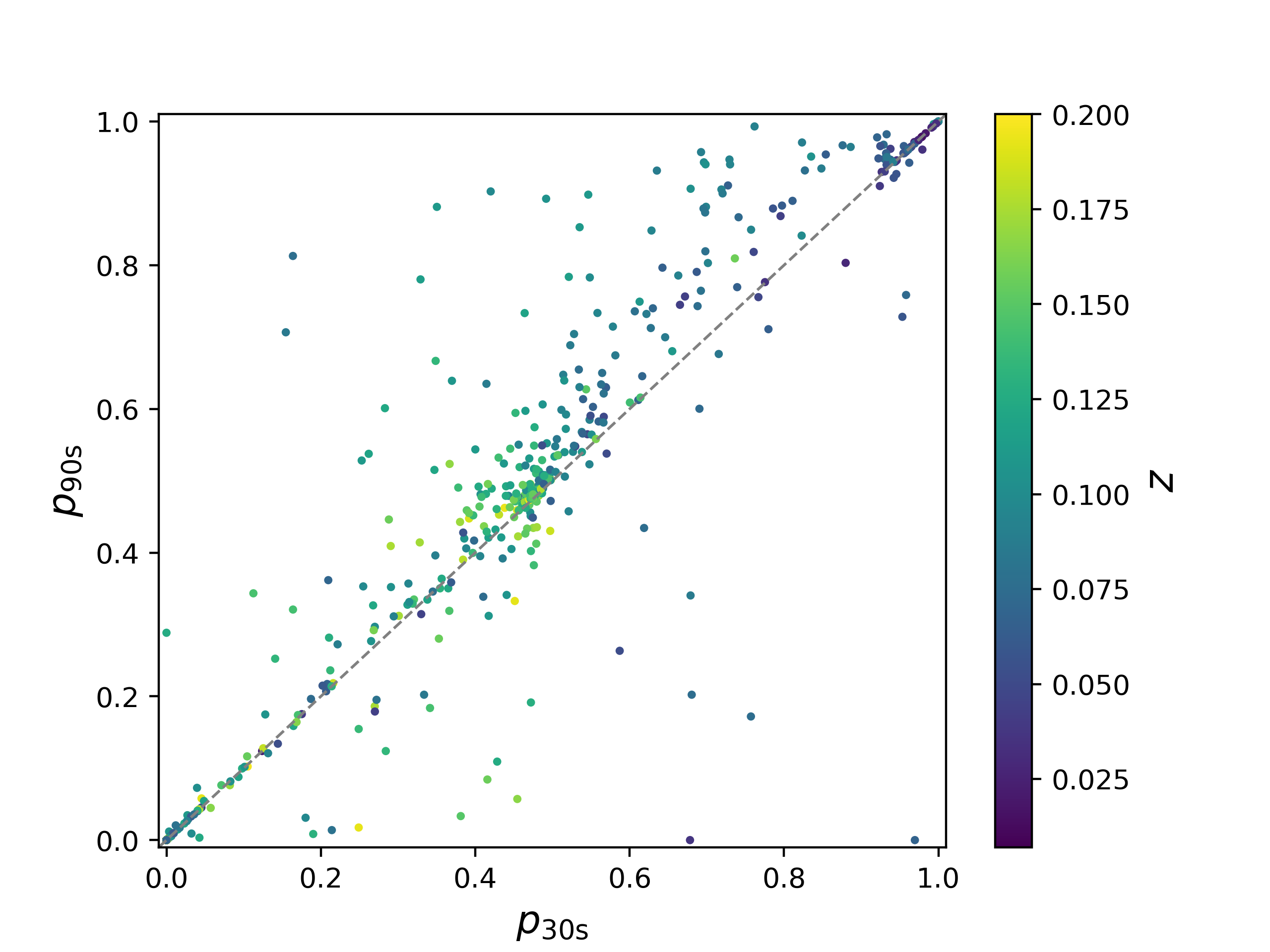}
}
\caption{Comparison of the probabilities of observing kilonovae with 30s and 90s exposure times in the case of bsk21 model and $g\&r$ multi-band searching. The left and right panels show the distributions of BNS and NSBH mergers, respectively.}
\label{figure_bns_exp}
\end{figure*}

For exposure time, due to a larger distribution for NSBH mergers' $\Delta\Omega$, it can be seen from the Figure \ref{figure_bns_exp} that, although the 90s exposure time still has a higher detection rate on the whole, more NSBH mergers have $p_\mathrm{30s}>p_\mathrm{90s}$ compared with BNS systems. This suggests that for NSBH mergers, a shorter exposure time of 30s needs to be more favourably considered.

\begin{table}
\renewcommand\arraystretch{1.5} 
\begin{center}
\begin{tabular}{cccccccc}
\hline\hline
&&&$g$&$r$&$i$&$g\&i$&$g\&r$\\
\hline
BNS&LHV&30s& 10.9& 12.9& 13.5& 9.9&9.9\\
&&60s& 12.0& 13.6& 14.2& 10.8& 10.8\\
&&90s& 12.5& 14.0& 14.5& 11.1 & 11.1\\
&LHVKI&30s& 15.8& 21.0& 23.2& 14.3& 14.3\\
&&60s& 18.4& 23.7& 25.6& 16.6& 16.6\\
&&90s& 19.8& 24.9& 26.6& 17.8& 17.8\\
\hline
NSBH, low $\chi_\mathrm{BH}$&LHV&30s& 0.70& 0.71& 0.72& 0.66& 0.66\\
&&60s& 0.76& 0.77& 0.78& 0.70& 0.70\\
&&90s& 0.79& 0.80& 0.81& 0.72& 0.71\\
&LHVKI&30s& 1.26& 1.29& 1.34& 1.19& 1.19\\
&&60s& 1.48& 1.51& 1.55& 1.40& 1.40\\
&&90s& 1.57& 1.59& 1.62& 1.49& 1.48\\
\hline
NSBH, high $\chi_\mathrm{BH}$&LHV&30s& 7.5& 7.5& 7.6& 6.9& 6.9\\
&&60s& 8.1& 8.0& 8.1& 7.3& 7.3\\
&&90s& 8.3& 8.2& 8.3& 7.4& 7.4\\
&LHVKI&30s& 11.9& 11.7& 12.0& 10.9& 10.8\\
&&60s& 13.6& 13.5& 13.8& 12.6& 12.5\\
&&90s& 14.3& 14.2& 14.4& 13.2& 13.2\\
\hline
\end{tabular}
\end{center}
\caption{Multi-messenger observation rates of WFST at different bands for kilonovae per year when the NS EOS is ms1 model.}
\label{kilonova_psum1}
\end{table}

\begin{table}
\renewcommand\arraystretch{1.5} 
\begin{center}
\begin{tabular}{cccccccc}
\hline\hline
&&&$g$&$r$&$i$&$g\&i$&$g\&r$\\
\hline
BNS&LHV&30s& 11.3& 12.9& 13.4& 10.3&10.3\\
&&60s& 12.3& 13.6& 14.0& 11.1& 11.1\\
&&90s& 12.8& 13.9& 14.3& 11.3 & 11.4\\
&LHVKI&30s& 16.7& 21.1& 22.9& 15.2& 15.2\\
&&60s& 19.0& 23.6& 25.0& 17.2& 17.2\\
&&90s& 20.5& 24.6& 26.0& 18.4& 18.4\\
\hline
NSBH, low $\chi_\mathrm{BH}$&LHV&30s& 0.04& 0.04& 0.04& 0.04& 0.04\\
&&60s& 0.04& 0.04& 0.05& 0.04& 0.04\\
&&90s& 0.05& 0.05& 0.05& 0.05& 0.05\\
&LHVKI&30s& 0.07& 0.07& 0.07& 0.07& 0.07\\
&&60s& 0.07& 0.07& 0.07& 0.07& 0.07\\
&&90s& 0.07& 0.07& 0.07& 0.07& 0.07\\
\hline
NSBH, high $\chi_\mathrm{BH}$&LHV&30s& 4.7& 4.6& 4.7& 4.3& 4.3\\
&&60s& 5.0& 5.0& 5.0& 4.6& 4.5\\
&&90s& 5.1& 5.1& 5.1& 4.6& 4.6\\
&LHVKI&30s& 7.7& 7.6& 7.8& 7.2& 7.1\\
&&60s& 8.7& 8.7& 8.9& 8.1& 8.1\\
&&90s& 9.2& 9.1& 9.3& 8.5& 8.5\\
\hline
\end{tabular}
\end{center}
\caption{Multi-messenger observation rates of WFST at different bands for kilonovae per year when the NS EOS is bsk21 model.}
\label{kilonova_psum2}
\end{table}

\begin{table}
\renewcommand\arraystretch{1.5} 
\begin{center}
\begin{tabular}{cccccccc}
\hline\hline
&&&$g$&$r$&$i$&$g\&i$&$g\&r$\\
\hline
BNS&LHV&30s& 11.3& 12.7& 13.1& 10.3&10.3\\
&&60s& 12.2& 13.4& 13.8& 11.0& 11.1\\
&&90s& 12.7& 13.7& 14.1& 11.3 & 11.4\\
&LHVKI&30s& 16.6& 20.4& 22.0& 15.2& 15.2\\
&&60s& 18.8& 22.7& 24.2& 17.1& 17.1\\
&&90s& 20.4& 23.8& 25.2& 18.4& 18.6\\
\hline
NSBH, low $\chi_\mathrm{BH}$&LHV&30s& 0& 0& 0& 0& 0\\
&&60s& 0& 0& 0& 0& 0\\
&&90s& 0& 0& 0& 0& 0\\
&LHVKI&30s& 0& 0& 0& 0& 0\\
&&60s& 0& 0& 0& 0& 0\\
&&90s& 0& 0& 0& 0& 0\\
\hline
NSBH, high $\chi_\mathrm{BH}$&LHV&30s& 2.2& 2.2& 2.3& 2.1& 2.1\\
&&60s& 2.4& 2.4& 2.4& 2.2& 2.2\\
&&90s& 2.5& 2.5& 2.5& 2.2& 2.2\\
&LHVKI&30s& 4.0& 4.0& 4.1& 3.8& 3.7\\
&&60s& 4.6& 4.6& 4.7& 4.3& 4.3\\
&&90s& 4.8& 4.8& 4.8& 4.5& 4.5\\
\hline
\end{tabular}
\end{center}
\caption{Multi-messenger observation rates of WFST at different bands for kilonovae per year when the NS EOS is wff2 model.}
\label{kilonova_psum3}
\end{table}

\subsection{The applications in $H_0$ constraints}
Assuming that the other parameters of the $\Lambda$CDM model are determined, according to the propagation of uncertainty formula \citep{2003psa..book.....W}, the measurement error of the Hubble constant can be written in the following form,
\begin{equation}
    (\Delta H_0)^2=\left[\frac{\partial H_0}{\partial(\log d_L)}\right]^2(\Delta\log d_L)^2 + \left(\frac{\partial H_0}{\partial z}\right)^2(\Delta z)^2,
\label{eq_DH}
\end{equation}
where $\Delta\log d_L$ comes from the measurement uncertainty of the GW detector networks, $\Delta z$ mainly comes from the peculiar velocity of the host galaxy, $v_\mathrm{p}$. In this work, we adopt $\sigma_{v_\mathrm{p}}\sim 300\ \mathrm{km}\ \mathrm{s}^{-1}$. It is worthwhile to note that observations of kilonovae can place constraints on the inclination angle \citep{2019MNRAS.489.5037B, 2020ApJ...888...67D, 2023arXiv230405779Z}, and then improve then improve the measurement for $d_L$. However, these constraints on $\iota$ are highly dependent on models of kilonovae. Therefore, in this work we ignore the improvement of the $d_L$ measurement by observations of kilonovae. For $N$ multi-messenger events, their constraining ability on $H_0$ is
\begin{equation}
    \Delta H_{0}=\left(\sum_{i=1}^{N}\frac{p}{\Delta H^2_{0,\ i}}\right)^{-1/2},
\label{eq_H0}
\end{equation}
where $p$ represents the probability that the $i$-th event's kilonova counterpart being observed.

Through Eq. (\ref{eq_H0}), we obtain the total constraints on the Hubble constant for these 10000 BNS/NSBH merger samples. When the statistical uncertainties are dominant, the uncertainty of $H_0$ is proportional to $\propto1/\sqrt{N}$, where $N$ is the number of the GW events. After adopting the local merging rates of $R_\mathrm{BNSmergers,0}=157.2\ \mathrm{Gpc}^{-3}\ \mathrm{yr}^{-1}$ and $R_\mathrm{NSBHmergers,0}=48.7\ \mathrm{Gpc}^{-3}\ \mathrm{yr}^{-1}$ and the redshift distribution model from \cite{2021ApJ...916...54Y}, we predict that there will be 3000 and 900 $z<0.2$ BNS, NSBH mergers, respectively, in a five years' observation time. The results normalized to these event numbers are listed in Table \ref{kilonova_H0_ms1}, \ref{kilonova_H0_bsk21}, and \ref{kilonova_H0_wff2}. For BNS mergers, if the GW detector array is LHV and the kilonovae are searched in the $r$ band for an exposure time of 90s, the $H_0$ can be constrained to $\Delta H_0 \sim 2.8\ \mathrm{km}\ \mathrm{s}^{-1}\ \mathrm{Mpc}^{-1}$. If the kilonovae need to be identified by multi-band observations, then the constraints become $\Delta H_0 \sim 3.0\ \mathrm{km}\ \mathrm{s}^{-1}\ \mathrm{Mpc}^{-1}$. In the case of the LHVKI array, due to the improved localization capability and $d_L$ constraints, these two results become $\Delta H_0 \sim 1.7\ \mathrm{km}\ \mathrm{s}^{-1}\ \mathrm{Mpc}^{-1}$ and $\sim 1.9\ \mathrm{km}\ \mathrm{s}^{-1}\ \mathrm{Mpc}^{-1}$, respectively. Therefore, multi-messenger observations of BNS mergers will provide an important observational method for solving the Hubble tension.\par

\cite{2018Natur.562..545C} also predicted the ability of the 2G detector arrays to constrain the Hubble constant. They predicted that the LHV array could constrain the $H_0$ to $\sim2\%$ with two years of BNS multi-messenger observations, and $\sim1\%$ by adding two years of LHVKI's observations. After considering the effects of local merging rate, observation duration and duty cycle (0.5 for the LHV array and 0.3 for LHVKI in \citealt{2018Natur.562..545C}), our constraints are still about three times looser. There are two main factors that account for this. First, \cite{2018Natur.562..545C} assumed that all EM counterparts of BNS mergers detected by GW detectors would be observed. However, we find in our simulations that due to the influence of various factors, such as the large GW localization region, the Sun, the Moon, the Milky Waty, and the geographical location of the telescope, WFST can only be able to observe about 20\%-25\% kilonovae through the follow-up search. Secondly, \cite{2018Natur.562..545C} predicted that the $H_0$ uncertainties would scale roughly as $15\%/\sqrt{N}$ and $13\%/\sqrt{N}$ for the LHV and LHVKI networks, respectively, where $N$ is the detected BNS mergers number. In our calculation, these two results are $32\%/\sqrt{N}$ and $26\%/\sqrt{N}$. This is due to \cite{2018Natur.562..545C} chose 400 Mpc as the BNS detection threshold. In our calculations, we simulate the BNS mergers with $z<0.2$ and find the LHVKI network can detect BNS mergers up to $z\sim0.18$. Considering that the redshift distribution of BNS merger is larger in our simulation, therefore the scaled $H_0$ uncertainties in our simulation are about two times larger. After taking these two factors into account, our results are roughly consistent with \cite{2018Natur.562..545C}.\par

For NSBH mergers, it is difficult to observe the kilonova emission in the case of low BH spin, so no effective constraints can be applied to $H_0$. In the case of high spin, the results are significantly influenced by the NS EOS. For example, for the stiffer EOS ms1, the constraints on $H_0$ for the LHV and LHVKI are $\Delta H_0 \sim 4\ \mathrm{km}\ \mathrm{s}^{-1}\ \mathrm{Mpc}^{-1}$ and $\sim 2.1\ \mathrm{km}\ \mathrm{s}^{-1}\ \mathrm{Mpc}^{-1}$ over a five -year observation period.The results with the LHVKI array are roughly comparable to those of BNS mergers. For the bsk21 model, the constraints become worse for $\Delta H_0 \sim 5.4\ \mathrm{km}\ \mathrm{s}^{-1}\ \mathrm{Mpc}^{-1}$ and $\sim 2.8\ \mathrm{km}\ \mathrm{s}^{-1}\ \mathrm{Mpc}^{-1}$ in the cases of LHV and LHVKI arrays, respectively. And for the wff2 model, these two constarints are only $\Delta H_0 \sim 8.2\ \mathrm{km}\ \mathrm{s}^{-1}\ \mathrm{Mpc}^{-1}$ for LHV and $\sim 4\ \mathrm{km}\ \mathrm{s}^{-1}\ \mathrm{Mpc}^{-1}$ for LHVKI. The results for bsk21 and wff2 models are significantly worse than the corresponding results for BNS. It shows that the contribution of multi-messenger observations of NSBH mergers to the Hubble constant constraints can reach the level of BNS mergers only if the BH spins are all very large and the EOS of the NSs are very stiff. However, this also means that multi-messenger observations of NSBH mergers can impose strong constraints on BH spins and NS EOS. Therefore, the search for kilonovae produced by NSBH mergers will remain an important scientific goal for WFST in the future.\par

\cite{2021PhRvL.126q1102F} also investigated future multi-messenger observations for NSBH mergers. With the A+ upgrade \citep{ligo_A+}, they predicted that about 2500 NSBH events will be detected by 2030. With the assumption of the DD2 NS EOS model, 99 of them will have sufficient ejecta ($>0.01\ M_\odot$) to produce observable EM counterparts. \cite{2021PhRvL.126q1102F} predicted that multi-messenger observations of these NSBH mergers could reach 1.5–2.4\% $H_0$ constraints. Considering that the mass of the NSBH merged ejecta is greatly influenced by the BH spin, mass ratio and the NS EOS, in some cases our results (low spin and/or soft EOS) differ dramatically from those in \cite{2021PhRvL.126q1102F}. Under the hard EOS model and high-spin assumptions, our results are roughly similar. Our results together with those of \cite{2021PhRvL.126q1102F} highlight the importance of improving our knowledge about the NSBH population. 

It is worthwhile to note that for the BHs in NSBH mergers, we use a power law + peak mass distribution following the GWTC-3 results \citep{2021arXiv211103634T}, which is mainly obtained from the BBH merging events. However, some recent studies, such as \cite{2021MNRAS.508.5028B} show that the BHs in NSBH systems may have a different mass distribution from those in BBH systems. Uncertainties in the mass distribution may introduce additional errors in the constraints of the Hubble constant. On the other hand, an accurate binary mass distribution model can be used to constrain cosmological parameters\citep{2021arXiv211103634T}. Thus, It is important to obtain an accurate binary population model through future observations. We leave this to a future work.

\begin{table}
\renewcommand\arraystretch{1.5} 
\begin{center}
\begin{tabular}{cccccccc}
\hline\hline
&&&$g$&$r$&$i$&$g\&i$&$g\&r$\\
\hline
BNS&LHV&30s& 3.14& 2.86& 2.80& 3.27& 3.27\\
&&60s& 2.96& 2.80& 2.76& 3.10 & 3.10\\
&&90s& 2.90& 2.79& 2.75& 3.07 & 3.07\\
&LHVKI&30s& 2.10& 1.78& 1.70& 2.23& 2.23\\
&&60s& 1.88& 1.70& 1.66& 1.98& 1.97\\
&&90s& 1.82& 1.68& 1.65& 1.92& 1.91\\
\hline
NSBH, high $\chi_\mathrm{BH}$&LHV&30s& 3.94& 3.94& 3.92& 4.08& 4.10\\
&&60s& 3.86& 3.86& 3.85& 4.02& 4.02\\
&&90s& 3.85& 3.85& 3.85& 4.04& 4.05\\
&LHVKI&30s& 2.15& 2.16& 2.14& 2.24& 2.24\\
&&60s& 2.07& 2.08& 2.07& 2.15& 2.15\\
&&90s& 2.06& 2.06& 2.06& 2.13& 2.13\\
\hline
\end{tabular}
\end{center}
\caption{$\Delta H_0(\mathrm{km}\ \mathrm{s}^{-1}\ \mathrm{Mpc}^{-1})$ with five years' multi-messenger observations and ms1 model.}
\label{kilonova_H0_ms1}
\end{table}

\begin{table}
\renewcommand\arraystretch{1.5} 
\begin{center}
\begin{tabular}{cccccccc}
\hline\hline
&&&$g$&$r$&$i$&$g\&i$&$g\&r$\\
\hline
BNS&LHV&30s& 3.05& 2.86& 2.81& 3.15&3.18\\
&&60s& 2.92& 2.81& 2.79& 3.03 & 3.03\\
&&90s& 2.88& 2.80& 2.78& 3.02 & 3.02\\
&LHVKI&30s& 2.00& 1.78& 1.71& 2.10& 2.10\\
&&60s& 1.87& 1.70& 1.68& 1.96& 1.95\\
&&90s& 1.80& 1.69& 1.67& 1.89& 1.89\\
\hline
NSBH, high $\chi_\mathrm{BH}$&LHV&30s& 5.26& 5.27& 5.24& 5.40& 5.41\\
&&60s& 5.17& 5.16& 5.14& 5.34& 5.35\\
&&90s& 5.14& 5.13& 5.12& 5.36& 5.36\\
&LHVKI&30s& 2.82& 2.83& 2.81& 2.90& 2.90\\
&&60s& 2.72& 2.72& 2.71& 2.79& 2.80\\
&&90s& 2.70& 2.70& 2.69& 2.77& 2.77\\
\hline
\end{tabular}
\end{center}
\caption{$\Delta H_0/(\mathrm{km}\ \mathrm{s}^{-1}\ \mathrm{Mpc}^{-1})$ with five years' multi-messenger observations and bsk21 model.}
\label{kilonova_H0_bsk21}
\end{table}

\begin{table}
\renewcommand\arraystretch{1.5} 
\begin{center}
\begin{tabular}{cccccccc}
\hline\hline
&&&$g$&$r$&$i$&$g\&i$&$g\&r$\\
\hline
BNS&LHV&30s& 3.16& 2.88& 2.86& 3.23&3.26\\
&&60s& 2.98& 2.84& 2.81& 3.07 & 3.07\\
&&90s& 2.91& 2.81& 2.79& 3.05 & 3.03\\
&LHVKI&30s& 2.06& 1.81& 1.77& 2.14& 2.15\\
&&60s& 1.92& 1.74& 1.71& 2.00& 2.00\\
&&90s& 1.82& 1.72& 1.69& 1.89& 1.88\\
\hline
NSBH, high $\chi_\mathrm{BH}$&LHV&30s& 7.99& 7.99& 7.95& 8.25& 8.26\\
&&60s& 7.81& 7.80& 7.78& 8.14& 8.15\\
&&90s& 7.77& 7.77& 7.76& 8.20& 8.20\\
&LHVKI&30s& 4.05& 4.04& 3.99& 4.15& 4.16\\
&&60s& 3.87& 3.86& 3.85& 3.97& 3.97\\
&&90s& 3.84& 3.84& 3.83& 3.94& 3.94\\
\hline
\end{tabular}
\end{center}
\caption{$\Delta H_0(\mathrm{km}\ \mathrm{s}^{-1}\ \mathrm{Mpc}^{-1})$ with five years' multi-messenger observations and wff2 model.}
\label{kilonova_H0_wff2}
\end{table}

\subsection{Uncertainties from fitting formulas}

In this section, we use several fitting formulas to calculate the mass of ejecta when BNS/NSBH merged. Since these formulas are fitted from tens of numerical relatively results with different NS EOS and merger parameters, and lack the constraints of the real observations, the final results depend on those fitting parameters and may have rather large uncertainties. In order to investigate the effect of the fitting parameters on the results, in this subsection we adopt several different fitting formulas for comparison.

For BNS mergers, we adopt the $m_\mathrm{dyn}$ formula from \cite{2017CQGra..34j5014D},
\begin{equation}
\begin{split}
    \frac{M_\mathrm{dyn}}{10^{-3}M_\odot}=&\left [a\left(\frac{M_2}{M_1}\right)^{1/3}\left( \frac{1-2C_1}{C_1}\right)+b\left (\frac{M_2}{M_1}\right )^n \right. \\
    &\left.\quad c\left(1-\frac{M_1}{M^\mathrm{b}_{1}}\right)\right]M^{\mathrm{b}}_1+(1\leftrightarrow2)+d,
\end{split}
\end{equation}
where fitting parameters are $a=-1.35695,\ b=6.11252,\ c=-49.43355,\ d=16.1144,\ n=-2.5484$, and the $m_\mathrm{disk}$ fitting from \cite{2018ApJ...869..130R}, 
\begin{equation}
    \log_{10}\left (\frac{M_\mathrm{disk}}{M_\odot}\right )=\max\left[-3,a\left\{1+b\tanh\left(\frac{c-M/M_\mathrm{thr}}{d}\right)\right\}\right],
\end{equation}
with $a=-31.335$, $b=-0.9760$, $c=1.0474$, $d=0.05957$. As for NSBH mergers, we use the dynamical ejecta fitting formula from \cite{2020PhRvD.101j3002K} as a  comparison, 
\begin{equation}
    \frac{M_\mathrm{dyn}}{M^{*}_\mathrm{NS}}=a_1Q^{n_1}\frac{1-2C_\mathrm{NS}}{C_\mathrm{NS}}-a_2 Q^{n_2}\tilde{R}_\mathrm{ISCO}+a_4,
\end{equation}
with $a_1=0.007116$, $a_2=0.001436$, $a_4=-0.02762$, $n_1=0.8636$, and $n_2=1.6840$, and the numerical error is $\Delta M_\mathrm{dyn}=\sqrt{(0.1M_\mathrm{dyn})^2+(0.01M_\odot)^2}$. 

After replacing these fitting formulas, we repeat the above steps. For BNS mergers, most of the difference in ejecta mass is around 10\%, with a fraction of mergers differing more in fitting results due to a lager mass ratio or NS mass. For NSBH mergers, the mass of the ejecta obtained by different fitting formulas is much more different due to large numerical error. Then take the case of a 30s exposure time for example, we show the multi-messenger observation rates and $H_0$ constraints with these fitting formulas in Table \ref{kilonova_psum4} and \ref{kilonova_H0_ms12}, respectively. For BNS mergers, the change in the multi-messenger observation rate is quite small, approximately a few percent smaller. However, for NSBH mergers, the difference in results is substantial, especially for ms1 and wff2, the two stiffest and softest models in our discussions with observation rates half as small and twice as larger, respectively. The difference under different fitting formulas further demonstrates the need for further studies of the kilonovae generated by NSBH mergers.

 {Reflecting on the $H_0$ constraints, the results of the BNS mergers are also nearly the same (about $\lesssim4\%$ looser),} but the constraints of NSBH mergers with ms1 models are $\sim1/3$ looser. For NSBH mergers with wff2 models, five years' multi-messenger observations with the LHV network can constrain $\Delta H_0\sim 6.8\ \mathrm{km}\ \mathrm{s}^{-1}\ \mathrm{Mpc}^{-1}$, which are 20\% tighter that the results in Tabel \ref{kilonova_H0_wff2}. As for the LHVKI array, the constraints are only $\sim10\%$ tighter, this is due to that the additional part of the kilonovae mainly located at higher redshifts and contributing less to the $H_0$ constraints.

\begin{table}
\renewcommand\arraystretch{1.5} 
\begin{center}
\begin{tabular}{cccccccc}
\hline\hline
&&&$g$&$r$&$i$&$g\&i$&$g\&r$\\
\hline
BNS&LHV&ms1& 10.5& 12.4& 13.2& 9.6&9.6\\
&&bsk21& 10.9& 12.4& 12.9& 9.9& 9.8\\
&&wff2& 11.0& 12.2& 12.7& 10.0 & 10.0\\
&LHVKI&ms1& 15.3& 19.2& 22.1& 13.7& 13.8\\
&&bsk21& 15.9& 19.5& 21.5& 14.5& 14.3\\
&&wff2& 15.9& 19.1& 20.5& 14.5& 14.5\\
\hline
NSBH, high $\chi_\mathrm{BH}$&LHV&ms1& 4.0& 3.9& 4.0& 3.6& 3.6\\
&&bsk21& 4.0& 4.0& 4.0& 3.7& 3.7\\
&&wff2& 2.5& 2.5& 2.6& 2.4& 2.4\\
&LHVKI&ms1& 6.4& 6.3& 6.4& 5.8& 5.8\\
&&bsk21& 6.8& 6.8& 6.9& 6.3& 6.2\\
&&wff2& 4.4& 4.4& 4.5& 4.2& 4.1\\
\hline
\end{tabular}
\end{center}
\caption{Multi-messenger observation rates of WFST at different band with a 30s exposure time, different NS EOS models, and replaced ejecta fitting formulas from \cite{2017CQGra..34j5014D}, \cite{2018ApJ...869..130R} and \cite{2020PhRvD.101j3002K}.}
\label{kilonova_psum4}
\end{table}

\begin{table}
\renewcommand\arraystretch{1.5} 
\begin{center}
\begin{tabular}{cccccccc}
\hline\hline
&&&$g$&$r$&$i$&$g\&i$&$g\&r$\\
\hline
BNS&LHV&ms1& 3.23& 2.89& 2.83& 3.31& 3.29\\
&&bsk21& 3.18& 2.91& 2.86& 3.26 & 3.29\\
&&wff2& 3.14& 2.95& 2.90& 3.29 & 3.28\\
&LHVKI&ms1& 2.12& 1.83& 1.73& 2.25& 2.22\\
&&bsk21& 2.06& 1.86& 1.77& 2.14& 2.16\\
&&wff2& 2.11& 1.88& 1.80& 2.20& 2.20\\
\hline
NSBH, high $\chi_\mathrm{BH}$&LHV&ms1& 5.14& 5.16& 5.14& 5.43& 5.44\\
&&bsk21& 5.32& 5.32& 5.30& 5.42& 5.42\\
&&wff2& 6.72& 6.72& 6.69& 6.83& 6.84\\
&LHVKI&ms1& 3.02& 3.02& 3.00& 3.19& 3.20\\
&&bsk21& 2.90& 2.91& 2.89& 2.99& 2.99\\
&&wff2& 3.75& 3.75& 3.73& 3.84& 3.84\\
\hline
\end{tabular}
\end{center}
\caption{ $\Delta H_0(\mathrm{km}\ \mathrm{s}^{-1}\ \mathrm{Mpc}^{-1})$ with five years' multi-messenger observations, a 30s exposure time, different NS EOS models, and replaced ejecta fitting formulas from \cite{2017CQGra..34j5014D}, \cite{2018ApJ...869..130R} and \cite{2020PhRvD.101j3002K}.}
\label{kilonova_H0_ms12}
\end{table}

\section{Dark Sirens}
\label{section_ds}
In the previous section, we discuss the applications of BNS/NSBH merger-kilonova sirens in cosmology in the era of 2G GW detections. However, in the actual observation, the search for kilonova is limited by many factors, such as the Sun, the Moon, the Milky Way, the weather, and the geographical location of the telescope. For a ground-based telescope, its average observable area each night of the year is about 50\% of the whole sky, and this percentage needs to be discounted considering the influence of weather, Moon and other factors. In particular, when a telescope is located at high latitudes, its observing sky area and observing hours in summer will be very limited due to the long hours of daylight. As a result, a significant fraction of kilonovae cannot be observed, even if they are quite bright or the GW signal is well localized. For example, in the simulations of the previous section, we find that for BNS mergers, the detection rate of their kilonova counterparts is only $\sim1/5$ of that of the LHVKI array.\par

In contrast, the observation of GW is hardly limited by these factors. When the GW detector is in operation, it can monitor GW signals coming from almost all directions. Therefore, for a dark siren, it is much less influenced by other factors than a bright siren. For example, the method of comparing GW events' localization area and a survey data is only affected by the localization accuracy of the GW signals and the completeness of the survey catalog in the localization region. Since the survey data are accumulated from long time observations, they are not affected by the constraints of the observation area, weather and other factors as much as follow-up observations. \par

In our pervious work \cite{2020MNRAS.498.1786Y}, we estimated the performance of the SBBH merger dark sirens by comparing the Sloan Digital Sky Survey data release 7 (SDSS DR7) group catalog \citep{2007ApJ...671..153Y} in the localization area. We found that for the 2G array, LHVKI, the effect of constraining the Hubble constant for the large mass SBBH dark sirens is close to that of the BNS bright sirens. However, for small-mass mergers, such as BNS, NSBH, and part of BBH systems, these events are difficult to be used as dark sirens to effectively constrain the Hubble constant in the era of 2G GW detection because of the weak GW signals produced. Only in the 3G era, with the significant increase in positioning capability, the BNS and NSBH dark sirens can play an essential role in cosmology. In this section, we will discuss the BNS and NSBH dark sirens and their applications.

\subsection{Group catalog}
Our group catalog is based on the high-resolution N-body Uchuu simulation \citep{2021MNRAS.506.4210I}. Uchuu evolves the distribution of $12800^3$ dark matter particles in a box of a side length of 2.0 $h^{-1}$ Gpc and particle mass of $3.27\times10^8\ h^{-1}\ M_\odot$. The redshift range is from $z=127$ to $z=0$. The cosmological parameters adopted by Uchuu are listed at the end of Section \ref{intro}.\par

For simplicity, we only use the halo/subhalo catalog obtained by the Uchuu simulation using the Rockstar halo finder \citep{2013ApJ...762..109B} for the redshift $z=0$ snapshot. We populate each subhalo in the catalog with a galaxy whose stellar mass is assigned according to the stellar to subhalo mass relation obtained in \cite{2012ApJ...752...41Y}. Once all the subhalos in the catalog are populated with galaxies of given stellar masses, we then calculate their {\it accumulative} stellar mass function. This accumulative stellar mass function is compared to the accumulative $z$-band luminosity function of galaxies obtained by \cite{2021ApJ...909..143Y} from the DESI Legacy imaging Surveys data release 9. Using the abundance matching between the stellar mass and luminosity, we assign each galaxy a $z$-band luminosity. By properly taking into account the average $k$-correction \citep{2021ApJ...909..143Y} and the redshift space distortion effect (e.g, \citealt{2004MNRAS.350.1153Y}), we construct a mock galaxy and group catalog with $z$-band apparent magnitude limit $z_{\rm limit}=21$ in a sphere with redshift $z\le 0.5$ for this study (see Gu et al. 2023 in preparation for more details). \par

\begin{figure}[htbp]
\centering
\includegraphics[width=9cm]{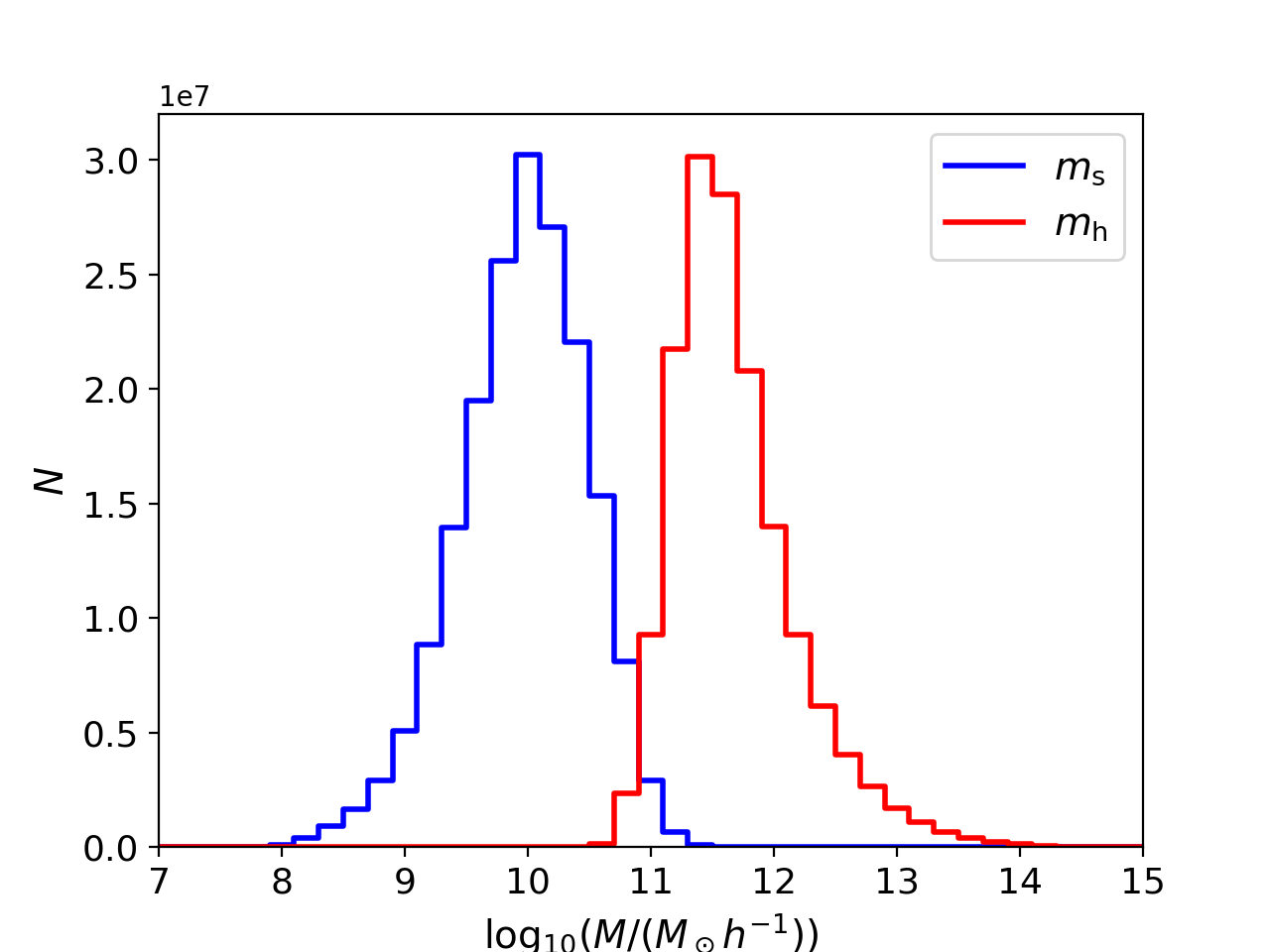}
\caption{The blue line represents the stellar mass distribution of galaxies in catalog and the red line represents the halo mass distribution of groups. }
\label{figure_m_distribution}
\end{figure}

In Figure \ref{figure_m_distribution} we show the stellar mass distribution of galaxies and halo mass distribution of groups in catalog. The average stellar mass of galaxies and the average halo mass of groups are $2.10\times10^{10}\ M_\odot/h$ and $1.94\times10^{12}\ M_\odot/h$, respectively.

To ensure the completeness of the group catalog, as in our previous work \cite{2020MNRAS.498.1786Y}, we choose halo mass $M_\mathrm{halo}>10^{12}\ M_\odot/h$ as a cutoff. These groups account for roughly 1/3 of the total stellar mass in the catalog. Under this halo mass cutoff, groups in the range $z\leq0.5$ are all roughly uniformly distribution in the comoving volume, which demonstrates the completeness of the catalog. We will discuss the reasonableness of adopting this cutoff at the end of this section.

\subsection{Localization volumes}

In our pervious work \cite{2020MNRAS.498.1786Y}, for a SBBH merger, we derived its covariance matrix $\mathbf{Cov}[\alpha,\delta,\log(d_L)]$ from the Fisher matrix and drew an ellipsoid in the parameter space of $(\alpha, \delta,\log(d_L))$, and converted the constraints on the source's luminosity distance $d_L$ to redshift $z$ under the $\Lambda$CDM model. Then we traversed group catalog to find the number of groups in it. We used this number, $N_\mathrm{in}$ to quantify the ability of identifying the source's host galaxy group. $N_\mathrm{in}=1$ means that the host galaxy group of this GW event can be identified directly without EM counterpart.\par

Though for the CE2ET network, most SBBH mergers with $z<0.1$ have $N_\mathrm{in}=1$ due to a small localization volume \citep{2020MNRAS.498.1786Y}, these results significantly rely on the cosmological model when tansforming $d_L$ to $z$. In contrast, the NSs' tidal effects in merging time provide a method of measuring GW sources' redshifts that does not rely on specific cosmological models. In this section, we discuss BNS and NSBH mergers respectively. For each group in catalog, we put an assumed BNS/NSBH merger at its center. Assuming that the EOS of NS has been tightly constrained from the X-ray plateaus \citep{2014PhRvD..89d7302L}, kilonovae \citep{2021MNRAS.505.3016N}, etc., we obtain the covariance matrix $\mathbf{Cov}[\alpha, \delta, z]$ from 10-parameters' Fisher matrix, $(\alpha,\delta, \psi,\iota, M, \eta, t_{c}, \varphi_{c}, \log(d_{L}), z)$, paint the error ellipsoid in the parameter space of $(\alpha,\beta,z)$ and traverse group catalog to count its $N_\mathrm{in}$. \par
\begin{figure}[htbp]
\centering
\includegraphics[width=9cm]{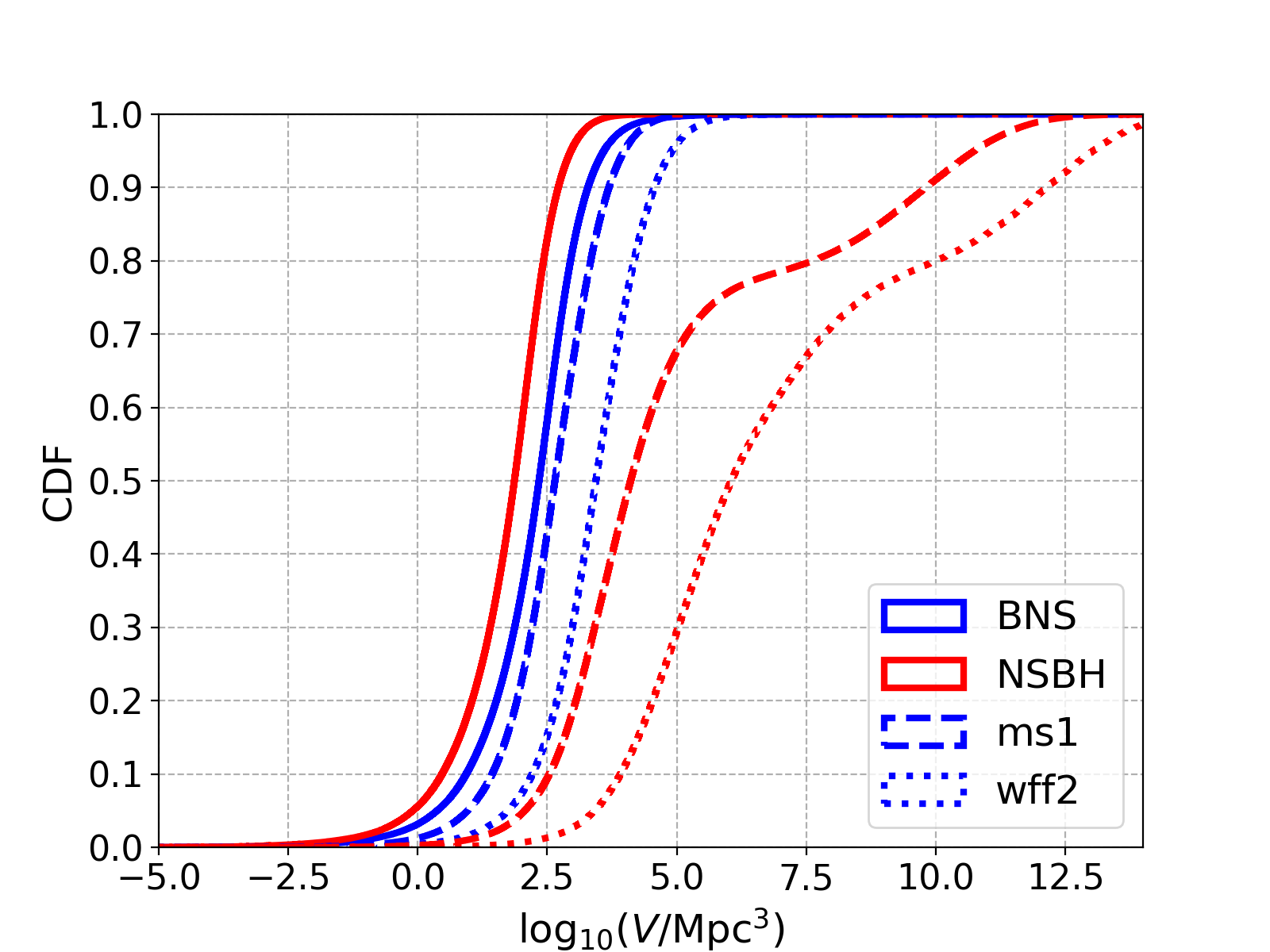}
\caption{The CDF of localization volumes for with BNS samples (blue line) and NSBH samples (red line). The soild line represents the case of $\mathbf{Cov}[\alpha, \delta, \log(d_L)]$. The dash and dot lines represent the cases of $\mathbf{Cov}[\alpha,\delta,z]$ with ms1 and wff2 models, respectively.}
\label{figure_vin1}
\end{figure}

In Figure \ref{figure_vin1}, we show the cumulative distribution function (CDF) of localization volumes. We choose the cases of ms1 and wff2 as representatives, which have the smallest and largest localization volumes. As a comparison, we also show the CDF of localization volumes from the covariance matrix $\mathbf{Cov}[\alpha, \delta, \log(d_L)]$. \par
\begin{figure*}[htbp]
\centering
\subfigure[BNS]{
	\includegraphics[width=8.5cm]{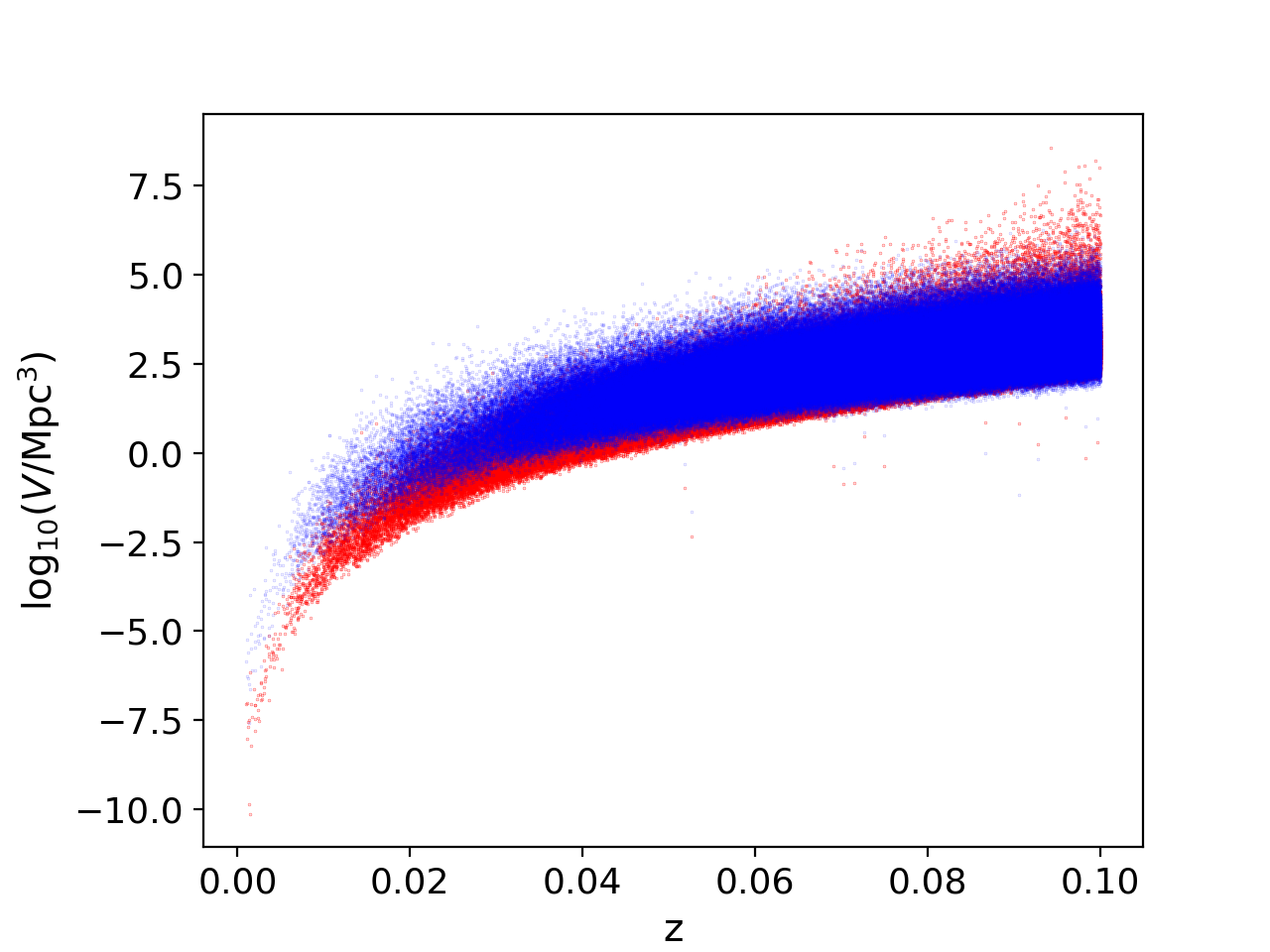}
}
\hspace{1.0pt}
\subfigure[NSBH]{
	\includegraphics[width=8.5cm]{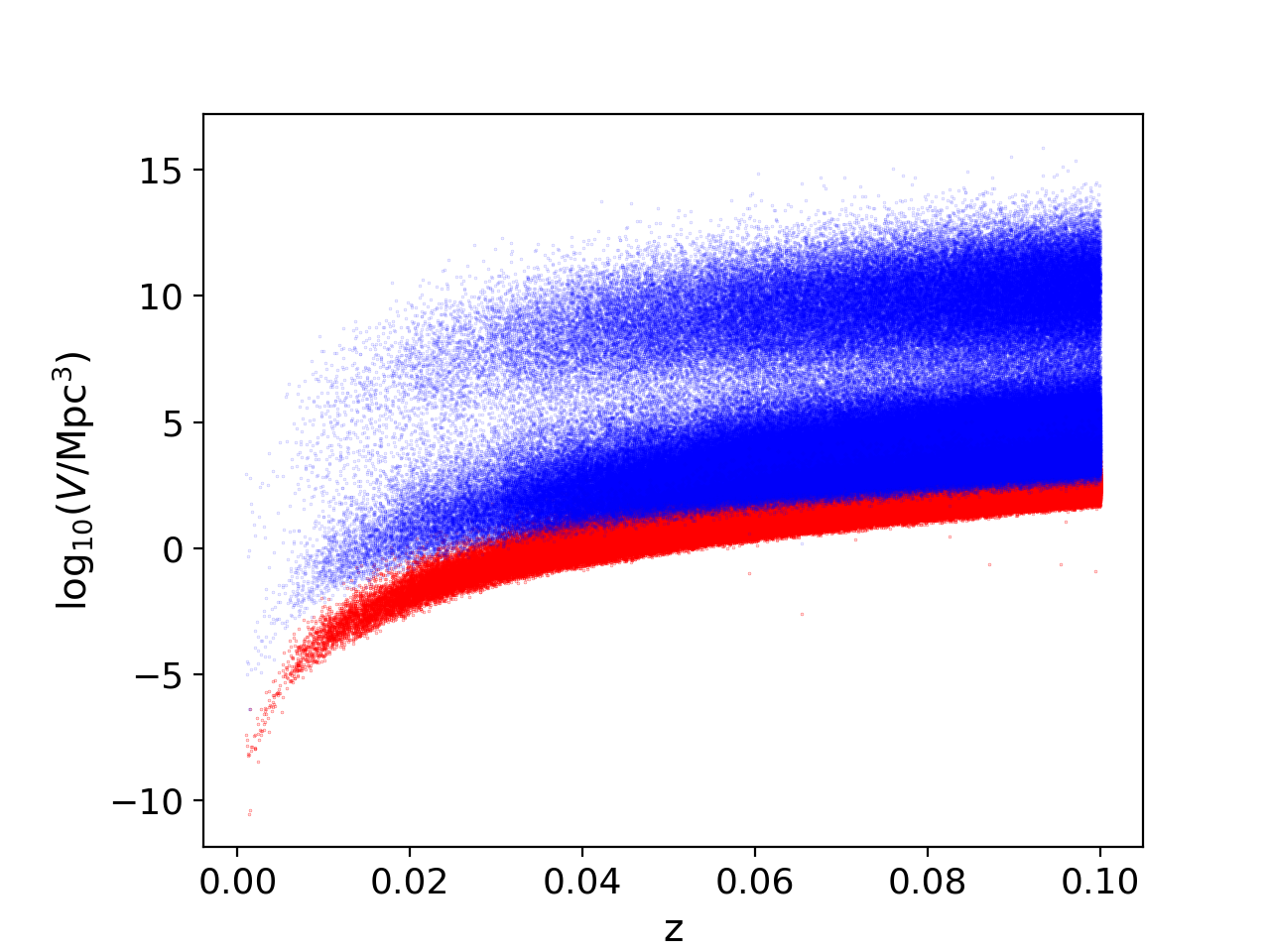}
}
\caption{The distributions of localization volumes for different cases. The left-hand panel is for the case of BNS mergers and the right-hand panel is for the case of NSBH mergers. The red and blue dots represent the cases of $\mathbf{Cov}[\alpha, \delta, \log(d_L)]$ and $\mathbf{Cov}[\alpha,\delta,z]$ with ms1 model, respectively.}
\label{figure_vin2}
\vspace{4em} 
\end{figure*}

For BNS mergers, the typical localization volumes from $\mathbf{Cov}[\alpha, \delta, \log(d_L)]$ are around $(10-10^3)$ Mpc$^3$. In the case of $\mathbf{Cov}[\alpha,\delta,z]$ and ms1, the localization volumes are slightly larger. And for wff2, the typical localization volumes are approximately 10 times larger. In the left panel of Figure \ref{figure_vin2}, we show the distributions of BNS mergers' redshifts and their localization volumes obtained by different locating methods. At low redshift ($z\lesssim0.05$), the volumes from $\mathbf{Cov}[\alpha, \delta, \log(d_L)]$ are obviously smaller than those from $\mathbf{Cov}[\alpha, \delta, z]$. As the redshift increases, the tidal deformation term $\Phi_\mathrm{tidal}(f)$ will become larger. Thus, the localization capability of the observations of tidal effect will gradually catch up with the method of luminosity distance measurements. \par
For NSBH mergers, the typical localization volumes from $\mathbf{Cov}[\alpha, \delta, \log(d_L)]$ are a few times smaller than BNS mergers, since they would produce stronger GW signals. However, in NSBH systems, the tidal effect is significantly weaker than in BNS systems due to the larger mass of the BH and its lack of tidal deformation. Therefore, the localization volumes from $\mathbf{Cov}[\alpha,\delta,z]$ are much larger. The distributions of NSBH mergers' redshifts and their localization volumes are showed in the right panel of Figure \ref{figure_vin2}. Due to the existence of substructure in mass distribution of BHs around $34\ M_\odot$, the localization volumes from $\mathbf{Cov}[\alpha,\delta,z]$ are divided into two parts. NSBH mergers with massive BHs will be located in a volume several magnitudes larger because of a weaker tidal field.

\subsection{The distributions of $N_\mathrm{in}$}
\label{subsection_nin}
In each case of EOS, we calculate $N_\mathrm{in}$ for the assumed GW events and get a distribution of $N_\mathrm{in}$. In Table \ref{table_bns} and \ref{table_nsbh}, we list the fractions of $N_\mathrm{in}=1$, $N_\mathrm{in}\leq2$, $N_\mathrm{in}\leq5$ and $N_\mathrm{in}\leq10$ for each case. \par
For BNS mergers, among these EOSs, ms1 has the highest fraction of $N_\mathrm{in}=1$ due to smaller localization volumes, which reaches 44.8\% and the fractions of $N_\mathrm{in}\leq10$ is 92.1\%. In the case of ms1b, 43.5\% BNS mergers' host galaxy groups can be identified directly through the observations of tidal effect. The results of other EOSs are worse than theirs, where in the cases of ap4, qmf40, sly, wff2 and pQCD800, the fractions of $N_\mathrm{in}=1$ are below 30\%. Meanwhile, in all cases, the fractions of $N_\mathrm{in}\leq10$ are larger than 70\%.\par
For NSBH mergers, the results are much worse. Only for three EOSs, the fractions of $N_\mathrm{in}=1$ exceed 10\%, and they are H4, ms1 and ms1b. In the case of ms1, the fractions of $N_\mathrm{in}=1$ and $N_\mathrm{in}\leq10$ are 16.1\% and 49.3\%, respectively. However, for wff2, only 3\% NSBH mergers have $N_\mathrm{in}=1$ and 15.4\% assumed events have $N_\mathrm{in}\leq10$.\par
\begin{figure*}[htbp]
\centering
\subfigure[ms1]{
	\includegraphics[width=8.5cm]{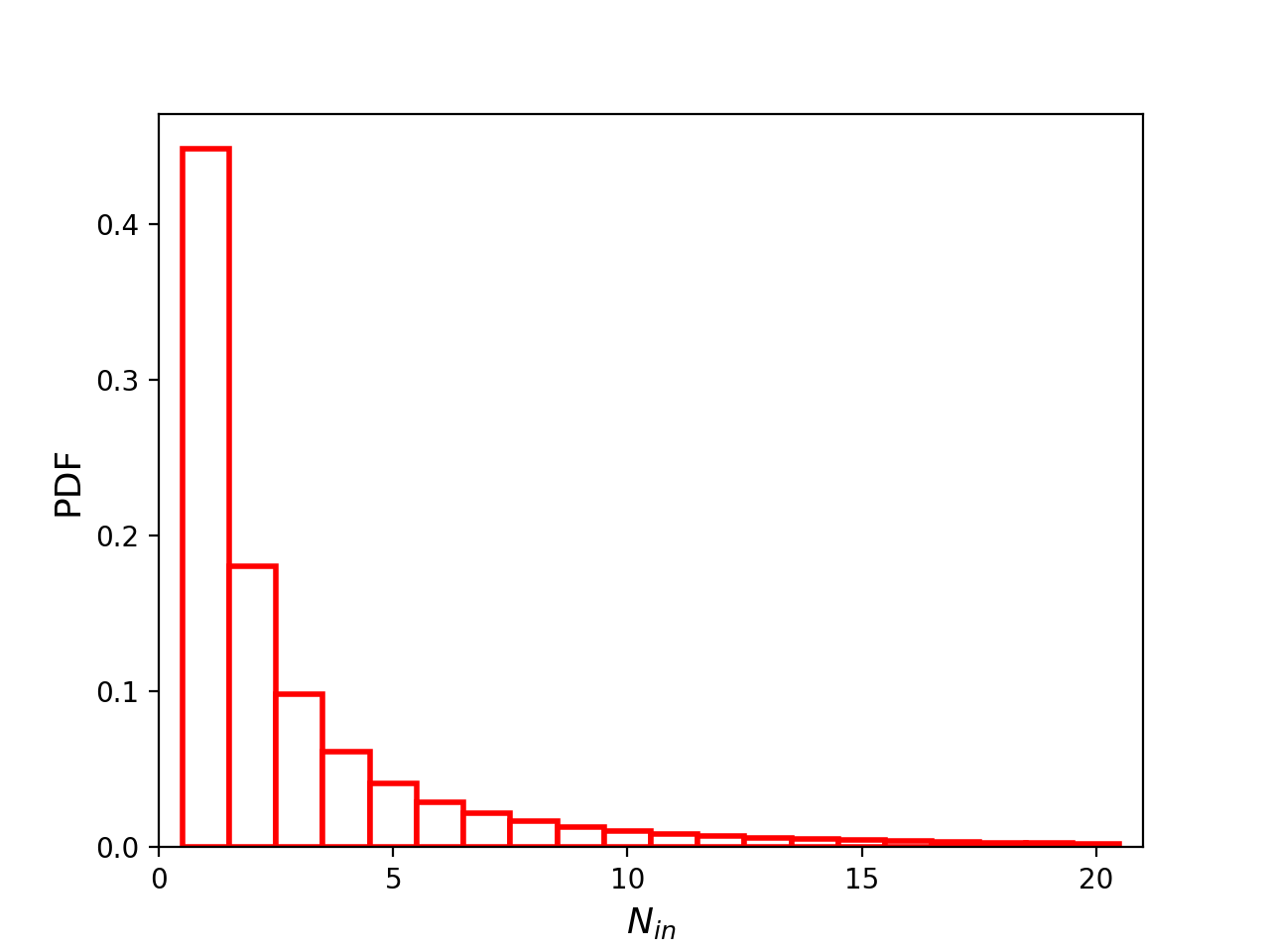}
}
\hspace{1.0pt}
\subfigure[$d_L$]{
	\includegraphics[width=8.5cm]{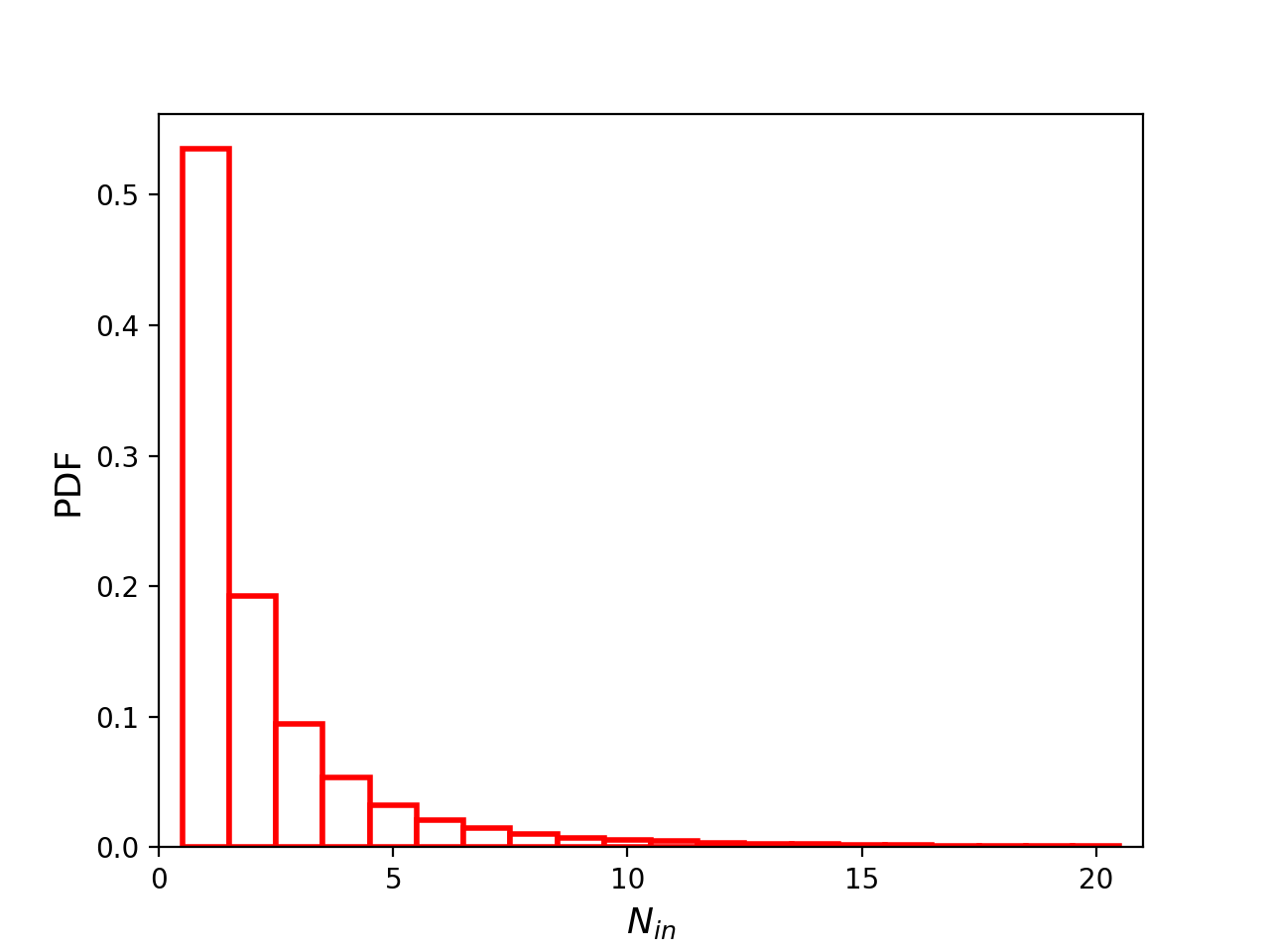}
}
\caption{The PDF of BNS samples' $N_\mathrm{in}$ from 1 to 20 with the CE2ET network. The left-hand panel shows the distribution with error ellipsoids from $\mathbf{Cov}[\alpha,\delta,z]$ and ms1 model. The right-hand panel shows the distribution with error ellipsoids from $\mathbf{Cov}[\alpha, \delta, \log(d_L)]$. }
\label{figure_bns}
\end{figure*}
\begin{figure*}[htbp]
\centering
\subfigure[ms1]{
	\includegraphics[width=8.5cm]{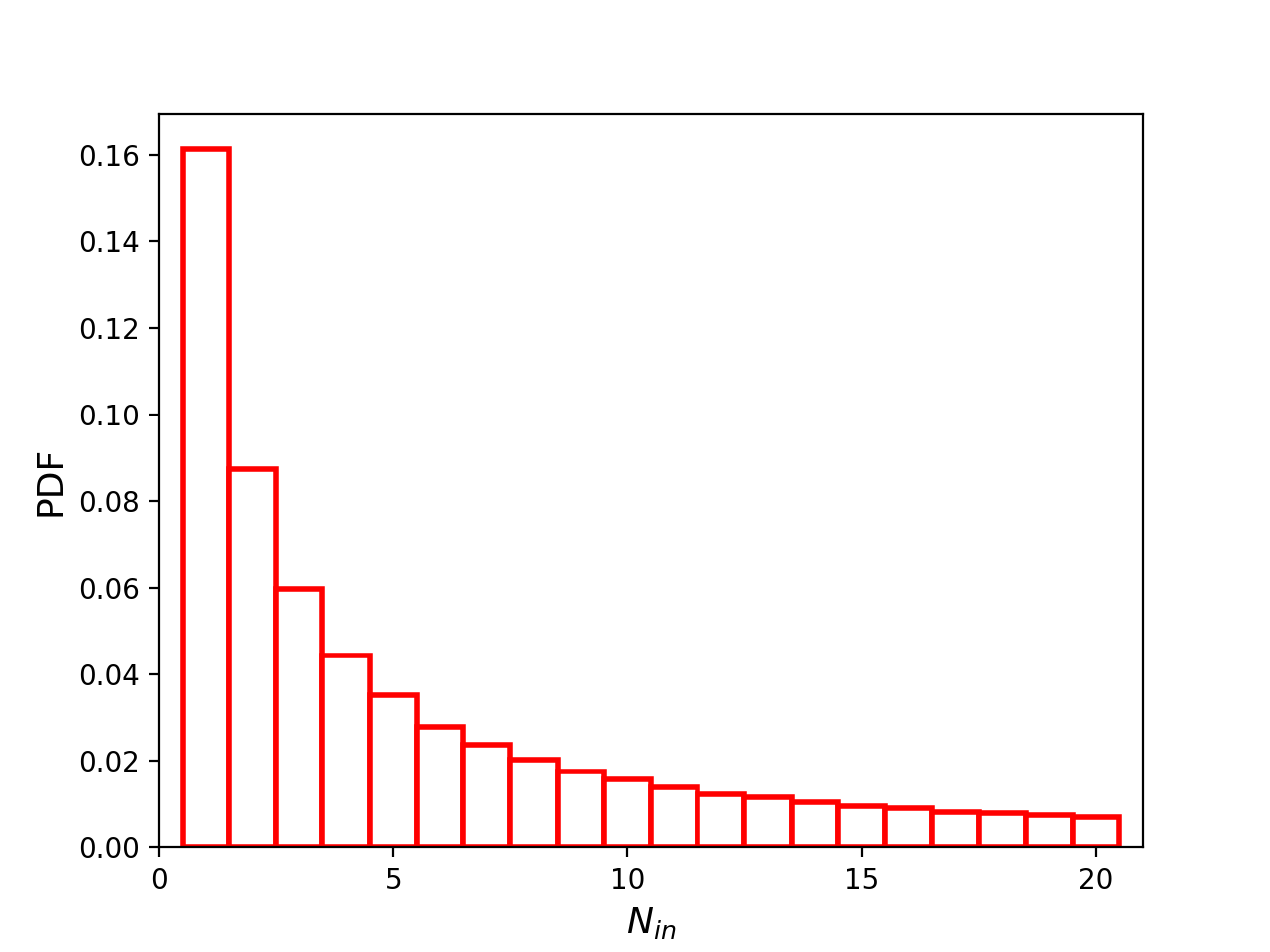}
}
\hspace{1.0pt}
\subfigure[$d_L$]{
	\includegraphics[width=8.5cm]{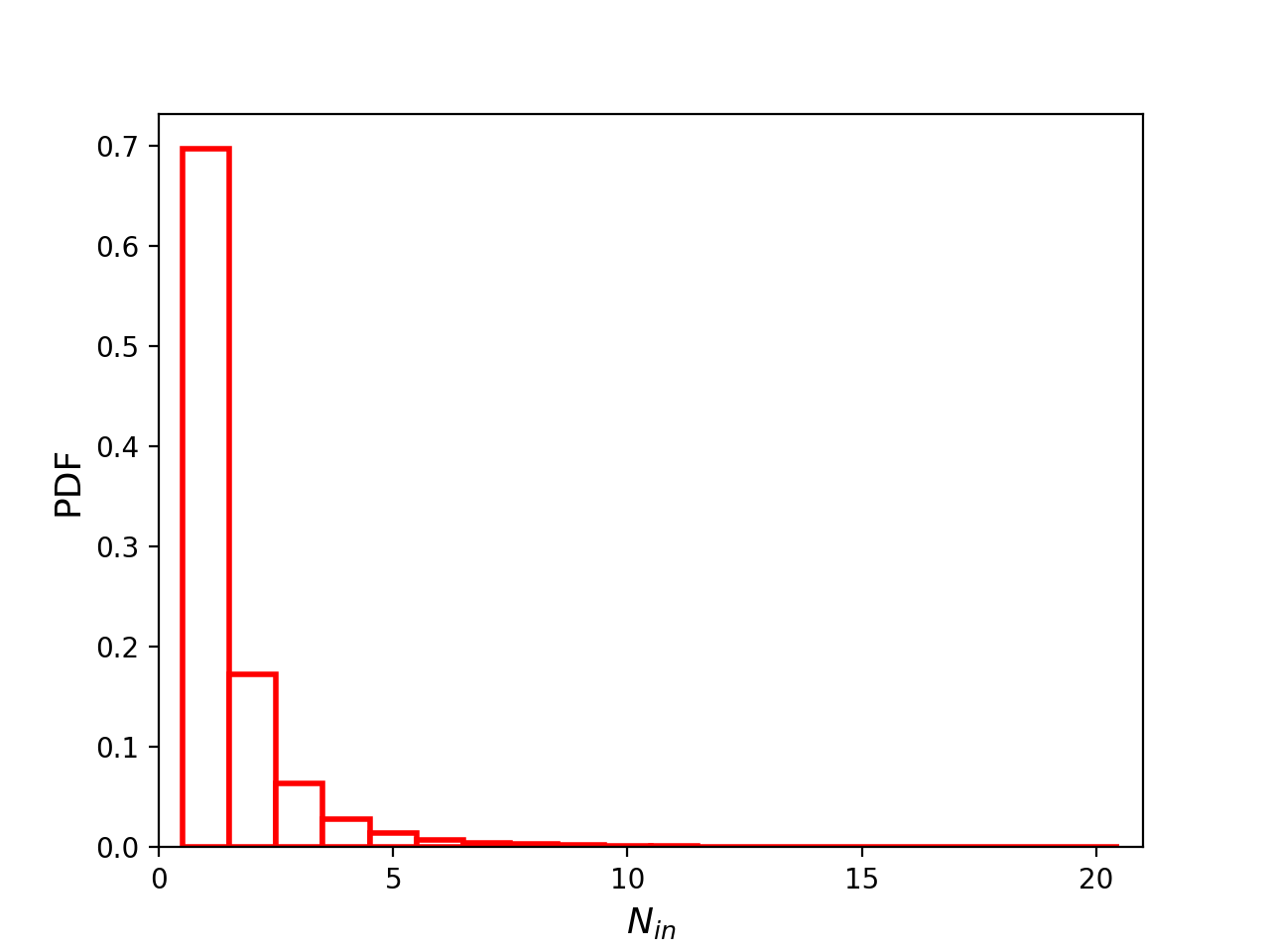}
}
\caption{Same as Figure \ref{figure_bns}, but with NSBH samples.}
\label{figure_nsbh}
\end{figure*}

As a comparison, we also list the fractions of $N_\mathrm{in}$ with  error ellipsoids from $\mathbf{Cov}[\alpha, \delta, \log(d_L)]$ in Table \ref{table_dL}. Almost all BNS, NSBH and BBH mergers have $N_\mathrm{in}\leq10$. And the fractions of $N_\mathrm{in}=1$ are 53.5\%, 69.7\%, 93.2\% for BNS, NSBH and BBH mergers, respectively. We compare their probability density function (PDF) of  $N_\mathrm{in}$ with the cases of error ellipsoids from $\mathbf{Cov}[\alpha, \delta, z]$ in Figure \ref{figure_bns} and \ref{figure_nsbh}. The ability to identify the BNS merger's host galaxy group through tidal effects with ms1 model is similar to the method of luminosity distance measurements. Thus for low redshift BNS mergers, when the EM counterparts are not observable, tidal effect observations will be a reliable and model-independent method of identifying the sources' host galaxy groups. However, for NSBH mergers, this method is not useful due to weaker tidal effect. \par
\begin{figure*}[htbp]
\centering
\subfigure[BNS, $\mathbf{Cov}(\alpha, \delta,\log(d_L))$]{
	\includegraphics[width=8cm]{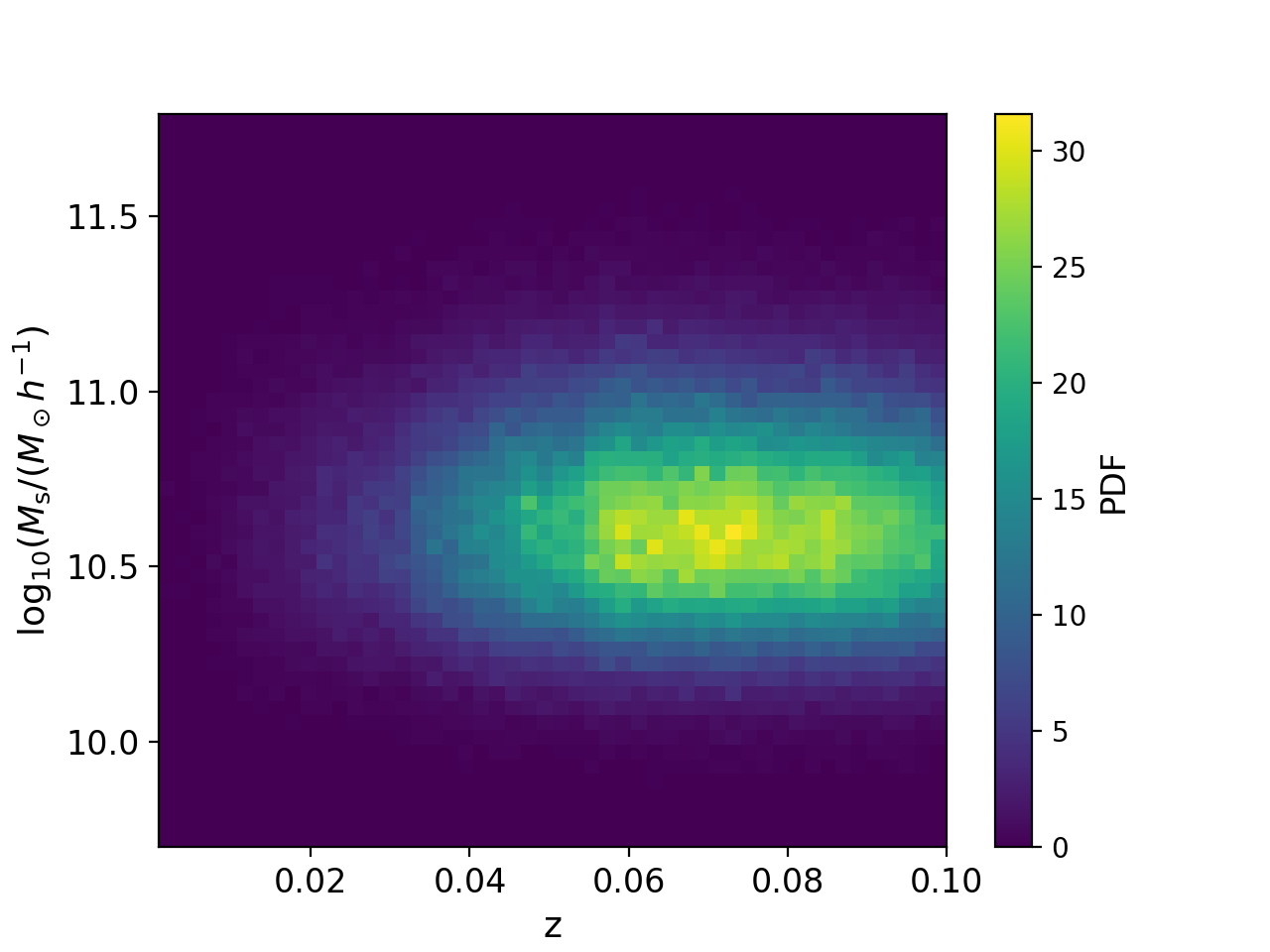}
}
\hspace{1.0pt}
\subfigure[NSBH, $\mathbf{Cov}(\alpha, \delta, \log(d_L))$]{
	\includegraphics[width=8cm]{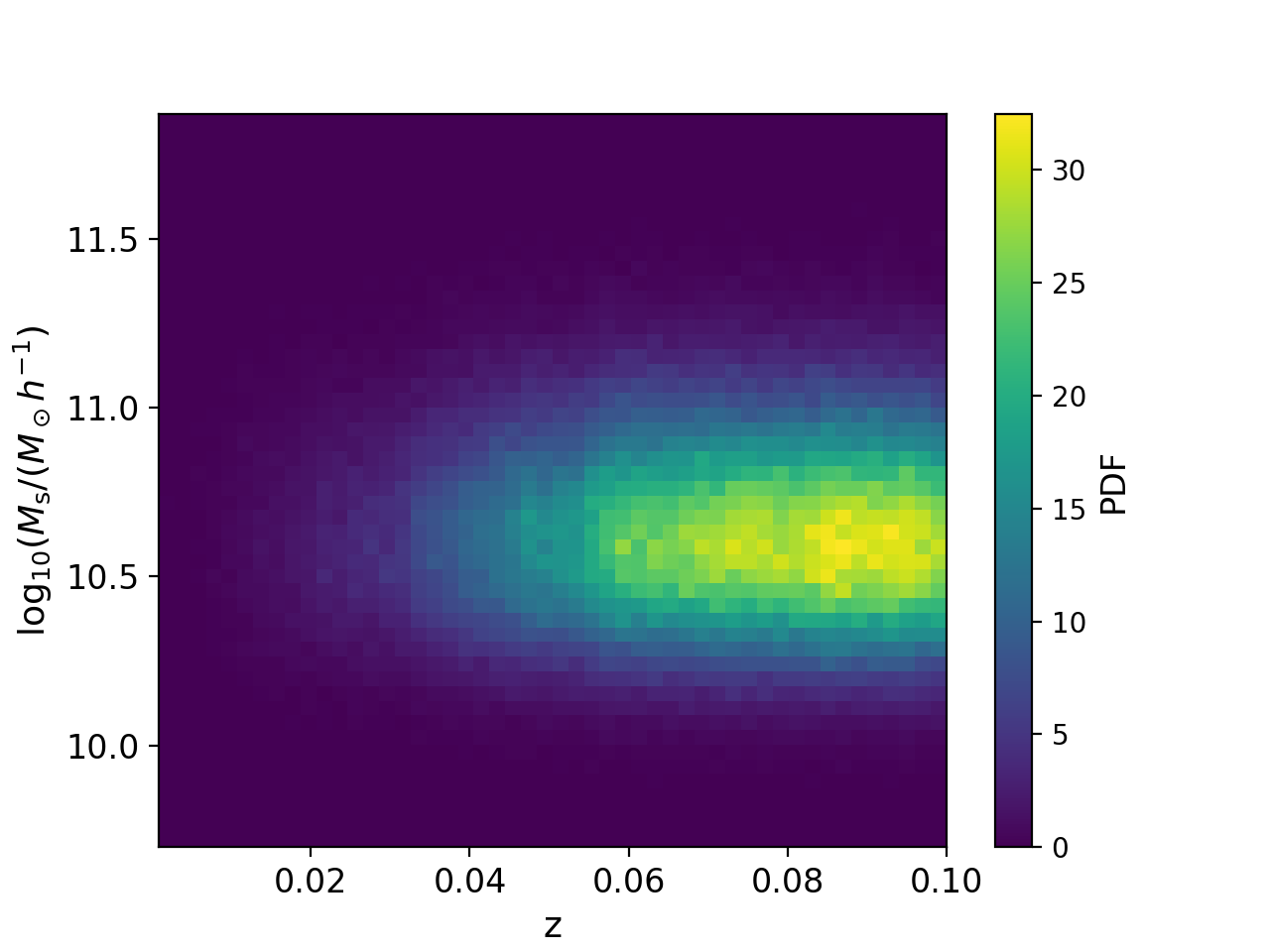}
}
\hspace{1.0pt}
\subfigure[BNS, ms1, $\mathbf{Cov}(\alpha,\delta,z)$]{
	\includegraphics[width=8cm]{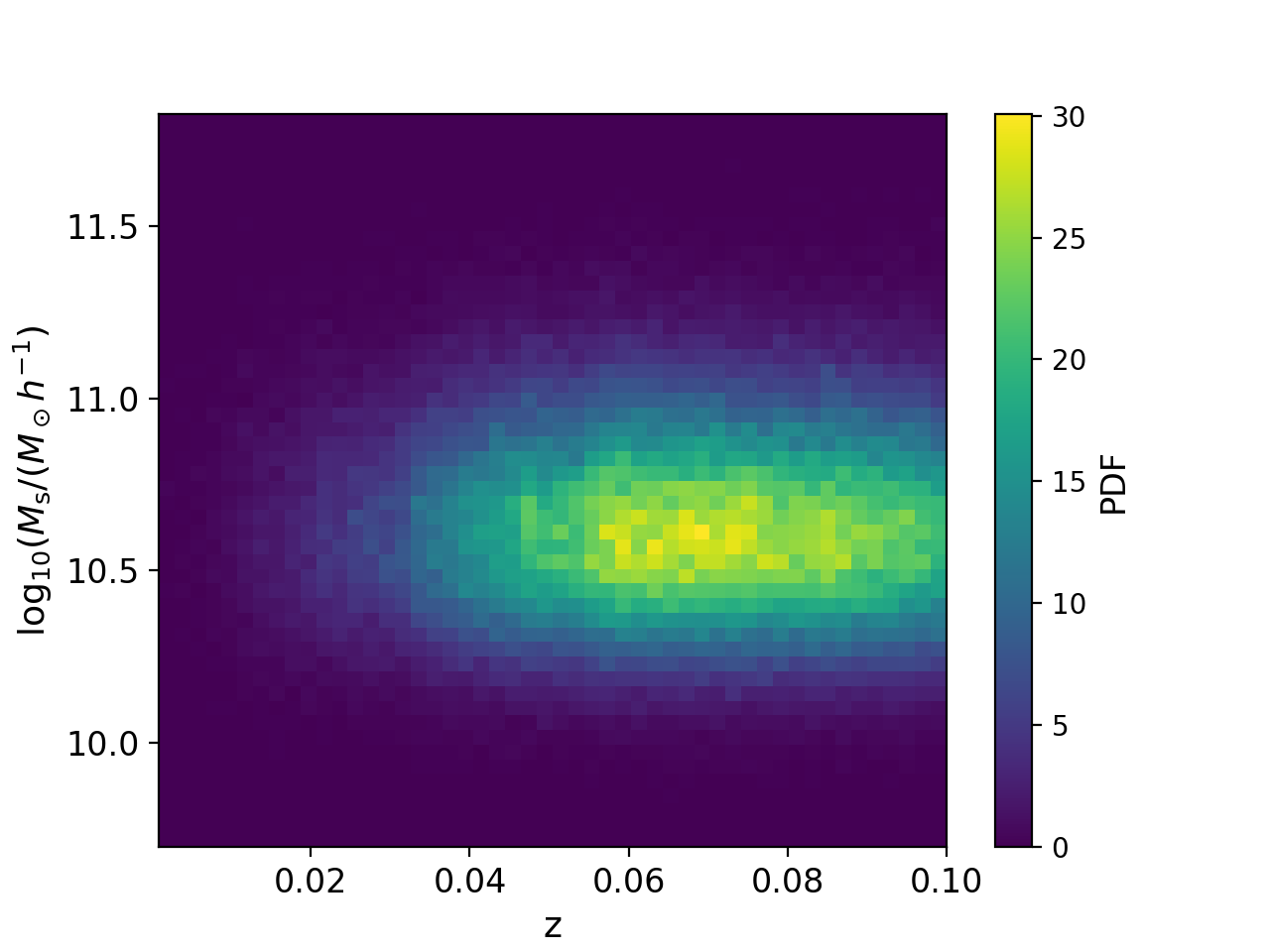}
}
\hspace{1.0pt}
\subfigure[NSBH, ms1, $\mathbf{Cov}(\alpha,\delta,z)$]{
	\includegraphics[width=8cm]{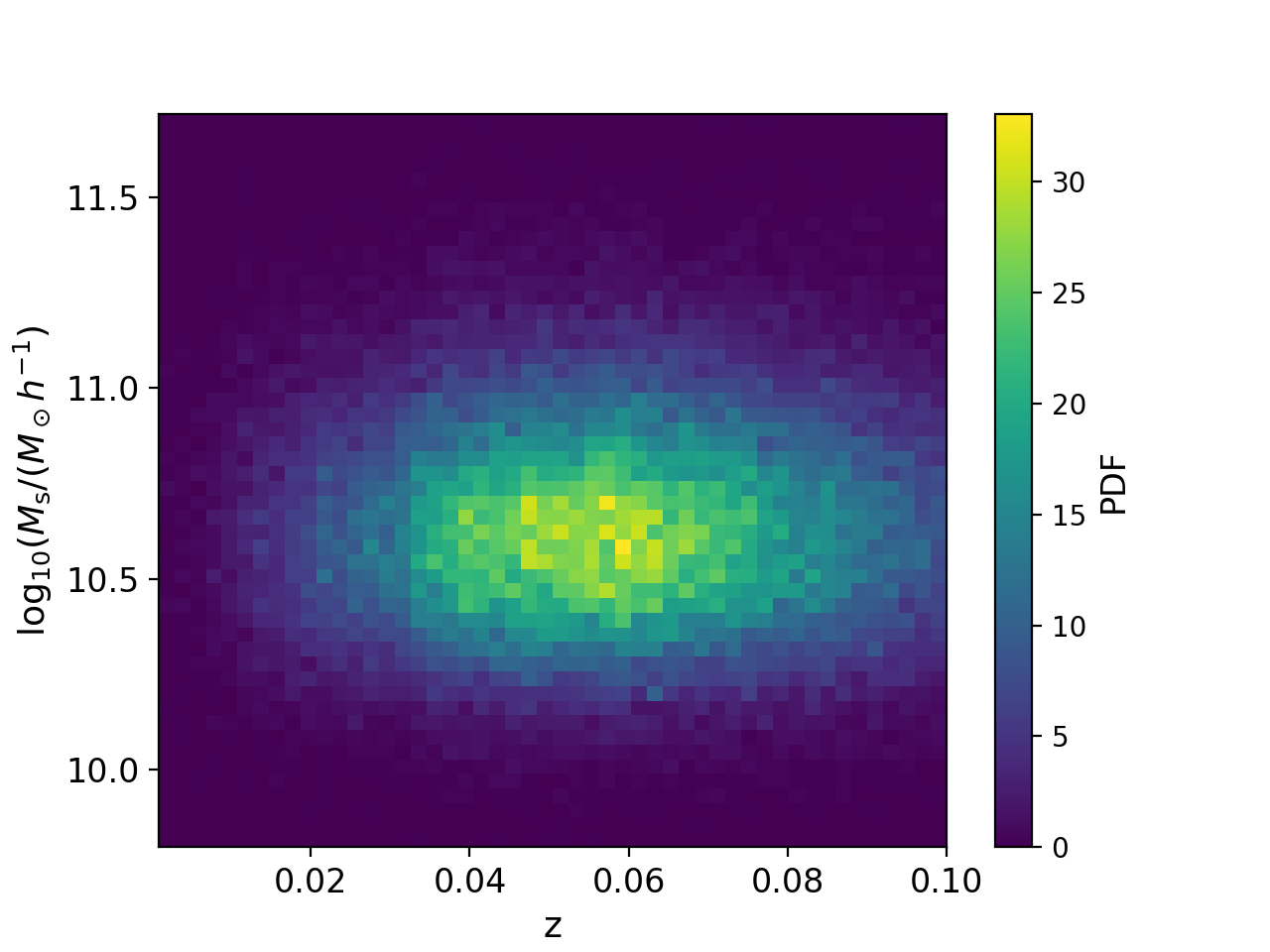}
}
\subfigure[BNS, wff2, $\mathbf{Cov}(\alpha,\delta,z)$]{
	\includegraphics[width=8cm]{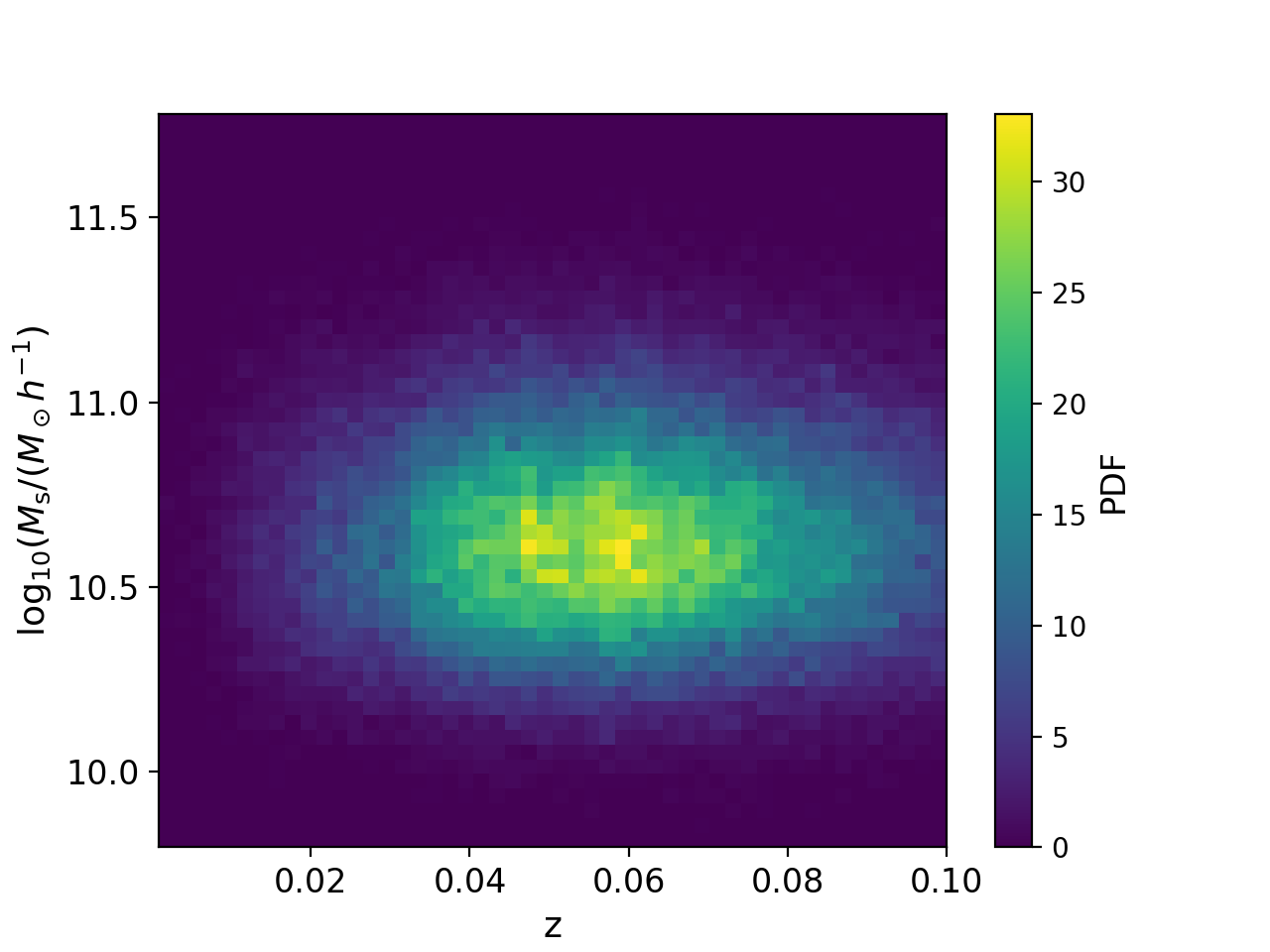}
}
\hspace{1.0pt}
\subfigure[NSBH, wff2, $\mathbf{Cov}(\alpha,\delta,z)$]{
	\includegraphics[width=8cm]{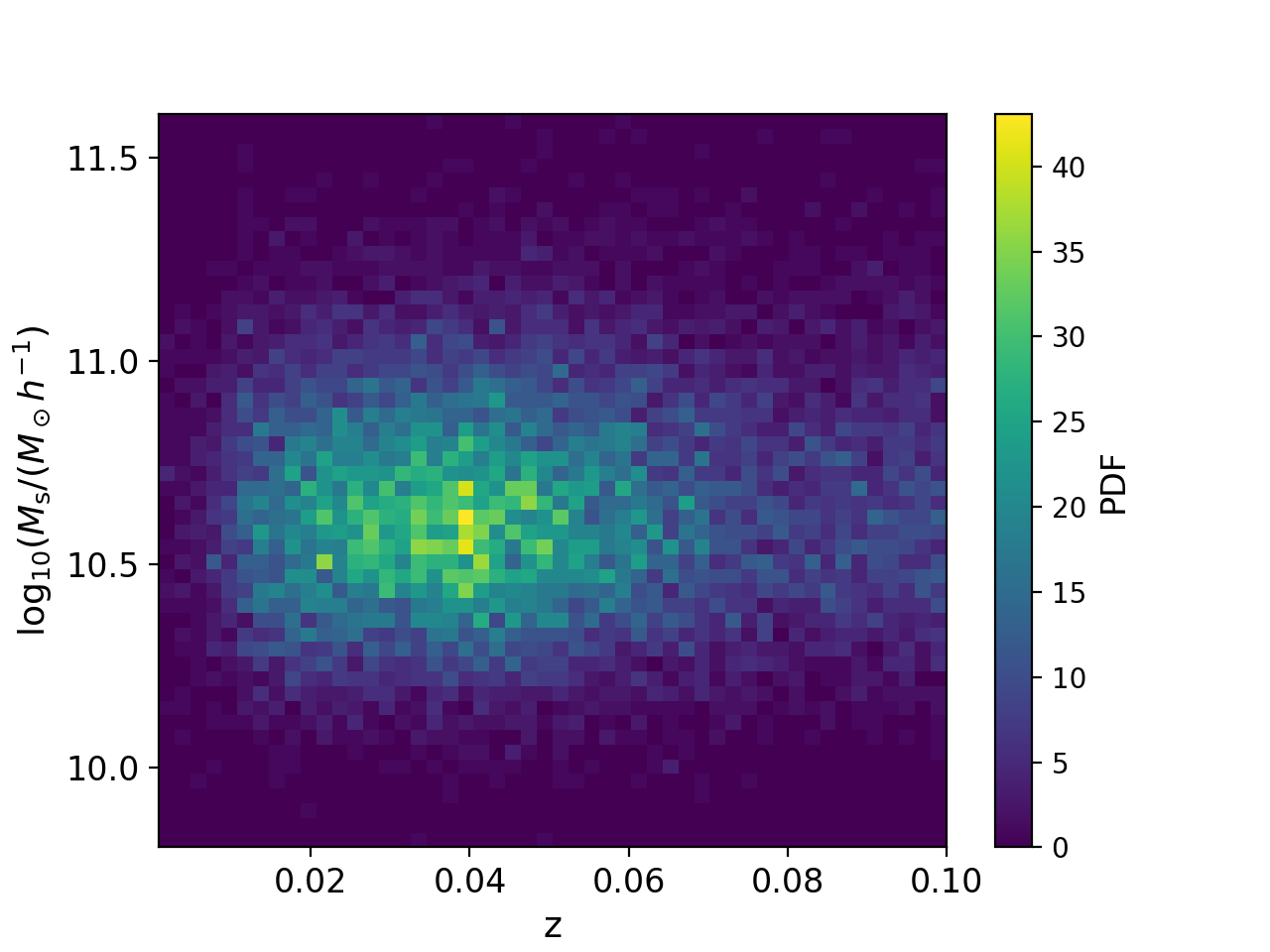}
}
\caption{Redshift and stellar mass distributions of BNS mergers' host groups with $N_\mathrm{in}=1$. The left and right panels represent the cases of BNS and NSBH mergers, respectively. The upper panels show the constraints from $\mathrm{Cov}(\alpha, \delta, \log(d_L))$. The middle and lower panels show the constraints from $\mathrm{Cov}(\alpha, \delta, z)$.}
\label{figure_m_s}
\end{figure*}

In Figure \ref{figure_m_s}, we compare the host groups' redshift and stellar mass distributions for those $N_\mathrm{in}$ events.  From the left panels, we can see that BNS mergers around $z\sim0.07$ have the largest directly identified rate through the method of $\mathbf{Cov}[\alpha, \delta, \log(d_L)]$. In the case of tidal effect method with ms1 model, the distributions are almost same. However, in the case of wff2, the maximum rate occurs at $z\sim0.06$ due to a worse localization ability. For NSBH mergers, because of stronger SNR, the maximum $N_\mathrm{in}=1$ rate in the case of $\mathbf{Cov}[\alpha, \delta, \log(d_L)]$ occurs at $z\sim0.09$. For the cases of $\mathbf{Cov}[\alpha, \delta, z]$ with ms1 and wff2 models, redshifts corresponding to the maximum rate are reduced to $z\sim0.06$ and $\sim0.04$, respectively.

\begin{table}
\renewcommand\arraystretch{1.5} 
\begin{center}
\begin{tabular}{ccccc}
\hline\hline
&1&$\leq2$&$\leq5$&$\leq10$\\
$N_\mathrm{in}$&(per cent)&(per cent)&(per cent)&(per cent)\\
\hline
alf2&37.7&54.7&76.1&87.4\\
ap3&30.4&45.6&67.1&80.5\\
ap4&25.9&39.7&60.8&74.9\\
bsk21&34.0&50.1&71.7&84.2\\
eng&30.4&45.7&67.3&80.6\\
H4&35.7&52.2&73.8&85.7\\
mpa1&32.8&48.7&70.3&83.0\\
ms1&44.8&62.8&82.9&92.1\\
ms1b&43.5&61.3&81.7&91.3\\
qmf40&29.2&44.0&65.5&79.1\\
qmf60&30.6&45.8&67.4&80.7\\
sly&28.0&42.6&64.0&77.8\\
wff2&24.3&37.65&58.3&72.7\\
MIT2&34.8&51.2&72.7&85.0\\
MIT2cfl&34.9&51.4&72.9&85.1\\
pQCD800&28.6&43.3&64.7&78.5\\
sqm3&31.0&46.4&68.0&81.2\\
\hline
\end{tabular}
\end{center}
\caption{The fractions of the BNS samples with different values of $N_\mathrm{in}$ and EOS.}
\label{table_bns}
\end{table}

\begin{table}
\renewcommand\arraystretch{1.5} 
\begin{center}
\begin{tabular}{ccccc}
\hline\hline
&1&$\leq2$&$\leq5$&$\leq10$\\
$N_\mathrm{in}$&(per cent)&(per cent)&(per cent)&(per cent)\\
\hline
alf2&9.9&16.0&26.9&36.6\\
ap3&5.4&8.8&16.1&23.8\\
ap4&3.5&5.9&11.1&17.3\\
bsk21&7.3&12.0&21.0&29.8\\
eng&5.4&8.9&16.2&24.0\\
H4&12.3&19.4&31.6&41.8\\
mpa1&6.6&10.9&19.3&27.8\\
ms1&16.1&24.9&38.8&49.3\\
ms1b&14.9&23.1&36.6&47.0\\
qmf40&4.8&8.0&14.6&22.0\\
qmf60&5.5&9.0&16.3&24.1\\
sly&4.3&7.2&13.4&20.3\\
wff2&3.0&5.0&9.7&15.4\\
MIT2&7.9&12.8&22.3&31.4\\
MIT2cfl&8.0&13.0&22.6&31.8\\
pQCD800&4.6&7.6&14.0&21.2\\
sqm3&5.7&9.4&16.9&24.9\\
\hline
\end{tabular}
\end{center}
\caption{Same as Table \ref{table_bns}, but with NSBH mergers.}
\label{table_nsbh}
\end{table}

\begin{table}
\renewcommand\arraystretch{1.5} 
\begin{center}
\begin{tabular}{ccccc}
\hline\hline
&1&$\leq2$&$\leq5$&$\leq10$\\
$N_\mathrm{in}$&(per cent)&(per cent)&(per cent)&(per cent)\\
\hline
BNS&53.5&72.7&90.8&96.7\\
NSBH&69.7&87.0&97.6&99.5\\
BBH&93.2&98.6&99.9&99.97\\
\hline
\end{tabular}
\end{center}
\caption{The fractions of $N_\mathrm{in}$ with error ellipsoids from $\mathbf{Cov}[\alpha, \delta, \log(d_L)]$.}
\label{table_dL}
\end{table}

\section{Applications of Dark Sirens in $H_0$ Measurement}
\label{section_dsH0}
By comparing the galaxy/group catalog with GW signal's localization region, we can get GW event's redshift distribution from the redshift statistics of its probable hosts. This approach was proposed by \cite{2012PhRvD..86d3011D} and \cite{2018Natur.562..545C}, and has been applied in \cite{2019ApJ...871L..13F}, \cite{2019ApJ...876L...7S}, \cite{2020PhRvD.101l2001G}, \cite{2020MNRAS.498.1786Y}, \cite{2020ApJ...900L..33P}, \cite{2020ApJ...905L..28B}, \cite{2021JCAP...08..026F}, \cite{2021arXiv211103604T}, \cite{2021arXiv211106445P}, \cite{2022arXiv220303643M}, \cite{2022SCPMA..6559811Z} and \cite{2022MNRAS.512.1127G}. Generally, these analyses use the constraints on $\{\alpha,\delta,\log(d_L)\}$ given by the GW detector network, take specific value of Hubble constant $H_0$ as priors, transform the luminosity distance measurements into redshift space, compare them with the galaxy/group catalog, and calculate the likelihood of $H_0$. Then, through the Bayes' theorem, we can obtain the posterior probability distribution of $H_0$. Since this method relies on the prior of $H_0$, the existence of galaxies/groups with different redshifts in the localized area will have a significant impact on the posterior of $H_0$.\par 
Observations of tidal effects, on the other hand, provide an independent method of measuring the redshifts of GW events. Thus, the search for potential hosts through constraints on $\{\alpha,\delta,z\}$ will not rely on the prior of specific cosmological models. We will compare these two methods of dark sirens' redshift measurement in this section.\par

\subsection{Constraints from $\mathbf{Cov}[\alpha, \delta, \log(d_L)]$}

\begin{figure}[htbp]
\centering
\includegraphics[width=9cm]{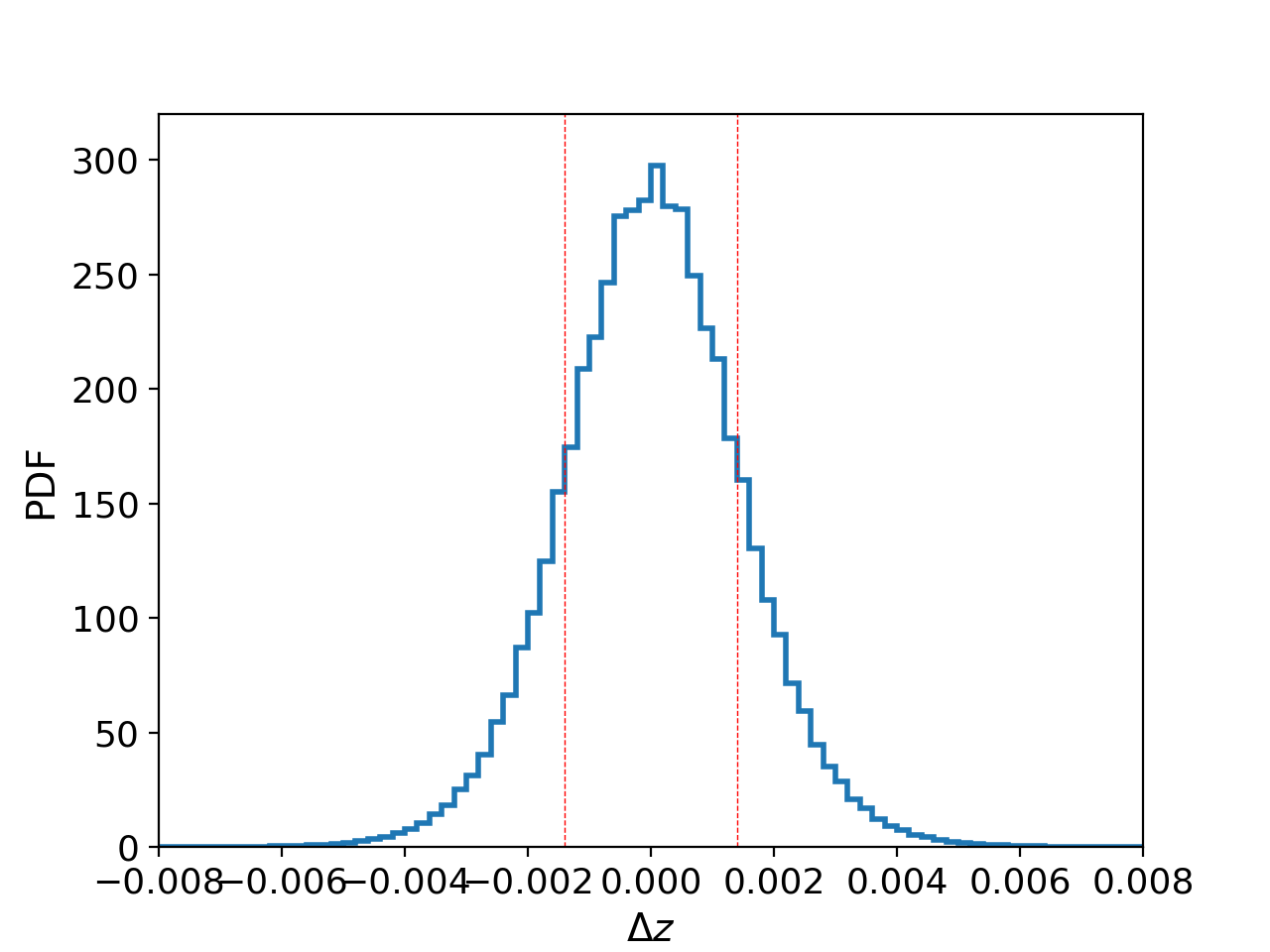}
\caption{The PDF of $\Delta z=z_\mathrm{obs}-z_\mathrm{com}$.}
\label{figure_z_o}
\end{figure}

Following the Bayes' theorem, the posterior of $H_0$ can be written as 
\be
	p(H_0|d_\mathrm{GW},d_\mathrm{EM})\propto p(d_\mathrm{GW},d_\mathrm{EM}|H_0)p(H_0),
\ee
where $d_\mathrm{GW}$ and $d_\mathrm{EM}$ represent the GW signals observed by GW detector network and the data from galaxy group catalog, respectively. The term $p(H_0)$ represents the prior of $H_0$. In this section, we discuss a prior with uniform distribution in the interval $[20,140]\ \mathrm{km}\ \mathrm{s}^{-1}\ \mathrm{Mpc}^{-1}$. Following \cite{2018Natur.562..545C}, the likelihood $p(d_\mathrm{GW},d_\mathrm{EM}|H_0)$ can be written as an integral over the solid angle $\Omega$ and redshift $z$
\be
\begin{split}
	p(d_\mathrm{GW},d_\mathrm{EM}|H_{0})&\propto\frac{1}{\beta(H_{0})}\int p(d_\mathrm{GW}|d_{L}(z,H_{0}),\Omega)\times\\
	&p(d_\mathrm{EM}|z,\Omega)p_{0}(z,\Omega)\dif\Omega \dif z,
\label{eq_likelihood}
\end{split}
\ee
where the GW likelihood $p(d_\mathrm{GW}|d_{L}(z,H_{0}),\Omega)$ could be obtained from $\mathbf{Cov}[\alpha,\delta,\log(d_L)]$. We ignore the uncertainty of groups’ positions and assume their redshifts follow a Gaussian distribution $\mathcal{N}(z_\mathrm{obs}, \sigma_z|z)$, where $z_\mathrm{obs}$ is the group's observed redshift, $\sigma_z$ is uncertainty of $z_\mathrm{obs}$. In Figure \ref{figure_z_o}, we show the distribution of the difference between the groups' comoving redshift $z_\mathrm{com}$ and observed redshift $z_\mathrm{obs}$. The distribution of $\Delta z=z_\mathrm{obs}-z_\mathrm{com}$ follows a Gaussian distribution with $\sigma=0.0014$, hence we set $\sigma_z=0.0014$. Thus, the EM likelihood $p(d_\mathrm{EM}|z,\Omega)$ could be written as
\be
	p(d_\mathrm{EM}|z,\Omega)\propto\sum_{i}w_i\delta(\Omega-\Omega_i)\mathcal{N}(z_{\mathrm{obs},i}, \sigma_{z}|z),
\label{eq_emlikelihood}
\ee
where $(z_{\mathrm{obs},i}, \Omega_i)$ represent the redshift and position of $i$-th possible host groups, and $w_i$ is the weight of each group, which represent the probability that different groups host a GW events. This weight is related to many factors, such as the stellar mass $M_\mathrm{s}$, morphologies, and metallicities \citep{2018MNRAS.474.4997C}. In this section, we choose two different weight function, $w_i=1$ and $w_i\propto M_\mathrm{s}$, as a comparison. \par
For the group prior $p_0(z,\Omega)$, we assume the groups are isotropic distributed on large scales, and $p_0(z,\Omega)$ could be written as \citep{2018Natur.562..545C, 2019ApJ...871L..13F}
\be
	p_0(z,\Omega)\propto\sum_i \mathcal{N}(z_{\mathrm{obs},i}, \sigma_{z}|z),
\label{eq_prior}
\ee
where the summation symbol $\sum$ is applied to the whole catalog. There is another approach that assume the galaxies/groups are uniformly distributed in comoving volume $V$. Therefore, $p_0(z, \Omega)$ could be obtain from \citep{2019ApJ...876L...7S, 2023AJ....166...22G}
\begin{equation}
    p_0(z, \Omega)\propto \frac{\mathrm{d}^2 V}{\mathrm{d}z\mathrm{d}\Omega} \propto \frac{\chi^2(z)}{H(z)},
\end{equation}
where $\chi(z)$ is the comoving distance and $H(z)$ is Hubble parameter with respect to redshift. In our previous work \cite{2020MNRAS.498.1786Y}, we discussed both approaches with the SDSS DR7 group catalog and found when the catalog is complete and the number density of the groups in comoving space does not vary with redshift, the results obtained by the two methods are almost the same.  In this work, we are mainly concerned with the dark sirens at $z<0.1$, a redshift range much smaller than the imcomplete redshift of our group catalog. Therefore, in this paper we only consider the former method. \par
Following \cite{2018MNRAS.474.4997C} and \cite{2019ApJ...871L..13F}, the normalization term $\beta(H_0)$ in Eq. (\ref{eq_likelihood}) could be written as
\be
\begin{split}
	\beta(H_{0})&=\int_{d_\mathrm{EM}>\mathrm{thresh}} \int_{d_\mathrm{GW}>\mathrm{thresh}}
	\int p(d_\mathrm{GW}|d_{L}(z,H_{0}),\Omega)\times\\
	&\quad\quad\quad\quad p(d_\mathrm{EM}|z,\Omega)p_{0}(z,\Omega)\dif\Omega \dif z \dif d_\mathrm{EM}\dif d_\mathrm{GW},\\
	&=\int P^\mathrm{GW}_\mathrm{det}(d_{L}(z,H_{0}),\Omega)P^\mathrm{EM}_\mathrm{det}(z,\Omega)p_{0}(z, \Omega)\dif\Omega \dif z\\
	&=\int^{z_{h}}_{0}\int P^\mathrm{GW}_\mathrm{det}(d_{L}(z,H_{0}),\Omega)p_{0}(z,\Omega)\dif\Omega \dif z,
\end{split}
\ee
where $P^\mathrm{GW}_\mathrm{det}(d_{L}(z,H_{0}),\Omega)$ and $P^\mathrm{EM}_\mathrm{det}(z,\Omega)$ are defined as
\be
	P^\mathrm{GW}_\mathrm{det}(d_{L}(z,H_{0}),\Omega) = \int_{d_\mathrm{GW}>\mathrm{thresh}} p(d_\mathrm{GW}|d_{L}(z,H_{0}),\Omega) \dif d_\mathrm{GW},
\ee
and 
\be
	P^\mathrm{EM}_\mathrm{det}(z,\Omega) = \int_{d_\mathrm{EM}>\mathrm{thresh}}p(d_\mathrm{EM}|z,\Omega) \dif d_\mathrm{EM}=\mathcal{H}(z-z_{h}),
\ee
where $\mathcal{H}$ is the Heaviside step function. For $d_\mathrm{EM}$, we set its thresold as $z_h<0.5$. For $d_\mathrm{GW}$, we set its threshold as SNR$>12$. Since we only consider low redshift ($z<0.1$) BNS/NSBH mergers, and the detection limit of 3G GW detector network is much farther than that \citep{2021ApJ...916...54Y}, thus for all $H_0\in[20,140]\ \mathrm{km}\ \mathrm{s}^{-1}\ \mathrm{Mpc}^{-1}$, the values of $\beta(H_0)$ are same.\par
For N events with sequence numbers 1, 2, 3, $\cdots$, $j$, the posterior distribution of $H_0$ has the following formulation
\be
\begin{split}
	p(H_0|d_\mathrm{GW},d_\mathrm{EM})\propto&\prod_{j=1}^{N}\frac{p(H_0)}{\beta(H_0)}\sum_{i}w_i\int P^\mathrm{GW}_\mathrm{det}(d_{L}(z,H_{0}),\Omega_{ij})\\
	&\times \mathcal{N}(z_{\mathrm{obs},ij}, \sigma_{z}|z)p_0(z,\Omega_{ij})\dif z.
\end{split}
\ee

\subsection{Constraints from $\mathbf{Cov}[\alpha,\delta,z]$}

Since the observations of tidal effect can directly give constraints on $z$, the PDF of source's redshift could be obtained from the Bayes' theorem
\begin{equation}
    P(z|d_\mathrm{GW}, d_\mathrm{EM})\propto p_0(z)p(d_\mathrm{GW}, d_\mathrm{EM}|z),
\end{equation}
where $p_0(z)$ is the redshift prior. This expression can be expanded as
\be
\begin{split}
	P(z|d_\mathrm{GW}, d_\mathrm{EM})&\propto \int p_0(z)p_0(\Omega)p(d_\mathrm{GW}, d_\mathrm{EM}|z,\Omega)\dif\Omega\\
    &=\int p_0(z,\Omega)p(d_\mathrm{GW}|z,\Omega)p(d_\mathrm{EM}|z,\Omega)\dif\Omega,
\end{split}
\label{eq_zpdf}
\ee
where $p_0(\Omega)$ represents the probability of producing GW events at $\Omega$ and $p_0(z, \Omega)\equiv p_0(z) p_0(\Omega)$. For this term, we set it to be proportional to the group distribution in catalog, so it can be obtained from Eq. (\ref{eq_prior}). Using Eq. (\ref{eq_emlikelihood}), Eq. (\ref{eq_zpdf}) can be rewritten as 
\begin{equation}
 P(z|d_\mathrm{GW}, d_\mathrm{EM})=\sum_i w_i p(d_\mathrm{GW}|z,\Omega_i)\times\mathcal{N}(z_{\mathrm{obs},i},\sigma_z|z)p_{0}(z,\Omega_i),
\label{eq_emlikeliehood2}
\end{equation}
where $p(d_\mathrm{GW}|z,\Omega_i)$ could be obtained from $\mathbf{Cov}[\alpha,\delta,z]$. As for the posterior distribution $P(H_0|d_\mathrm{GW}, d_\mathrm{EM})$, its CDF is
\begin{equation}
    P(H<H_0|d_\mathrm{GW}, d_\mathrm{EM})=\int\dif z\int_{0}^{d_{L}(z,H_0)}\dif \hat{d_{L}} p(z,\hat{d_{L}}|d_\mathrm{GW}, d_\mathrm{EM}),
\end{equation}
after differentiation of $H_0$ and combining the $\Delta\log(d_L)$ from the Fisher matrix method, we can obtain the posterior distribution of $H_0$ from
\be
\begin{split}
	P(H_0|d_\mathrm{GW}, d_\mathrm{EM})=&\int P(z|d_\mathrm{GW}, d_\mathrm{EM}) P(d_L|d_\mathrm{GW}) \\
    &\quad\times\left. \frac{\dif d_L(z, H)}{\dif H}\right |_{H=H_0}\dif z.
\end{split}
\ee

\subsection{Results}
\subsubsection{$w_i=1$}
\begin{figure*}[htbp]
\centering
\subfigure[BNS, $\mathbf{Cov}(\alpha, \delta,\log(d_L))$]{
	\includegraphics[width=8.5cm]{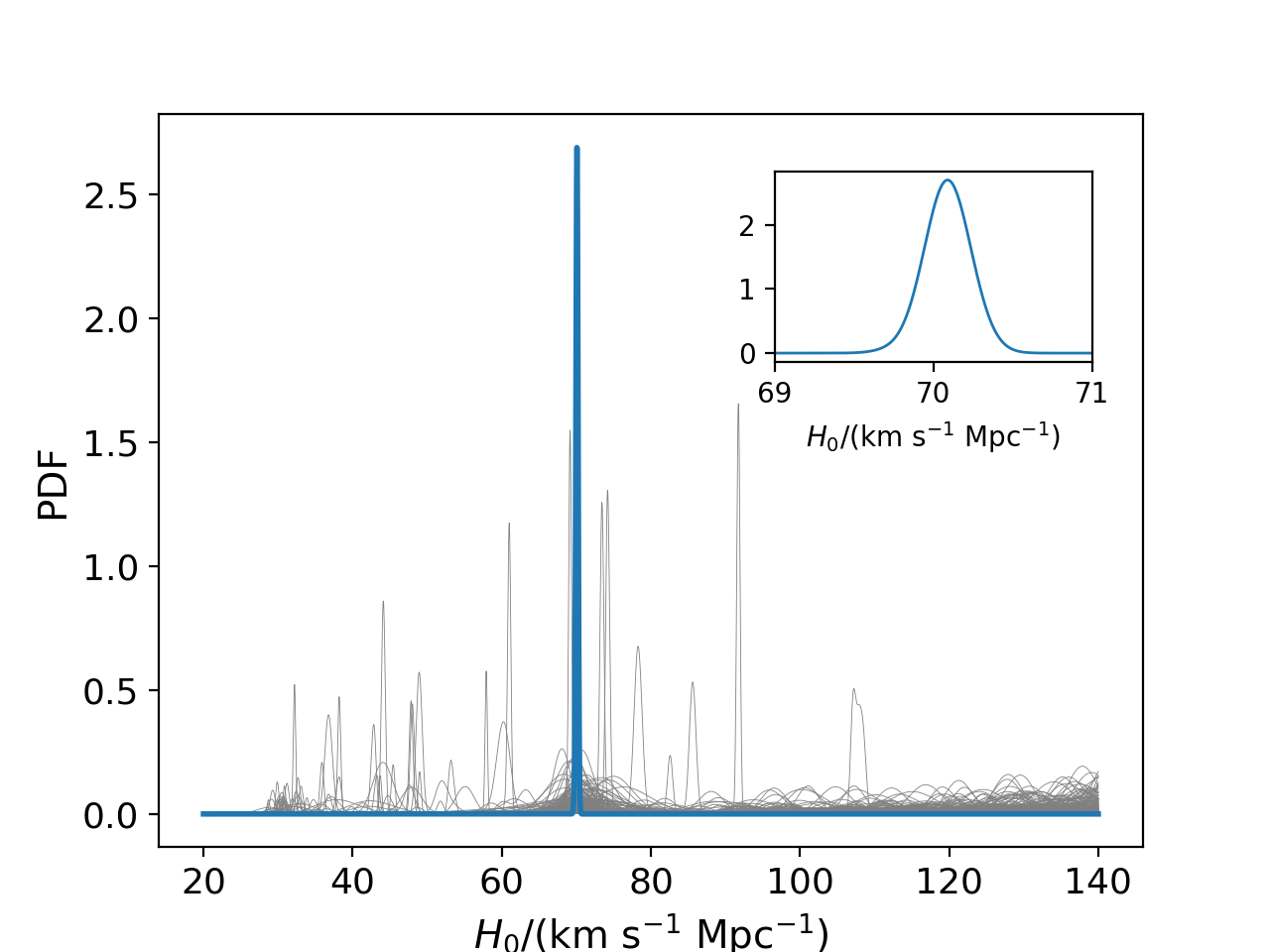}
}
\hspace{1.0pt}
\subfigure[NSBH, $\mathbf{Cov}(\alpha, \delta, \log(d_L))$]{
	\includegraphics[width=8.5cm]{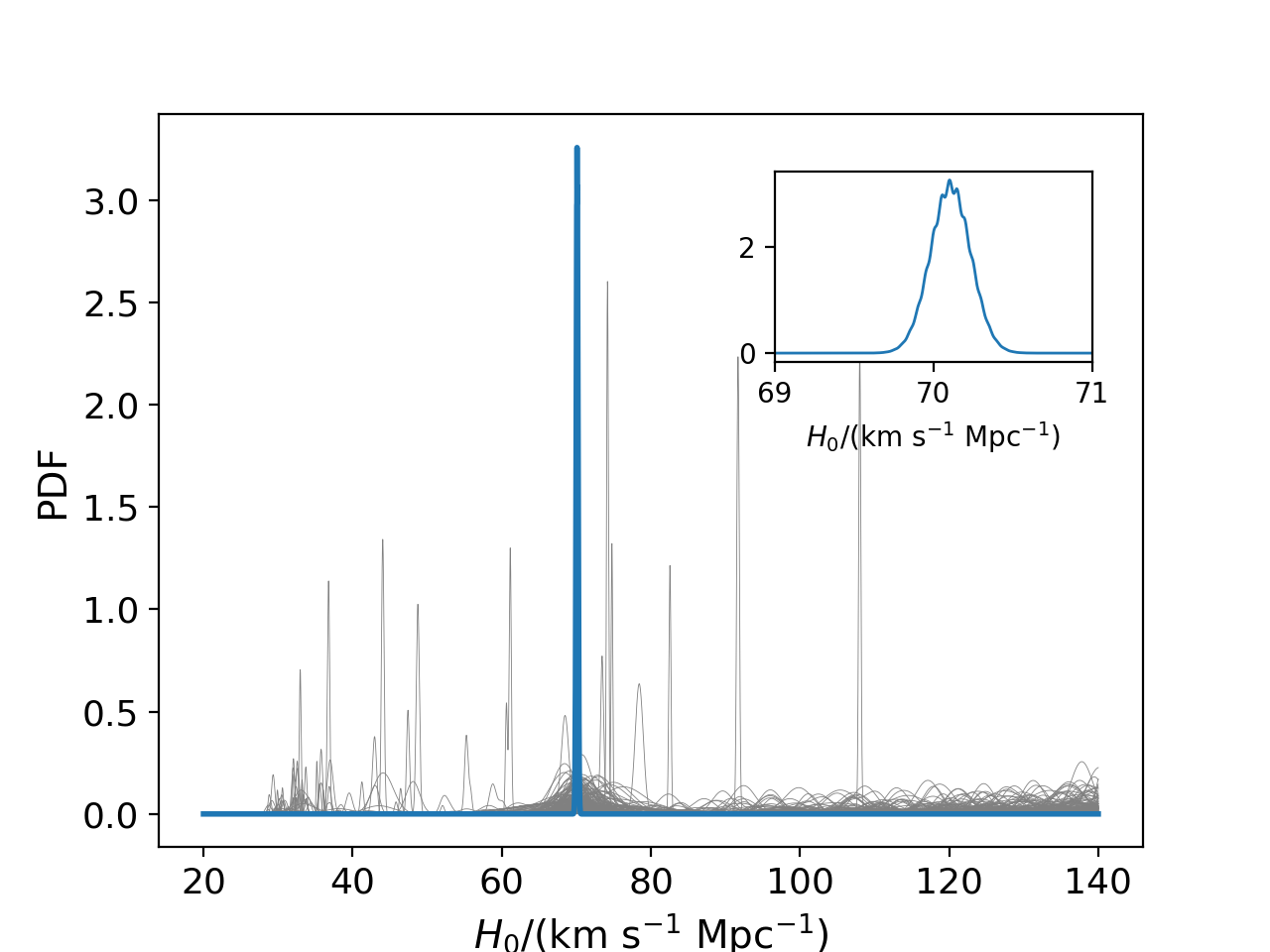}
}
\hspace{1.0pt}
\subfigure[BNS, ms1, $\mathbf{Cov}(\alpha,\delta,z)$]{
	\includegraphics[width=8.5cm]{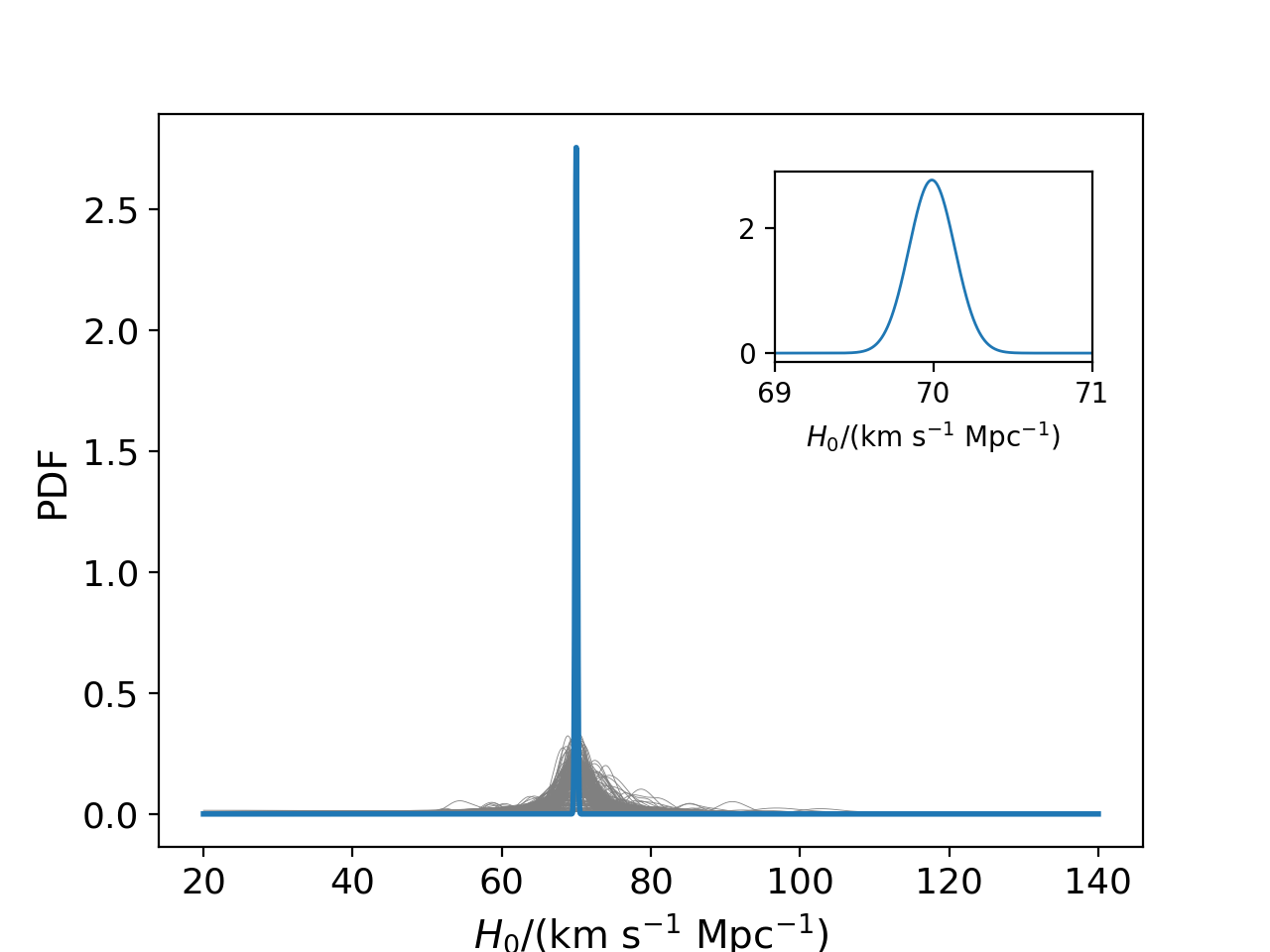}
}
\hspace{1.0pt}
\subfigure[NSBH, ms1, $\mathbf{Cov}(\alpha,\delta,z)$]{
	\includegraphics[width=8.5cm]{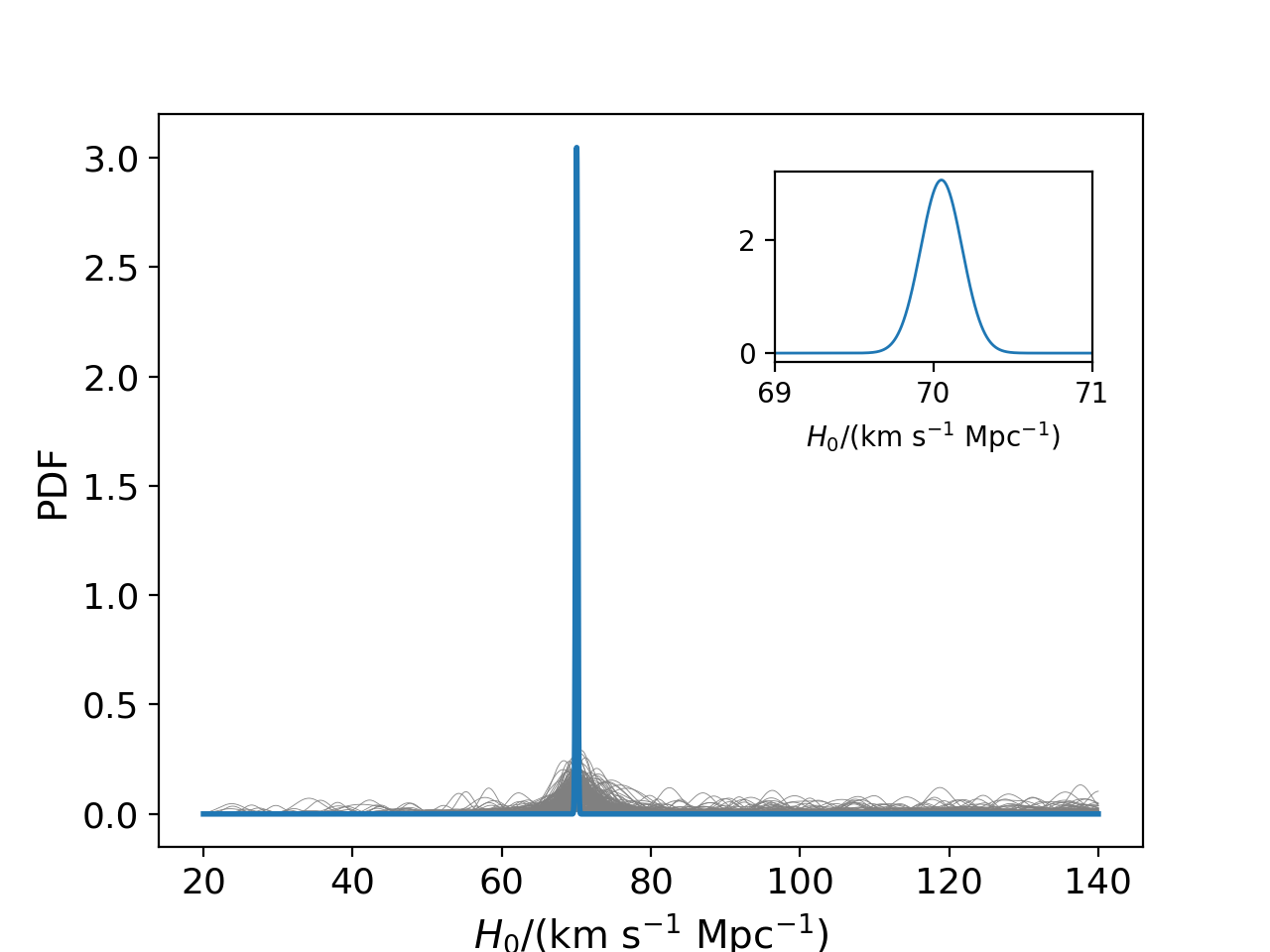}
}
\subfigure[BNS, wff2, $\mathbf{Cov}(\alpha,\delta,z)$]{
	\includegraphics[width=8.5cm]{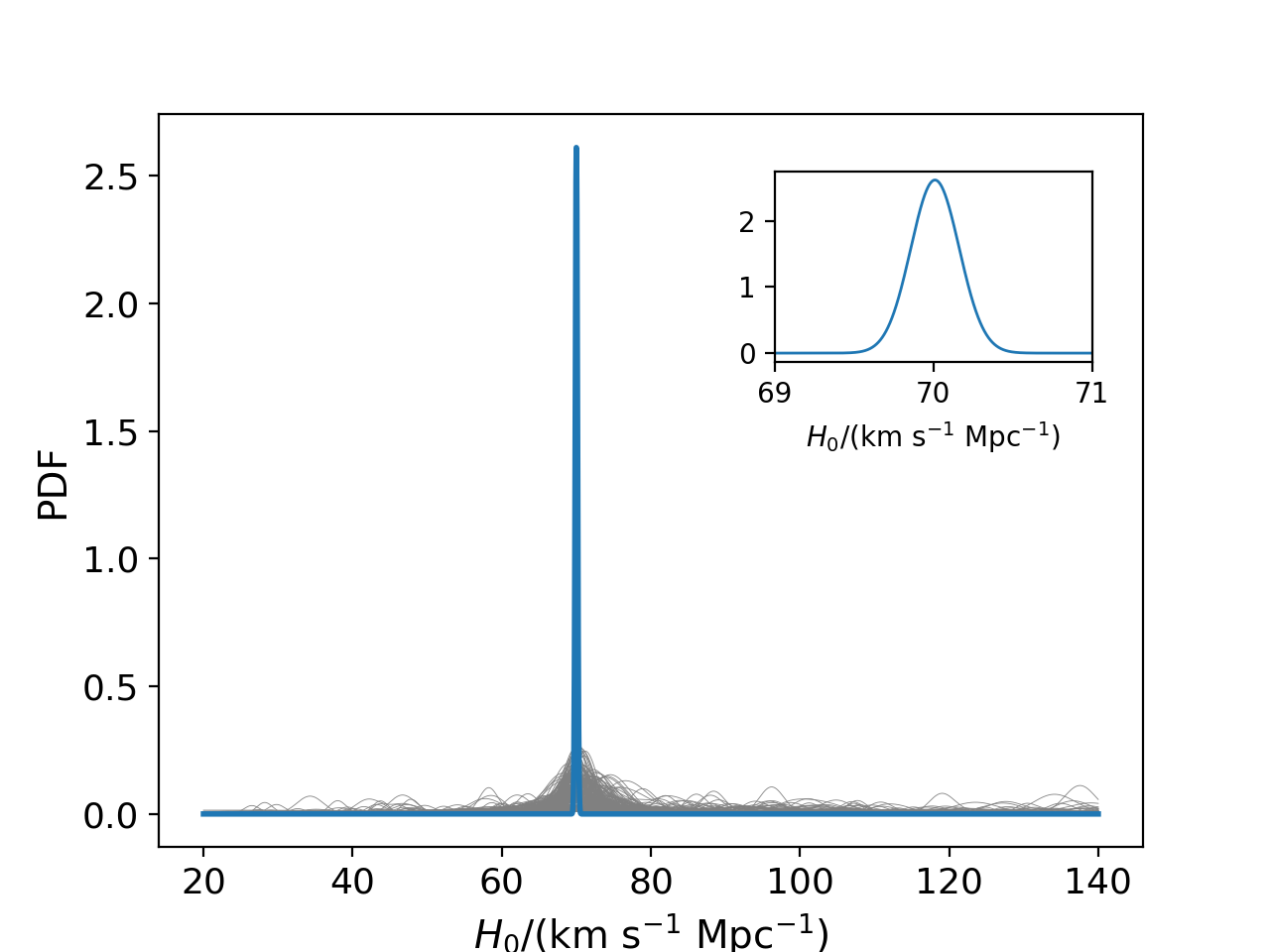}
}
\hspace{1.0pt}
\subfigure[NSBH, wff2, $\mathbf{Cov}(\alpha,\delta,z)$]{
	\includegraphics[width=8.5cm]{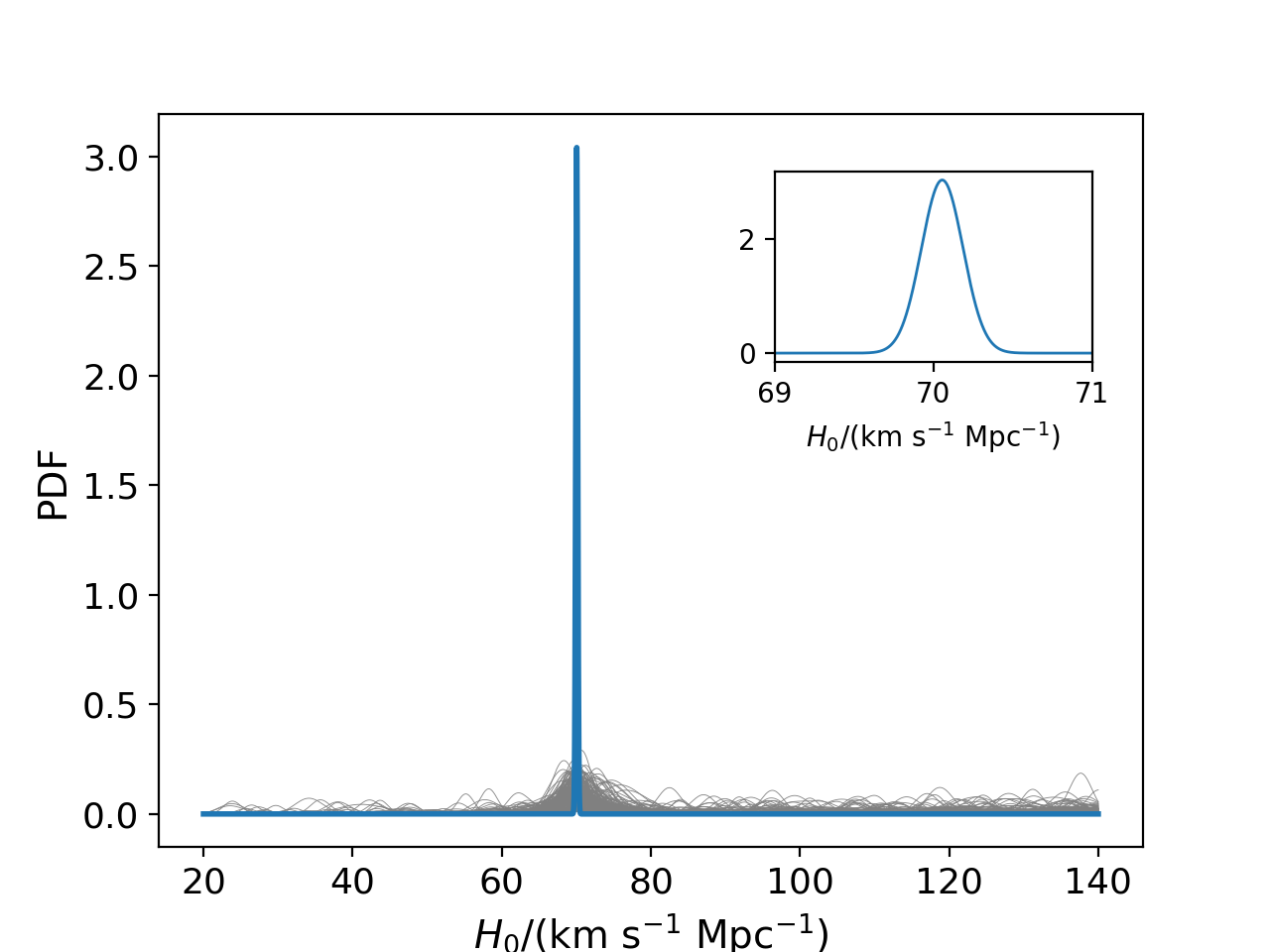}
}
\caption{The posterior distributions of Hubble constant $H_0$ from 200 BNS/NSBH samples with  $z<0.1$. The left and right panels represent the cases of BNS and NSBH mergers, respectively. The upper panels show the constraints from $\mathbf{Cov}(\alpha, \delta,\log(d_L))$. The middle and lower panels show the constraints from $\mathbf{Cov}(\alpha,\delta,z)$. The grey lines show the constraints with a single event on $H_0$ and the blue line is the sum of them.}
\label{figure_LH0}
\end{figure*}

In order to compare the abilities of these two methods in the Hubble constant constraints, we sample 200 assumed BNS/NSBH mergers at the center of the low redshift ($z<0.1$) galaxy groups, where their redshifts equal to the groups comoving redshift $z_\mathrm{com}$ and $d_L$ equal to $d_L(z_\mathrm{com},H_0=70\ \mathrm{km}\ \mathrm{s}^{-1}\ \mathrm{Mpc}^{-1})$. In this section, to be consistent with the mock group catalog, we adopt the $\Lambda$CDM model and use the fixed value of $\Omega_m$ mentioned in Section \ref{intro}. First, we consider the case of $w_i=1$. The posterior distributions of $H_0$ from these two methods in this case are showed in Figure \ref{figure_LH0} and Table \ref{table_H0}. The grey and blue lines in Figure \ref{figure_LH0} represent the constraints on $H_0$ for a single event and for 200 events, respectively. \par

In the upper left panel of Figure \ref{figure_LH0}, we show the constraints from the method of $\mathrm{Cov}(\alpha,\delta,\log{d_L})$ and BNS mergers. It can be seen that since this method relies on the prior of $H_0$, the posterior distribution given by a single event will exhibit a significant peak at values different from $H_0=70\ \mathrm{km}\ \mathrm{s}^{-1}\ \mathrm{Mpc}^{-1}$ when there are other groups in the localization region. Therefore, when there are only a few events, these peaks may cause the measurement of the Hubble constant to deviate from the true value. However, as the number of events increases, the posterior distribution will approach a normal distribution with a central value of $70\ \mathrm{km}\ \mathrm{s}^{-1}\ \mathrm{Mpc}^{-1}$. As a result, the constraint from these 200 BNS mergers is $H_0=70.09_{-0.15}^{+0.15}\ \mathrm{km}\ \mathrm{s}^{-1}\ \mathrm{Mpc}^{-1}$. For NSBH samples, the result is $H_0=70.10_{-0.12}^{+0.14}\ \mathrm{km}\ \mathrm{s}^{-1}\ \mathrm{Mpc}^{-1}$, which is slightly better than the case of BNS mergers due to a smaller error ellipsoid. The posterior distribution in this case is presented in the upper right panel of Figure \ref{figure_LH0}.\par

Now, we turn to the method of $\mathrm{Cov}(\alpha,\delta,z)$. The results in the case of ms1 model are show in the middle panels of Figure \ref{figure_LH0}. For BNS mergers, since the identification of the host groups does not depend on the $H_0$, the posterior distribution of all events is centered around $70\ \mathrm{km}\ \mathrm{s}^{-1}\ \mathrm{Mpc}^{-1}$. Thus, when the number of events is small, the method of tidal effect observations will give a more reliable measurement of the Hubble constant. Eventually, the sum of the posterior distributions is $H_0=69.99_{-0.14}^{+0.15}\ \mathrm{km}\ \mathrm{s}^{-1}\ \mathrm{Mpc}^{-1}$. This constraint is about $3\%$ better than the constraint from the method of $\mathrm{Cov}(\alpha,\delta,\log{d_L})$. In the middle right panel, since a part of the NSBH mergers with larger mass ratio have a poor constraint on redshift, some peaks deviate from $70\ \mathrm{km}\ \mathrm{s}^{-1}\ \mathrm{Mpc}^{-1}$. However, on the other hand, the stronger SNRs of the NSBH mergers lead to a smaller $\Delta \log(d_L)$. Thus, in total, the 200 NSBH mergers achieve a constraint of $70.05_{-0.13}^{+0.13}\ \mathrm{km}\ \mathrm{s}^{-1}\ \mathrm{Mpc}^{-1}$ for the Hubble constant, which is $\sim10\%$ better than BNS mergers. Since NSBH's local merging rate is one-third that of BNS's, after normalized, this constraint is $\sim1/3$ worse than the BNS mergers.\par

As a comparison, we also consider the EOS with the worst result in Section \ref{subsection_nin}, wff2. The results are showed in the bottom panels of Figure \ref{figure_LH0}. For BNS mergers, 200 events can constrain the Hubble constant to $H_0=70.01_{-0.15}^{+0.15}\ \mathrm{km}\ \mathrm{s}^{-1}\ \mathrm{Mpc}^{-1}$. This result is $\sim6\%$ worse than the case of ms1 model. For NSBH systems, the constraint is almost the same with the case of ms1 model, which is $H_0=70.06_{-0.13}^{+0.13}\ \mathrm{km}\ \mathrm{s}^{-1}\ \mathrm{Mpc}^{-1}$. This difference is significantly smaller than the difference in localization volume obtained by the two models. This is due to the fact that the measurement error in the $d_L$ dominates the $H_0$ constraints.\par


\begin{table}
\begin{center}
\begin{tabular}{ccc}
\hline\hline
&&$H_0$\\
&&($\mathrm{km}\ \mathrm{s}^{-1}\ \mathrm{Mpc}^{-1}$)\\
\hline
BNS&$\mathbf{Cov}[\alpha, \delta, \log(d_L)]$&$70.09_{-0.15}^{+0.15}$ $\left(_{-0.12}^{+0.12}\right)$\\\\
&$\mathbf{Cov}[\alpha, \delta, z]$, ms1&$69.99_{-0.14}^{+0.15}$ $\left(_{-0.11}^{+0.12}\right)$\\\\
&$\mathbf{Cov}[\alpha, \delta, z]$, wff2&$70.01_{-0.15}^{+0.15}$ $\left(_{-0.12}^{+0.12}\right)$\\\\
NSBH&$\mathbf{Cov}[\alpha, \delta, \log(d_L)]$&$70.10_{-0.12}^{+0.14}$ $\left(_{-0.17}^{+0.19}\right)$\\\\
&$\mathbf{Cov}[\alpha, \delta, z]$, ms1&$70.05_{-0.13}^{+0.13}$ $\left(_{-0.18}^{+0.18}\right)$\\\\
&$\mathbf{Cov}[\alpha, \delta, z]$, wff2&$70.06_{-0.14}^{+0.13}$ $\left(_{-0.18}^{+0.18}\right)$\\
\hline
\end{tabular}
\end{center}
\caption{The constraints of $H_0$ from different methods and EOSs with $w_i=1$. The numbers outside and inside the brackets denote the error bars for 200 mergers and normalized to five years of observation time, respectively. }
\label{table_H0}
\end{table}

\subsubsection{$w_i\propto M_\mathrm{s}$}
Usually, the choice of the weighting factor $w_i$ is a very important item in the dark standard siren method. When the number of groups in the localization volume is large, $w_i$ will have a large influence on the Hubble constant measurement and will make the results rely on the GW source population assumptions. \par
\begin{figure*}[htbp]
\centering
\subfigure[BNS, $\mathbf{Cov}(\alpha, \delta,\log(d_L))$]{
	\includegraphics[width=8.5cm]{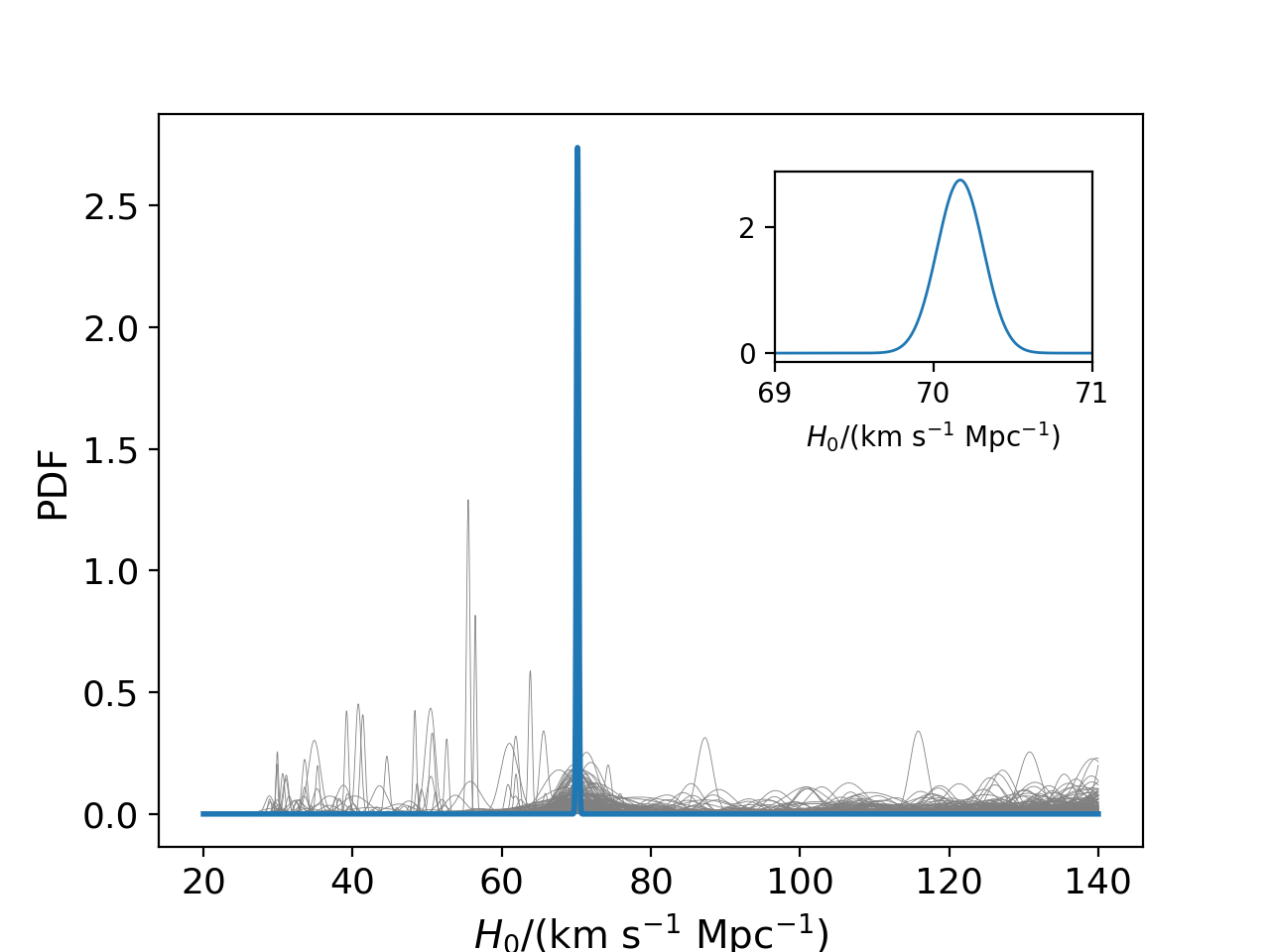}
}
\hspace{1.0pt}
\subfigure[NSBH, $\mathbf{Cov}(\alpha, \delta, \log(d_L))$]{
	\includegraphics[width=8.5cm]{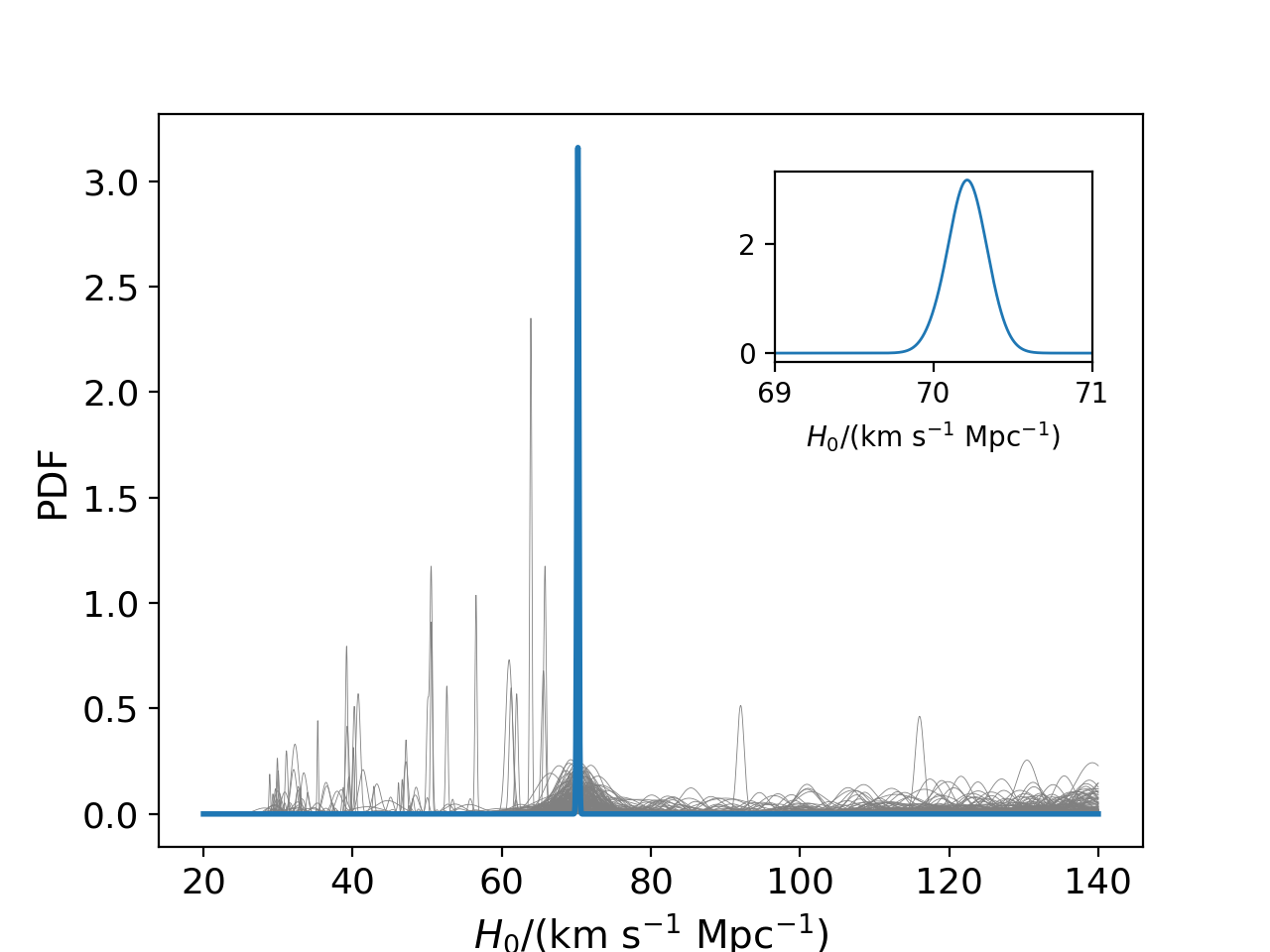}
}
\hspace{1.0pt}
\subfigure[BNS, ms1, $\mathbf{Cov}(\alpha,\delta,z)$]{
	\includegraphics[width=8.5cm]{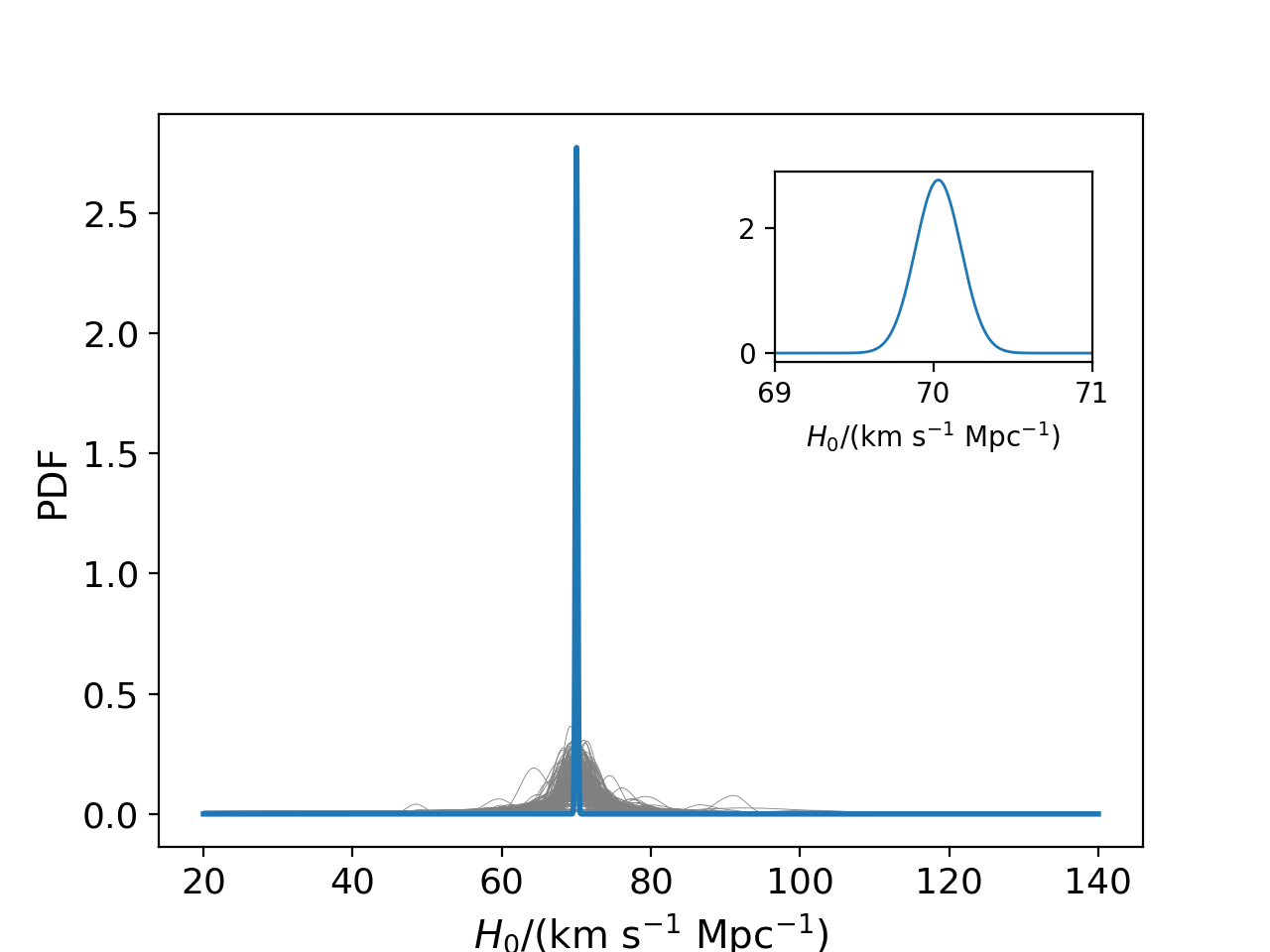}
}
\hspace{1.0pt}
\subfigure[NSBH, ms1, $\mathbf{Cov}(\alpha,\delta,z)$]{
	\includegraphics[width=8.5cm]{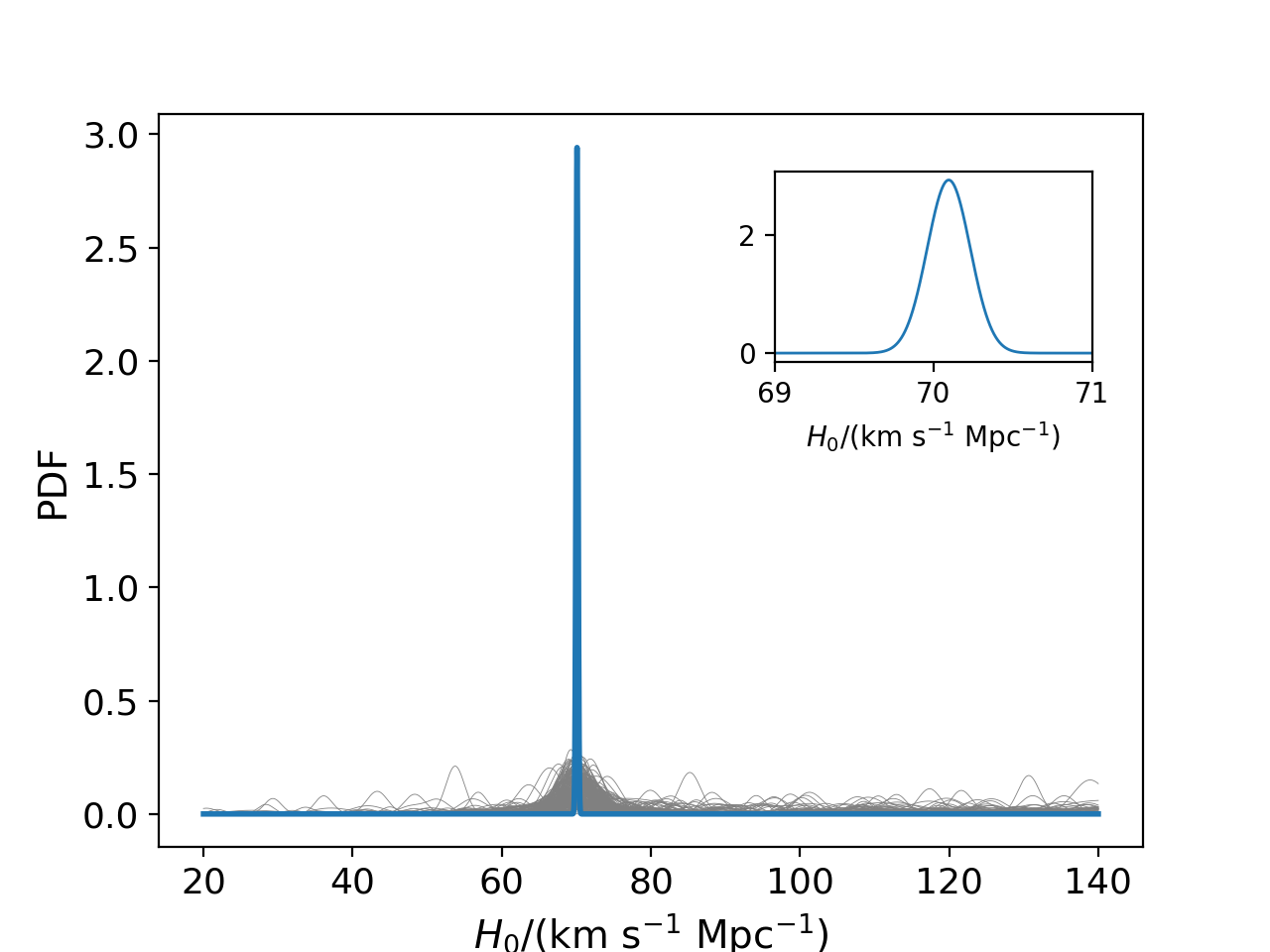}
}
\subfigure[BNS, wff2, $\mathbf{Cov}(\alpha,\delta,z)$]{
	\includegraphics[width=8.5cm]{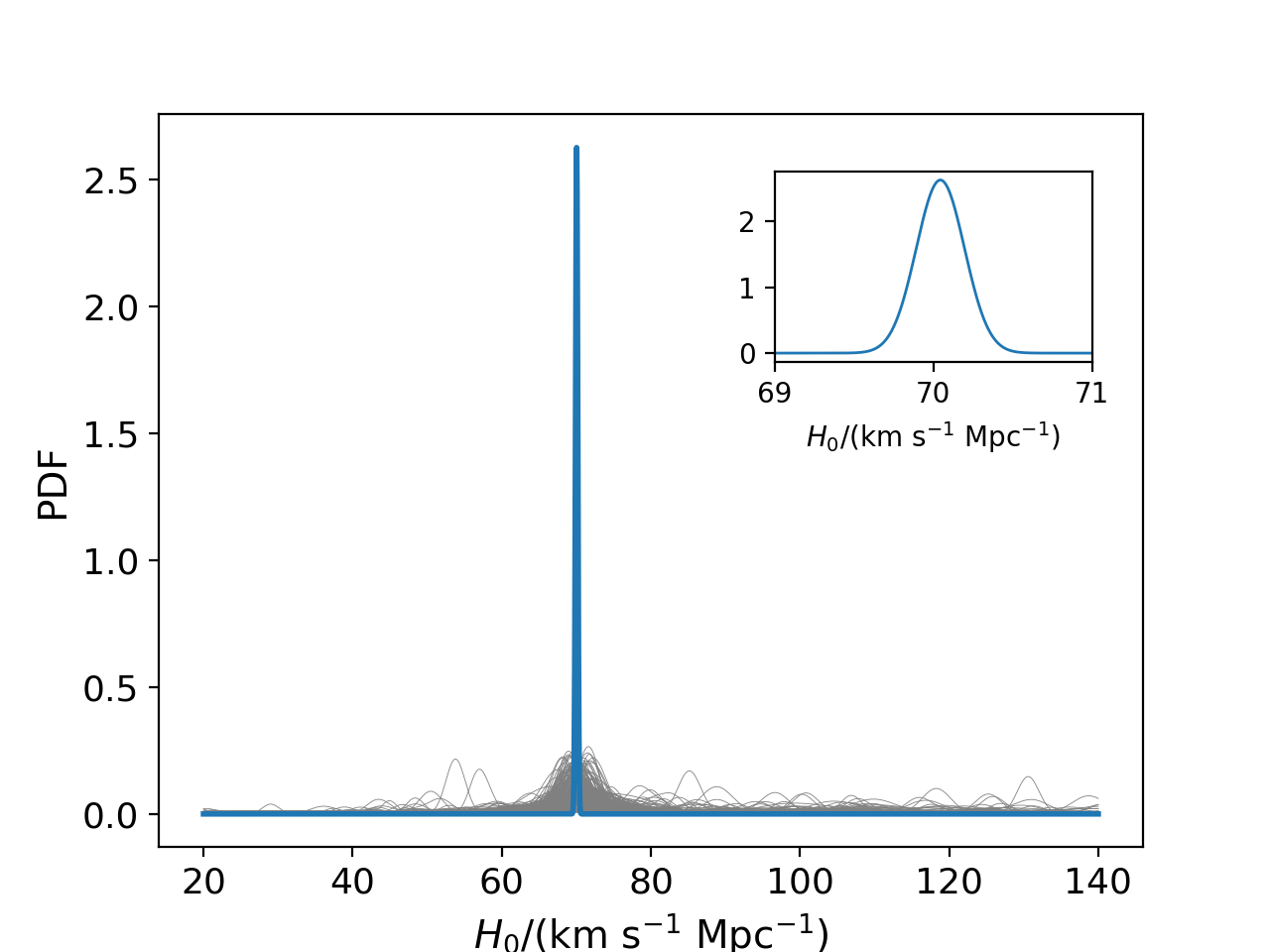}
}
\hspace{1.0pt}
\subfigure[NSBH, wff2, $\mathbf{Cov}(\alpha,\delta,z)$]{
	\includegraphics[width=8.5cm]{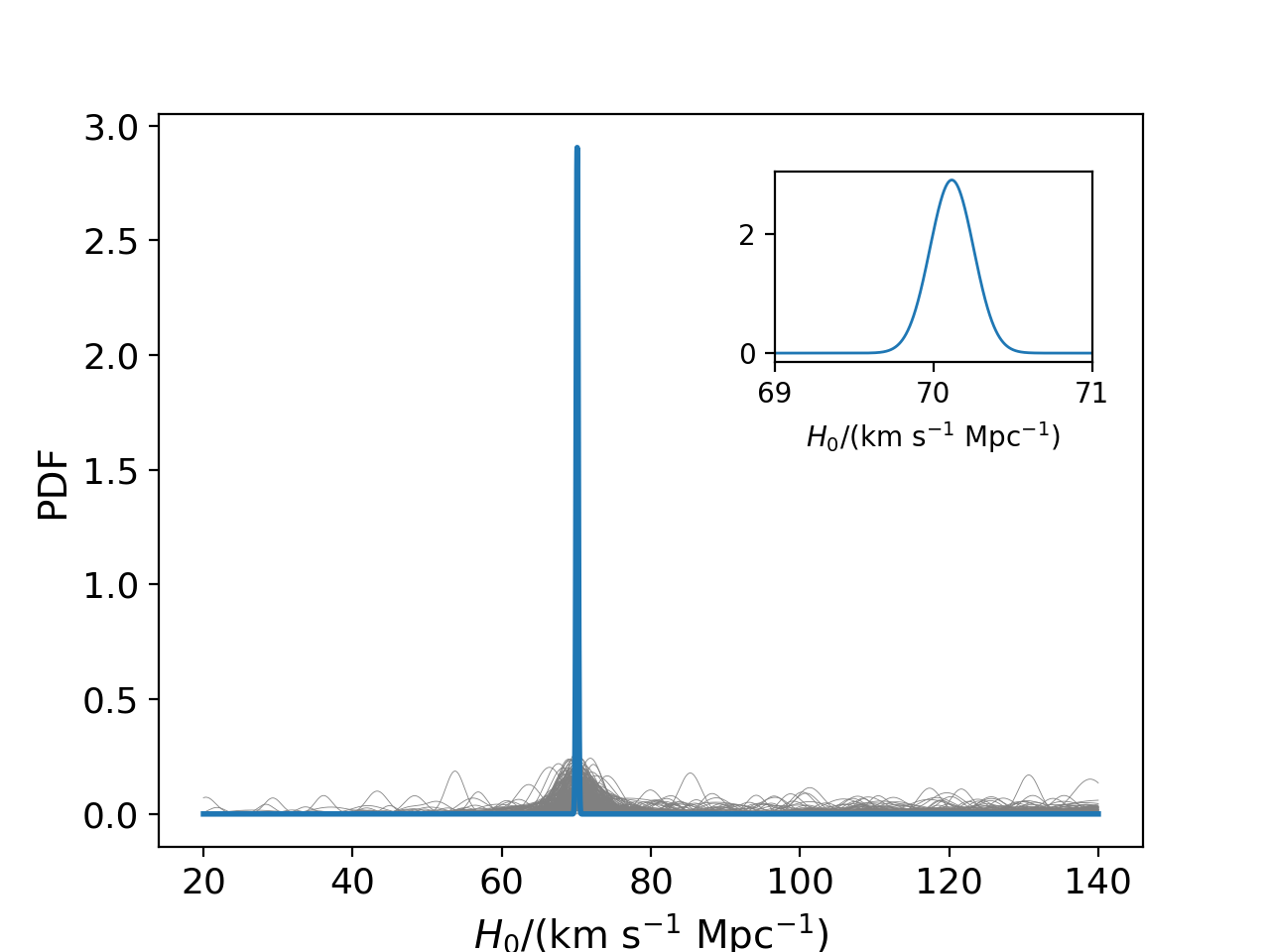}
}
\caption{The same with Figure \ref{figure_LH0}, but with $w_i\propto M_\mathrm{s}$.}
\label{figure_LH0_ms}
\end{figure*}

To investigate the effect of the weighting factor, similar to our previous work \cite{2020MNRAS.498.1786Y}, we consider a simple case where $w_i$ is proportional to group's stellar mass $M_\mathrm{s}$. We resample 200 BNS/NSBH mergers and repeat above calculations with $w_{i}\propto M_\mathrm{s}$. In addition, for comparison, we calculate the results for $w=1$ with the same samples. The results are showed in Figure \ref{figure_LH0_ms} and Table \ref{table_H0_2}. It can be seen that the choice of two different weights has almost no effect on the error bar and only slightly changes the peak of the BNS mergers' constraints. This result is quite different from that of \cite{2020PhRvD.101l2001G}. The main reason for this is that they discussed the 2G detector array, whereas we mainly focus on the 3G arrays. There is a several-order-of-magnitude difference between the localization volumes and the number of potential host galaxies/groups for GW events detected by the 2G and 3G arrays \citep{2020MNRAS.498.1786Y}. When there are many potential host galaxies/groups, an incorrect binary population may cause a large bias in the posterior of the $H_0$. \par
However, in our work, since most of the mergers have a small number of probable host groups, the contribution of these potential groups to the posterior of $H_0$ appears as a stacking of one or several isolated peaks, with the peak around $H_0=70\ \mathrm{km}\ \mathrm{s}^{-1}\ \mathrm{Mpc}^{-1}$ corresponding to the `true' host group, and the values of $w_i$ affecting the relative height of these peaks. For those `golden' dark sirens with $N_\mathrm{in}=1$, the choice of $w_i$ naturally has no influence, and due to the accuracy of the CE2ET array, they can provide very narrow constraints around $H_0=70\ \mathrm{km}\ \mathrm{s}^{-1}\ \mathrm{Mpc}^{-1}$. For other cases, the constraints given by those `golden' dark sirens are equivalent to adding an extremely narrow prior to them around $H_0=70\ \mathrm{km}\ \mathrm{s}^{-1}\ \mathrm{Mpc}^{-1}$. Therefore the peaks from `fake' host groups, which correspond to different $H_0$, would be strongly suppressed or even eliminated. Since $w_i$ only affects the relative heights between each peak, it is a very minor quantity compared to the `prior' produce by other samples. Therefore, we find roughly the same results for $w_i \propto M_\mathrm{s}$ as for $w_i = 1$.

\begin{table}
\begin{center}
\begin{tabular}{cccc}
\hline\hline
&&$w_i\propto M_\mathrm{s}$&$w_i=1$\\
&&$H_0$($\mathrm{km}\ \mathrm{s}^{-1}\ \mathrm{Mpc}^{-1}$)&$H_0$($\mathrm{km}\ \mathrm{s}^{-1}\ \mathrm{Mpc}^{-1}$)\\
\hline
BNS&$\mathbf{Cov}[\alpha, \delta, \log(d_L)]$&$70.17_{-0.15}^{+0.15}$ $\left(_{-0.11}^{+0.11}\right)$&$70.14_{-0.15}^{+0.15}$ $\left(_{-0.11}^{+0.12}\right)$\\\\
&$\mathbf{Cov}[\alpha, \delta, z]$, ms1&$70.03_{-0.14}^{+0.15}$ $\left(_{-0.11}^{+0.11}\right)$&$70.02_{-0.14}^{+0.14}$ $\left(_{-0.11}^{+0.11}\right)$\\\\
&$\mathbf{Cov}[\alpha, \delta, z]$, wff2&$70.04_{-0.15}^{+0.16}$ $\left(_{-0.12}^{+0.12}\right)$&$70.03_{-0.15}^{+0.15}$ $\left(_{-0.12}^{+0.12}\right)$\\\\
NSBH&$\mathbf{Cov}[\alpha, \delta, \log(d_L)]$&$70.21_{-0.13}^{+0.13}$ $\left(_{-0.18}^{+0.18}\right)$&$70.22_{-0.13}^{+0.13}$ $\left(_{-0.18}^{+0.18}\right)$\\\\
&$\mathbf{Cov}[\alpha, \delta, z]$, ms1&$70.10_{-0.14}^{+0.14}$ $\left(_{-0.19}^{+0.19}\right)$&$70.10_{-0.14}^{+0.14}$ $\left(_{-0.19}^{+0.19}\right)$\\\\
&$\mathbf{Cov}[\alpha, \delta, z]$, wff2&$70.12_{-0.14}^{+0.14}$ $\left(_{-0.19}^{+0.20}\right)$&$70.12_{-0.14}^{+0.14}$ $\left(_{-0.19}^{+0.20}\right)$\\
\hline
\end{tabular}
\end{center}
\caption{The same with Table \ref{table_H0}, but resampled BNS/NSBH mergers with $w_i\propto M_\mathrm{s}$. The third and fourth columns represent the results when $w_i\propto M_\mathrm{s}$ and $w_i=1$ weights are used in the calculation of the posterior respectively.}
\label{table_H0_2}
\end{table}

\subsection{The Gaussian likelihood approximation}

In this paper, we use the Fisher matrix method, which is based on a linear approximation around the maximum likelihood and could forecast parameter errors in GW detection very simply and quickly. Recently, the reliability of the Fisher matrix method are widely discussed \citep{2008PhRvD..77d2001V, 2013PhRvD..88h4013R, 2014PhRvD..89d2004G, 2022ApJ...935..139M, 2022ApJ...941..208I, 2023arXiv230813103G}. They found that at low SNRs, since the linear approximation no longer holds, the likelihood function deviated from the Gaussian distribution. Thus the results obtained by Fisher's method are largely biased. However, when the SNR is very large (SNR $>25$), due to the central limit theorem, these works found that the Fisher matrix and the Markov chain Monte Carlo (MCMC) methods have nealy the same estimations. For GW events with $z<0.1$ in this section, their SNRs are much larger than 25, which guarantees the reliability of the localization area estimations. Therefore, to simplify the discussion, we consider that the results obtained by the Fisher matrix are the same as those of the MCMC approach. 

\begin{figure}[htbp]
\centering
\includegraphics[width=9cm]{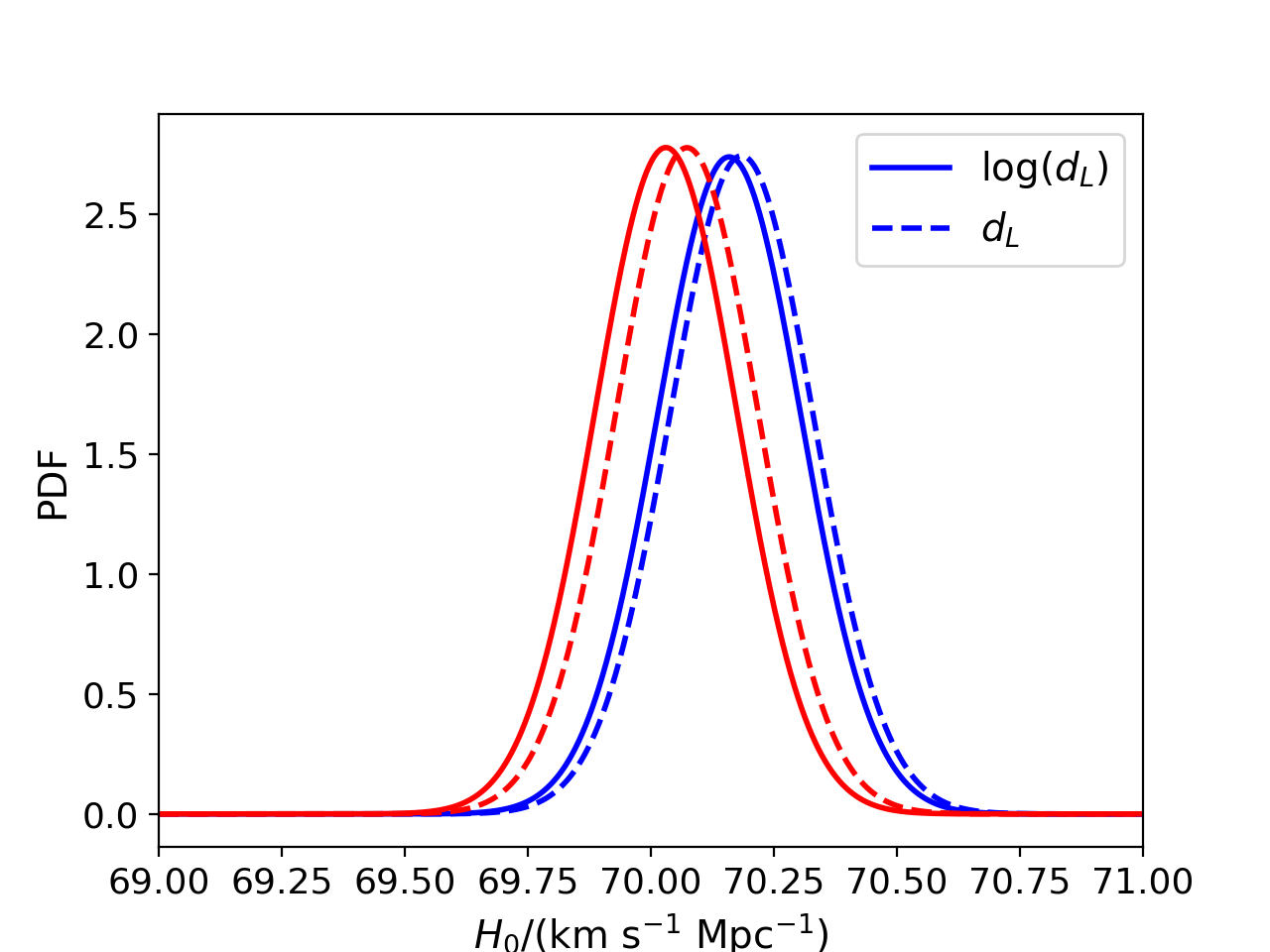}
\caption{The comparison of the results with two different likelihood functions. The blue, red lines represent the results from the methods of $\mathrm{Cov}(\alpha,\delta,\log(d_L))$ and $\mathrm{Cov}(\alpha,\delta,z)$, respectively. The solid and dashed lines reprsent the results of Gaussian likelihood functions about $\ln(d_L)$ and $d_L$, respectively. All the constraints are obtained with ms1 model, $w_i\propto m_\mathrm{s}$, and 200 BNS mergers. }
\label{figure_dL_compare}
\end{figure}

In addition, to verify the effect of the variations in the likelihood function on the results, we make a simple test in Figure \ref{figure_dL_compare}. It can be seen that the results for two different $d_L$ likelihood functions are almost the same, which somewhat indicate the reliability of the results.

\subsection{Halo mass cutoff}
In this section, we take a $10^{12}\ M_\odot/h$ group halo mass cutoff to ensure the completeness of catalog. To simplify the discussion, we assume that BNS/NSBH merging events are generated only in these groups. However, if a merging event happens in a small group outside of the catalog, the host group will be mislocated by the dark siren method, which may bias the $H_0$ measurements. Therefore, the reliability of the results needs to be tested.

For comparison, we continue assume that the distribution of mergers is exactly proportional to the stellar mass and resample the BNS/NSBH mergers with a cutoff throughout all groups without a halo mass cutoff. This way, there is a fraction of mergers whose host groups are not in the catalog. For a $10^{12}\ M_\odot/h$ group halo mass cutoff, this fraction is about 30\%. First, to test the effect of mislocalization on the results, we extend the localization region to ensure that all mergers have at least one potential host. Due to these false localized events, the $H_0$ measurements of 200 BNS tidal effect dark sirens become $70.65^{+0.15}_{-0.15}$ and $70.79^{+0.16}_{-0.16}\ \mathrm{km}\ \mathrm{s}^{-1}\ \mathrm{Mpc}^{-1}$, for ms1 and wff2 model, respectively. It can be seen that the bias has exceeded 4 confidence intervals CIs. The reason is that when the host group of the BNS mergers is not in the catalog, its redshift measures will be determined by one or a few nearby large groups. Since the high redshift region has larger comoving volume, those `false' host groups are more likely to occur at higher redshift, which lead to significantly larger results. And for NSBH systems, the biases are also lager than 4CIs.

Therefore, it is important to correct the bias caused by incomplete catalog. One method is to take use of the localization accuracy of the 3G detector network to exclude a part of mergers whose host groups are outside the catalog. For BNS mergers in the ms1 model, choosing a 90\% localization volume cutoff can correct the $H_0$ measurements to $70.12^{+0.17}_{-0.15}\ \mathrm{km}\ \mathrm{s}^{-1}\ \mathrm{Mpc}^{-1}$. However, in the case of wff2 model and NSBH mergers, the bias is almost unchanged due to the poor constraints in the redshift direction.

Another way is to complete the missing groups in the catalog. To account for the effects of these missing groups, we modify the EM likelihood term $p(d_\mathrm{EM}|z,\Omega)$ by adding a completeness factor $f_\mathrm{comp}$ \citep{2020PhRvD.101l2001G},
\begin{equation}
\begin{split}
    p(d_\mathrm{EM}|z,\Omega)=&f_\mathrm{comp}p(d_\mathrm{EM}|G, z,\Omega)\\
    +&(1-f_\mathrm{comp}) p(d_\mathrm{EM}|\bar{G}, z,\Omega),
\end{split}
\end{equation}
where $G$ and $\bar{G}$ represent the host group is, and is not in the catalog, respectively. The first term can be obtained from Eq. (\ref{eq_emlikelihood}). As for the latter term, it represents the contribution of these missing groups to the EM likelihood. Generally, this term is associated with the large-scale structure of the universe and the population of GW events. For simplicity, we neglect these effects and assume that $p(d_\mathrm{EM}|\bar{G}, z,\Omega)$ is uniformly distributed.

Since groups with $M_\mathrm{halo} > 10^{12}\ M_\odot/h$ occupy about 70\% stellar mass, we set $f_\mathrm{comp}=0.7$. For 200 BNS mergers with ms1/wff2 model, the results are $69.98^{+0.18}_{-0.18}/70.01^{+0.19}_{-0.19}\ \mathrm{km}\ \mathrm{s}^{-1}\ \mathrm{Mpc}^{-1}$. It can be seen that after considering the contribution of these missing groups, the bias is nearly completely eliminated. In addition, since a flatter component is added to the redshift PDF for each merger, the final constraints on $H_0$ become slightly looser. For NSBH mergers, the results are $70.20^{+0.16}_{-0.16}/70.20^{+0.17}_{-0.17}\ \mathrm{km}\ \mathrm{s}^{-1}\ \mathrm{Mpc}^{-1}$. The difference between the results and the fiducial value is still less than 1.5 CIs.

Since the completeness factor relates to the fraction of stellar mass contained in the catalog, for comparison, we relax the halo mass cutoff to $10^{11.5}\ M_\odot/h$, which makes the new catalog contain about 90\% of the stellar mass. So with $f_\mathrm{comp}=0.9$, the constraints become $70.07^{+0.16}_{-0.16}/70.08^{+0.17}_{-0.17}\ \mathrm{km}\ \mathrm{s}^{-1}\ \mathrm{Mpc}^{-1}$ and $70.21^{+0.15}_{-0.14}/70.20^{+0.15}_{-0.14}\ \mathrm{km}\ \mathrm{s}^{-1}\ \mathrm{Mpc}^{-1}$, for BNS and NSBH mergers, respectively. It can be seen that with more groups contained in the catalog, the missing groups contribute much less to the EM likelihood, and therefore the constraints are tighter compared with the case with a $10^{12}\ M_\odot/h$ halo mass cutoff.

In brief, although in the results of Table \ref{table_H0} and \ref{table_H0_2}, we neglected the mergers produced by small groups, which may lead to a bias in the measurement of $H_0$, this bias can be well corrected by simply including the contribution of this part of missing groups in the EM likelihood. After adding these corrections, the mergers from the missing groups only slightly increase the error bar. Overall, a halo mass cutoff will not cause the results to be bias.



\section{conclusion}
\label{section_conclusion}
The detection of GW signals opens up a brand new approach to study the universe. Since the luminosity distance of a compact binary merger can be measured directly from waveforms, when its redshift is measured by other methods, this event can be used as a standard siren for cosmological research. \par

Among the various methods of redshift measurement, one of the most straightforward methods is to search for the EM counterpart of a GW event and identify the host galaxy to obtain the redshift information. In this paper, we use the Possis package to simulate the light curves of kilonova counterparts of low redshift BNS/NSBH mergers, then simulate multi-messenger observations of these samples by the 2G GW array together with WFST. For BNS mergers, WFST is expected to observe about 11 kilonovae per year by follow-up observations of the triggered signal from the LHV array. Over a five-year observation duration, these BNS bright sirens can constrain the Hubble constant to $\Delta H_0\sim 3\ \mathrm{km}\ \mathrm{s}^{-1}\ \mathrm{Mpc}^{-1}$. As for the LHVKI array, since it can detect more BNS mergers and has higher localization accuracy, the multi-messenger detection rate of WFST is increased to about 18 per year, and the five-year observation time constraints are $\Delta H_0\sim 2\ \mathrm{km}\ \mathrm{s}^{-1}\ \mathrm{Mpc}^{-1}$.\par

For NSBH mergers, its ejecta mass is mainly related to the relative size of the tidal disrupted radius and the ISCO radius, and is therefore strongly influenced by the mass ratio, NS EOS, and $\chi_\mathrm{BH}$. Under the assumption that BH has high spin ($\chi_\mathrm{BH}\sim\mathcal{N}[0.85, 0.15^2]$) as well as the NS EOS is the stiff ms1 model, NSBH mergers will have larger tidal disrupted radius, so the multi-messenger observation rates can reach $\sim7$ and $\sim13$ per year for the LHV and LHVKI arrays, respectively. For a middle EOS bsk21, these two rates reduce to $\sim5$ and $\sim 8$ per year. For the softest EOS mentioned in the paper, wff2, the multi-messenger detection rates are only $\sim2$ and $\sim 4$ per year left. For these three EOSs, the five-year observation duration constraints are $\Delta H_0\sim 2.1,\ 2.8,\ 3.9\ \mathrm{km}\ \mathrm{s}^{-1}\ \mathrm{Mpc}^{-1}$ with LHV, and $\Delta H_0\sim 4.0,\ 5.4,\ 8.2\ \mathrm{km}\ \mathrm{s}^{-1}\ \mathrm{Mpc}^{-1}$ with LHVKI, respectively. However, for the case of low BH spin, NSBH mergers can hardly produce bright kilonova emissions, thus WFST can hardly detect their EM counterparts. \par

In actual observations, most of the EM counterparts of compact binary mergers cannot be searched for due to various factors. Therefore, for these dark sirens, other methods are needed to measure the redshifts. Currently, a common method is to match the 3D localization area $[\alpha,\delta,\log(d_L)]$ with a survey catalog to identify the host galaxies/groups of the GW source. In this method, it is necessary to use a $d_L-z$ relation to convert the constraints on $d_L$ into redshift. In the second half of this paper, we analyze the capability of the 3G GW detector array to identify the BNS/NSBH mergers' host group under the $\Lambda$CDM model with fixed parameters.  {With the $[\alpha,\delta,\log(d_L)]$ constraints obtained from the Fisher matrix method, about 54\% of the BNS and $70\%$ of the NSBH host groups with $z<0.2$ can be identified directly.} However this approach relies on a specific cosmological model when transform $d_L$ to $z$. Therefore, we then discuss the feasibility of using the tidal effect of NSs at merging time to measure the redshift. Using the localization accuracy of third generation gravitational wave detector arrays and the redshift constraints obtained by measuring tidal effects, about 24\% to 45\% of the host galaxy groups of BNS mergers with $z<0.2$ can be directly identified. The difference in this fraction is caused by the EOS of the NS, where the stiffer EOS corresponds to a higher fraction. Although the results are not as good as those of the $d_L$-constrained method, this still indicates the feasibility of the method in identifying the host groups of BNS mergers. However, for NSBH mergers, the tidal effect is much weaker, so this fraction is only about 3\%-16\%.\par

It is worthwhile to note that the method of $d_L$ constraints needs to rely on cosmological model. Therefore, its measured redshift cannot be directly used for the $H_0$ measurement, but needs to be given a prior for the Hubble constant and then constrained using the Bayesian method. In this process, the error ellipsoid will be greatly magnified in the distance direction and causes the $H_0$ posterior given by a single event to exhibit multiple peaks that deviate from the actual value. The tidal effect approach, on the other hand, can give redshift constraints without relying on cosmological models. Benefiting from this, the tidally BNS and NSBH dark siren can constrain the $H_0$ to 0.2\% and 0.3\% over five years of observations, respectively, which are very close to the results obtained from the former method.\par

\cite{2017PhRvD..95d3502D} and \cite{2021PhRvD.104h3528C} predicted that ET and CE can constrain the Hubble constant to $\sim8\%$ and $\sim2\%$ by measuring $10^3$ BNS mergers' tidal effect. In this work, we consider array of 3G GW detector rather than a single interferometer. Due to the strong localization capability of the array, the accuracy of the redshift measurements can be significantly enhanced by comparing the localization volume with the galaxy group catalog. Thus our constraints reach $\sim0.2\%$ in 5 years' observation time, which are substantially better than those of \cite{2017PhRvD..95d3502D} and \cite{2021PhRvD.104h3528C}. In addition, for NSBH mergers, since their tidal effect is much weaker than that of BNS mergers, the direct measurement of redshift will be  relatively poor. The method of comparing with catalog has greatly compensated this shortcoming and made the method of measuring the redshift of NSBH dark sirens by tidal effect become feasible.\par

An additional point to note is that in our simulations we use a mock catalog, which avoids the issues that exist in real surveys such as the edge effect. For the real catalog, if a part of the localized volume of the dark siren is outside the survey edge or the completeness range, those missing galaxies/groups may bias the $H_0$ measurement, so an integration of this part needs to be done (see \cite{2019ApJ...871L..13F} for details). In particular, when the majority of the localization area is all outside the catalog range, the dark siren is not able to make an effective constraint on $H_0$ at this time. However, during the same period as the 3G GW detections (2030s), several spectroscopic experiments have been proposed to progress to Stage-5 spectroscopy for cosmology experiments, such as the MegaMapper \citep{2019BAAS...51g.229S}, the Mauna Kea Spectroscopic Explorer (MSE; \citealt{2019BAAS...51g.126M}), and SpecTel \citep{2019BAAS...51g..45E}. We expect these experiments will provide a nearly complete group catalog, which will cover the majority of stellar mass at low redshift \citep{2022arXiv220903585S}. 

\acknowledgments
We appreciate the helpful discussions with Rui Niu, Lingfeng Wang and Shangjie Jin. This work is supported by the National SKA Program of China (Grant Nos.~2022SKA0110200 and 2022SKA0110203), the National Key R\&D Program of China (Grant Nos. 2020YFC2201602, 2021YFC2203100 and 2022YFC2204602),
the National Natural Science Foundation of China (Grant Nos.~11975072, 11835009, 12273035, 11833005, 11890692, 11621303), the Fundamental Research Funds for the Central Universities under Grant No. WK2030000036 and WK3440000004, and the 111 Project  (B23042, B20019). We also acknowledge the science research grants from the China Manned Space Project with No. CMS-CSST-2021-A02.

\end{document}